\documentclass[12pt]{article}
\usepackage{a4wide}
\usepackage{epsfig}
\newcommand{\eps}{\varepsilon}
\newcommand{\slk}{/\kern-6pt k}
\newcommand{\sll}{/\kern-4pt l}
\newcommand{\slp}{p\kern-5pt/}
\newcommand{\slq}{q\kern-5.5pt/}
\newcommand{\sls}{s\kern-5.5pt/}
\newcommand{\slv}{v\kern-5pt\raise1pt\hbox{$\scriptstyle/$}\kern1pt}
\newcommand{\tr}{{\rm tr}}

\newcommand{\pfrac}[2]{\left(\frac{#1}{#2}\right)}

\newcommand{\Li}{{\rm Li}}

\newcommand{\GeV}{{\rm\,GeV}}
\newcommand{\bbbone}{\hbox{\rm 1\kern-3pt l}}
\newcommand{\Frac}[2]{{\textstyle\frac{#1}{#2}}}
\newcommand{\sla}{{\sqrt\lambda}}
\newcommand{\slap}{{\sqrt{\lambda'}}}
\newcommand{\smu}{{\sqrt{\mu_1}}}

\newcommand{\real}{\mathop{\rm Re}\nolimits}

\begin{document}

\begin{center}
{\Large\bf First order electroweak radiative corrections\\[7pt]
to the decay of the polarised $W$ boson}\\[1.3cm]
{\Large Maria Naeem and Stefan Groote}\\[12pt]
{F\"u\"usika Instituut, Tartu \"Ulikool,
W.~Ostwaldi~1, EE-50411 Tartu, Estonia}
\end{center}
\vspace{1cm}
\begin{abstract}\noindent
Analytic results for the next-to-leading order electroweak radiative
corrections to the decay of the polarised $W$ boson into a pair of heavy
quarks are presented. Finite fermion masses are taken into account throughout
this calculation, and in performing the collinear limit for the final quark
states and all light fermions taking part in this process, the importance of
mass effects is demonstrated. This gives hints for the deviation of the total
decay rate and the branching ratios into massive fermions between experimental
measurements and theoretical predictions.
\end{abstract}

\newpage

\section{Introduction}
The $W$ boson as one of the two massive gauge bosons of the electroweak
theory has been subject to intense investigations recently. Besides the mass
discrepancy with the Standard Model (SM) value, challenged by a new
interpretation of old data from 2012 by the CDF collaboration that indicate a
deviation of approximately four standard deviations in 2022~\cite{CDF:2022hxs},
there are also less profound deviations reported in Ref.~\cite{Erler:2024},
for instance in the measured decay width of the $W$ decay of
$\Gamma_W=2.137\pm 0.032\GeV$~\cite{TevatronElectroweakWorkingGroup:2010mao,%
ATLAS:2024erm}, contrasted with the SM prediction of
$\Gamma_W=2.0892\pm 0.0008\GeV$~\cite{Denner:1990tx,Denner:1991kt} that is
$1.5\sigma$ smaller than the measured result. While measurements of the
leptonic branching ratios from ATLAS and CMS are in good agreement with lepton
universality~\cite{ATLAS:2020xea,CMS:2022mhs}, older, less precise data from
LEP~2 indicate a branching ratio for $W\to\tau+\bar\nu_\tau$ that is
$2.6\sigma$ larger than the electron--muon average~\cite{ALEPH:2013dgf}. This
and the experimentally measured larger decay rate might indicate that the
difference is due to mass effects and radiative corrections to the leading
order (LO) Born term result not taken properly into account. In this work we
deal with these issues.

Pairs of $W$ bosons can be produced at lepton or hadron colliders (cf.\ e.g.\
Refs.~\cite{CMS:2022uhn,Furusato:2024ghr}). However, as the main SM channel
for the single $W$ boson is the decay process $t\to W^++X_b$, in
Refs.~\cite{Fischer:1998gsa,Fischer:2000kx,Fischer:2001gp,Do:2002ky,%
Korner:2003zq} first order quantum chromodynamics (QCD) and electroweak (EW)
radiative corrections have been calculated in order to obtain next-to-leading
order (NLO) results for the decay processes $t\rightarrow b+W^+(\uparrow)$ and
the similar process $b\rightarrow c+W^-(\uparrow)$, flipped to the process at
hand. Note that due to the left-handed $V-A$ coupling of the electroweak
interaction, the $W$ boson is polarised~\cite{CMS:2020ezf}, and the
polarisation can be analysed by looking at the angular distribution of the
decay~\cite{Mirkes:1994eb,Aguilar-Saavedra:2015yza} and can be considered in
$WW$ scattering at the LHC~\cite{Maina:2017eig,Grossi:2020orx}. NLO EW
radiative corrections to the process $W\to\ell^++\nu_\ell$ have been
calculated in Ref.~\cite{Denner:1990tx} without taking into account the
polarisation of the decaying particle. This gap was closed by a master thesis
defended at the University of Tartu in 2009~\cite{Veermae:2009}. Taking quark
masses into account and using similarities to the process
$Z(\uparrow)\rightarrow q+\bar q$~\cite{Groote:1995yc,Groote:1995ky,%
Groote:1996nc}, NLO QCD radiative corrections for the process
$W^+(\uparrow)\rightarrow Q+\bar q$ were calculated in a master thesis at the
University of Tartu in 2010~\cite{Tuvike:2010,Groote:2012xr,Groote:2013hc,%
Groote:2013xt}. In order to discern from $W$ boson physics beyond the
SM~\cite{Cao:2003yk,Takano:2003qd,CMS:2024ndg} and in rare
decays~\cite{CMS:2020oqe,Gao:2023wde} it is inevitable to improve the
precision of the SM prediction by including NLO EW radiative
corrections~\cite{Denner:2023ehn} and by taking care of effects from the
masses of the decay products. This is the aim of the present publication.

The central object of interest is the angular dependence of the decay rate
for the polarised $W$ boson, as it can be inferred by looking at
Refs.~\cite{Groote:2012xr,Groote:2013hc,Groote:2013xt}. It contains
the spin-density matrix $\rho$ that describes the polarisation of the $W$
boson, the decay functions $H^{mm'}$, named here as helicity bilinears and
calculated for the decay channel of the $W$ boson, and the polar angle
dependence on $\theta$. The azimuthal angle dependence is found in the
References~\cite{Fischer:1998gsa,Fischer:2001gp}, the decay rate
function~\cite{Groote:2013xt} is
\begin{eqnarray}\label{Wtheta}
W(\theta)&=&\sum_{m=0,\pm1}\rho_{mm}H^{mm}(\theta)
  \ =\ \sum_{m,m'=0,\pm1}\rho_{mm}\,d^1_{mm'}(\theta)
  \,d^1_{mm'}(\theta)\,\,H^{m'm'}\ =\nonumber\\
  &=&\frac38(1+\cos^2\theta)\,\Big((\rho_{++}+\rho_{--})\,
  (H^{++}+H^{--})+2\rho_{00}H^{00}\Big)+\strut\nonumber\\&&\strut
  +\frac34\cos\theta\,\Big((\rho_{++}-\rho_{--})\,
  (H^{++}-H^{--})\Big)+\strut\nonumber\\&&\strut
  +\frac34\sin^2\theta\,\Big((\rho_{++}+\rho_{--})H^{00}
  +\rho_{00}(H^{++}+H^{--}-H^{00})\Big),
\end{eqnarray}
using the normalised spin density matrix elements $\rho_{mm'}$ with
$\rho_{++}+\rho_{00}+\rho_{--}=1$ and the helicity bilinears $H^{mm'}$
normalised in such a way that $H^{++}+H^{00}+H^{--}=1$ at the Born term level
for massless fermions. The results are given in the rest frame of the $W$
boson with the original direction of flight of the $W$ boson fixed in boosting
to this rest frame, while the polar angle $\theta$ is given as the angle
between this direction and the momentum of the quark. While the spin density
matrix elements $\rho_{00}=0.687(5)$, $\rho_{++}=0.0017(1)$ and
$\rho_{--}=0.311(5)$ obtained from the production of the $W$ boson in top
quark decay are inputs for our calculation taken from
Ref.~\cite{Czarnecki:2010gb}, here we have to calculate the helicity bilinears.

The work is divided up as follows: In Sections~2 and~3 we give the Born term
results and the tree and loop contributions to the EW radiative corrections to
the decay process $W^+\to c\bar b$ of an on-shell $W$ boson into charm and
anti-bottom quarks. Though contributing only with a small branching ratio to
the total decay rate due to the smallness of the Cabibbo--Kobayashi--Maskawa
(CKM) matrix element $V_{cb}$, this particular choice is made because for
on-shell $W$ bosons the masses of the two quarks are maximal and will most
prominently show the mass effects aimed to present in this paper.
The results are incorporated in the helicity bilinears in Section~4. In
Section~5 we give analytical results for the helicity bilinears, taking
also into account detailed counter terms for the cancellation of IR
singularities. A discussion of the result in terms of the angular dependence
of the decay rate is given in Section~6. Our conclusions are found in
Section~7, while detailed calculations and formulae are given in five
appendices.

\section{Tree contributions}
For the Born term, the squared absolute value of the matrix element is given by 
\begin{eqnarray}\label{Mruut}
|{\cal M}_0|^2&=&\sum_{s_1,s_2}{\cal M}_0^*{\cal M}_0\ =\nonumber\\
  &=&\frac{e^2}{8s_W^2}|V_{cb}|^2\tr\left((\slp_2-m_2)\gamma^\mu
  (1-\gamma_5)(\slp_1+m_1)\gamma^\nu(1-\gamma_5)\right)
  \eps_\mu^*(q,\lambda)\eps_\nu(q,\lambda)\ =\nonumber\\
  &=&\frac{e^2}{s_W^2}|V_{cb}|^2\left[p_2^\mu p_1^\nu+p_2^\nu p_1^\mu-g^{\mu\nu}
  (p_2p_1)-p_{2\kappa}p_{1\lambda}i\epsilon^{\kappa\lambda\mu\nu}\right]
  \eps_\mu^*(q,\lambda)\eps_\nu(q,\lambda),
\end{eqnarray}
with $s_W=\sin\theta_W$ the sine of the Weinberg angle. $p_1$ and $p_2$ are
the momenta of the quark and antiquark, respectively, with corresponding
masses $m_1$ and $m_2$. In later steps we will be more specific in choosing
$m_1=m_c$ and $m_2=m_b$. For an unpolarised $W$ boson we sum over the
polarisations $\lambda$, and choosing the unitary gauge $\xi_W=\infty$, we
obtain~\cite{Gallagher:2020ajd}
\begin{equation}\label{unitary}
  \sum_\lambda\eps_\mu^*(q,\lambda)\eps_\nu(q,\lambda)
  =-g_{\mu\nu}+\frac{q_\mu q_\nu}{m_W^2}.
\end{equation}
On mass shell we can use $q^2=m_W^2$, where $q=p_1+p_2$ is the momentum of the
$W$ boson. As the $W$ boson is the incoming particle, instead of summing over
the (three) polarisations $\lambda=+$, $-$ and $0$ we have to calculate the
mean value over these. This gives an additional factor $1/3$ to the decay
rate, in compliance with Refs.~\cite{Denner:1990tx,Groote:2013xt}.
Rearranging $q=p_1+p_2$ accordingly and squaring, one obtains
\begin{equation}
  p_1p_2=\frac{q^2}2(1-\mu_1-\mu_2),\qquad
  p_1q=\frac{q^2}2(1+\mu_1-\mu_2),\qquad
  p_2q=\frac{q^2}2(1-\mu_1+\mu_2),
\end{equation}
where $m_1^2=p_1^2=\mu_1 q^2$ and $m_2^2=p_2^2=\mu_2 q^2$ introduces two
dimensionless mass parameters $\mu_1$ and $\mu_2$. Therefore, summing over
the polarisations of the final states one has
\begin{eqnarray}\label{M02}
  \overline{|{\cal M}_0|}^2&=&\frac{e^2|V_{cb}|^2}{3s_W^2}
  \left(p_1^\mu p_2^\nu+p_2^\mu p_1^\nu-p_1p_2g^{\mu\nu}
  +ip_{1\kappa}p_{2\lambda}\epsilon^{\kappa\lambda\mu\nu}\right)
  \left(-g_{\mu\nu}+\frac{q_\mu q_\nu}{m_W^2}\right)\ =\nonumber\\
  &=&\frac{e^2|V_{cb}|^2m_W^2}{6s_W^2}
  \left(2-\mu_1-\mu_2-(\mu_1-\mu_2)^2\right).
\end{eqnarray}
In our approach, the Weinberg angle $\theta_W$ is fixed by the boson masses,
$\cos\theta_W=m_W/m_Z$. The first line of Eq.~(\ref{M02}) can be written as
\begin{equation}
\overline{|{\cal M}_0|}^2=\frac{e^2|V_{cb}|^2q^2}{3s_W^2}
  H^{\mu\nu}({\it Born\/})\left(-g_{\mu\nu}+\frac{q_\mu q_\nu}{m_W^2}\right),
\end{equation}
where $H^{\mu\nu}({\it Born\/})$ is the Born term hadron tensor. In
calculating the helicity bilinears $H^{\lambda_1\lambda_2}$, instead of
Eq.~(\ref{unitary}) one contracts with the product
$\eps_\mu^*(q,\lambda_1)\eps_\nu(q,\lambda_2)$ and normalises the result by
dividing by the sum of the unnormalised results for
$(\lambda_1,\lambda_2)=(+,+)$, $(-,-)$, and $(0,0)$. This will be exemplified
in Sec.~4.

\begin{figure}[t]\begin{center}
  \epsfig{figure=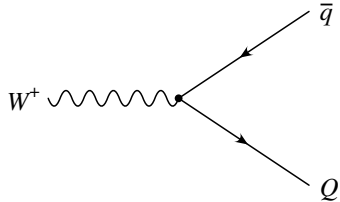, scale=0.35}\qquad
\caption{\label{wlagu0}Born-term contribution}
\end{center}
\end{figure}

\subsection{The two body decay}
For the decay rate, one has to integrate the absolute square of the matrix
element over the phase space. In order to calculate the phase space, we have
to make explicit the kinematics of the decay in the rest frame of the $W$
boson. In the two-body decay of the $W$ boson into up-type quark $Q(p_1)$ and
down-type antiquark $\bar q(p_2)$, the two quarks are produced back to back.
Here we can choose two frames. In the {\em quark frame\/} we take the positive
$z$ axis to be the direction of flight of the charm quark. One obtains
\begin{equation}
p_1=(E_1;0,0,|\vec p\,|),\qquad p_2=(E_2;0,0,-|\vec p\,|),
\end{equation}
where $p_1+p_2=q=(\sqrt{q^2};0,0,0)$ and
\begin{equation}
E_1=\frac12(1+\mu_1-\mu_2)\sqrt{q^2},\quad
E_2=\frac12(1-\mu_1+\mu_2)\sqrt{q^2},\quad
|\vec p\,|=\frac12\sqrt{\lambda(1,\mu_1,\mu_2)}\sqrt{q^2},
\end{equation}
with the K\"all\'en function given by $\lambda(a,b,c):=a^2+b^2+c^2-2ab-2ac-2bc$.
Taking into account that the axis of flight of the $W$ boson produced in e.g.\
the dominant process $t\to W^++b$, this defines another axis. Considering the
polarisation of the $W$ boson as important parameter for our analysis, the
polar angle $\theta$ between the quark frame and the axis of flight of the $W$
is of importance here. The rest frame of the $W$ boson with the direction of
flight pointing in positive $z$ direction is called the {\em $W$ frame\/} in
the following. The two-particle phase space depends on the polar angle and is
given by
\begin{equation}\label{dPS2}
  dPS_2=\frac1{16\pi q^2}\sqrt{\lambda(q^2,m_1^2,m_2^2)}\ d(\cos\theta)
  =\frac1{16\pi}\sqrt{\lambda(1,\mu_1,\mu_2)}\ d(\cos\theta).
\end{equation}
According to Fermi's golden rule, for the calculation of the decay rate we
have to combine the squared absolute value of the matrix element and the phase
space factor to obtain
\begin{equation}
  d\Gamma=\frac1{2m_W}|{\cal M}|^2dPS.
\end{equation}
After integration over the polar angle, the full unpolarised Born term decay
rate for the process shown in Fig.~\ref{wlagu0} with an on-shell $W$ boson
($q^2=m_W^2$) is given by
\begin{equation}\label{Gamma0}
  \Gamma_0=\frac{e^2m_W}{96\pi s_W^2}N_c|V_{cb}|^2
  \sqrt{\lambda(1,\mu_1,\mu_2)}\left(2-\mu_1-\mu_2-(\mu_1-\mu_2)^2\right).
\end{equation}

\begin{figure}[t]\begin{center}
  \epsfig{figure=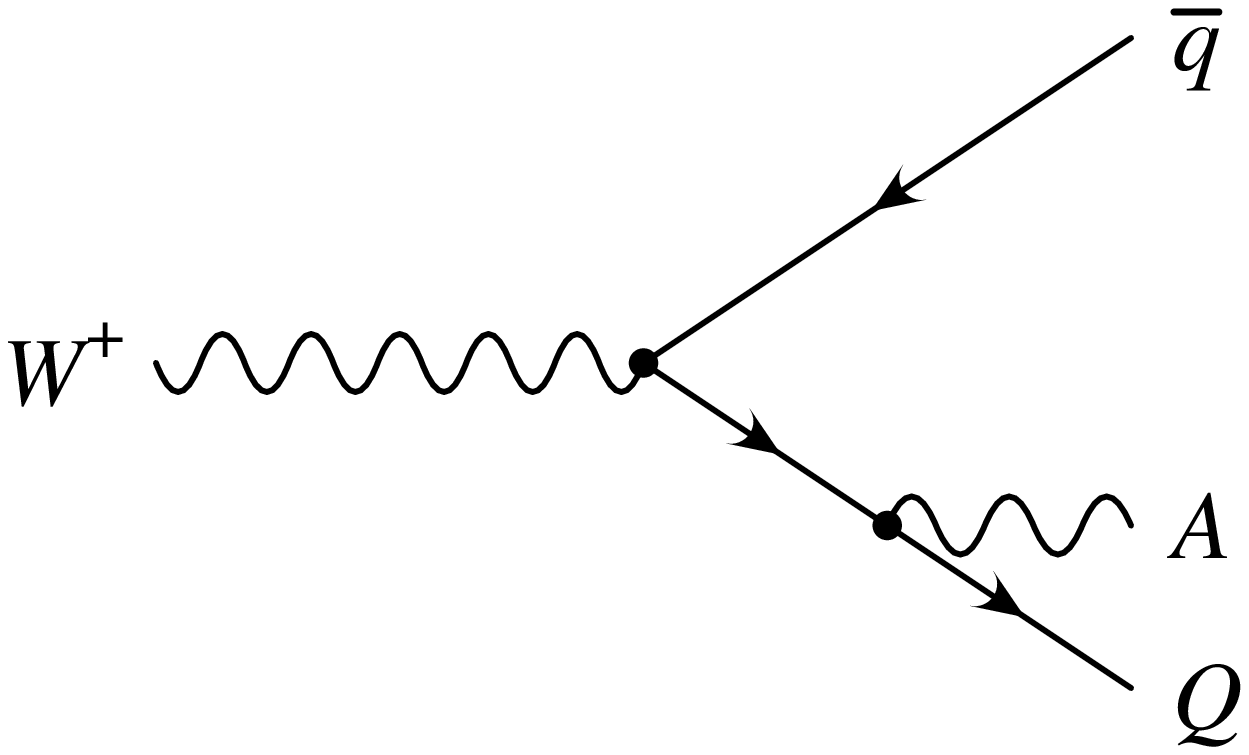, scale=0.35}\quad
  \epsfig{figure=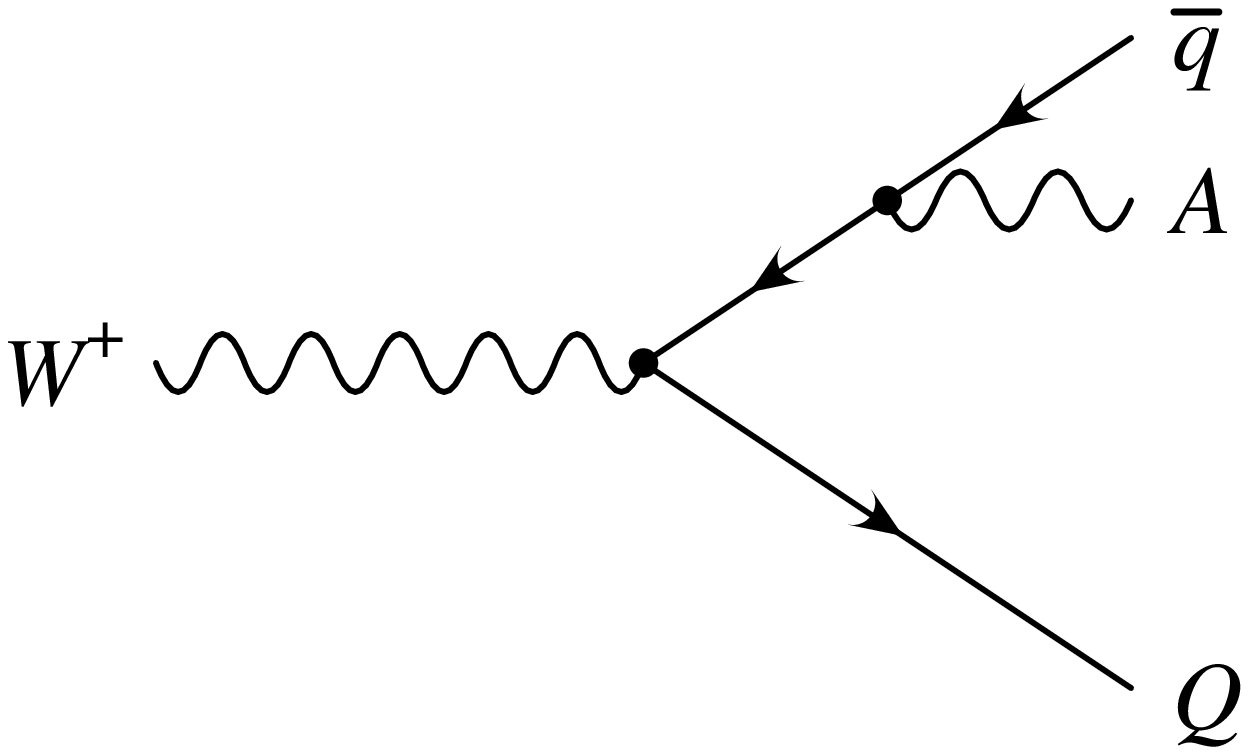, scale=0.35}\quad
  \epsfig{figure=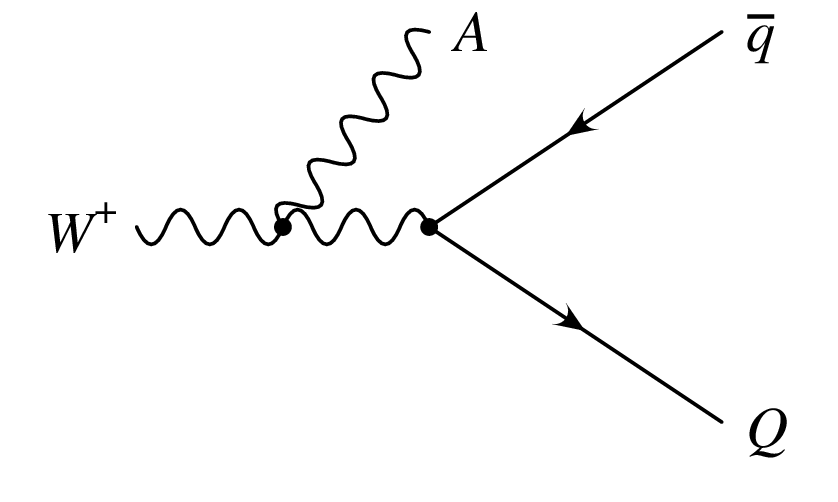, scale=0.35}\\
  (1)\kern137pt(2)\kern137pt(3)
  \caption{\label{wlagu}First order electroweak tree corrections}
\end{center}\end{figure}

\subsection{The three body decay}
\begin{figure}[ht]\begin{center}
\epsfig{figure=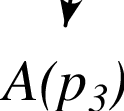, scale=0.6}
\caption{\label{theta2b}Kinematics of the three (and two) particle decay}
\end{center}\end{figure}

In this work we are dealing with NLO electroweak corrections. Taking into
account the tree corrections with a real photon emitted, we end up with the
three body decays shown in Fig.~\ref{wlagu}, with the kinematics shown in
Fig.~\ref{theta2b}. For the phase space integration it would be appropriate to
have the denominator factors in the easiest shape. For this we define
dimensionless quantities $y_1$ and $y_2$ by
\begin{eqnarray}
  y_1q^2&:=&(p_1+p_3)^2-m_1^2\ =\ p_1^2+2p_1p_3+p_3^2-m_1^2\ =\ 2p_1p_3+p_3^2,
  \nonumber\\
  y_2q^2&:=&(p_2+p_3)^2-m_2^2\ =\ p_2^2+2p_2p_3+p_3^2-m_2^2\ =\ 2p_2p_3+p_3^2
\end{eqnarray}
with $q^2=(p_1+p_2+p_3)^2=p_1^2+2p_1p_2+2p_1p_3+p_2^2+2p_2p_3+p_3^2$. As the
integration over the phase space will result in infrared (IR) divergencies,
we use mass regularisation by giving a small mass $m_A^2=p_3^2=\Lambda q^2$ to
the emitted photon with momentum $p_3$. The scalar products read
\begin{equation}
p_1p_2=\frac12\left(1-(\mu_1+y_1)-(\mu_2+y_2)+\Lambda\right)q^2,\quad
p_1p_3=\frac12(y_1-\Lambda)q^2,\quad
p_2p_3=\frac12(y_2-\Lambda)q^2.
\end{equation}
The general ansatz for the kinematics in the quark frame is given by
$p_1=(E_1;0,0,|\vec p_1|)$,
\begin{equation}
p_2=(E_2;|\vec p_2|\sin\theta_{12},0,|\vec p_2|\cos\theta_{12}),\qquad
p_3=(E_3;|\vec p_3|\sin\theta_{13},0,|\vec p_3|\cos\theta_{13})
\end{equation}
with
\begin{eqnarray}
  E_1=\frac12\left(1+\mu_1-(\mu_2+y_2)\right)\sqrt{q^2}&&
  |\vec p_1|=\frac12\sqrt{\lambda(1,\mu_1,\mu_2+y_2)}\sqrt{q^2}\nonumber\\
  E_2=\frac12\left(1-(\mu_1+y_1)+\mu_2\right)\sqrt{q^2}&&
  |\vec p_2|=\frac12\sqrt{\lambda(1,\mu_1+y_1,\mu_2)}\sqrt{q^2}\nonumber\\
  E_3=\frac12(y_1+y_2)\sqrt{q^2}&&
  |\vec p_3|=\frac12\sqrt{(y_1+y_2)^2-4\Lambda}\sqrt{q^2}
\end{eqnarray}
and the cosines and sines of the relative angles given by
\begin{eqnarray}
\cos\theta_{12}&=&\frac{y_1y_2+(1-\mu_1+\mu_2)y_1+(1+\mu_1-\mu_2)y_2-\lambda
  -2\Lambda}{\sqrt{\lambda(1,\mu_1,\mu_2+y_2)}
  \sqrt{\lambda(1,\mu_1+y_1,\mu_2)}},\nonumber\\
\sin\theta_{12}&=&\frac{2\sqrt{N(y_1,y_2)}}{\sqrt{\lambda(1,\mu_1,\mu_2+y_2)}
  \sqrt{\lambda(1,\mu_1+y_1,\mu_2)}},\nonumber\\
\cos\theta_{13}&=&\frac{-y_1(1-\mu_1+\mu_2+y_2)+(1+\mu_1-\mu_2-y_2)y_2
  +2\Lambda}{\sqrt{\lambda(1,\mu_1,\mu_2+y_2)}\sqrt{(y_1+y_2)^2-4\Lambda}},
  \nonumber\\
\sin\theta_{13}&=&\frac{-2\sqrt{N(y_1,y_2)}}{\sqrt{\lambda(1,\mu_1,\mu_2+y_2)}
  \sqrt{(y_1+y_2)^2-4\Lambda}},
\end{eqnarray}
with
\begin{eqnarray}
N(y_1,y_2)&=&(1-\mu_1-\mu_2)y_1y_2-y_1^2(\mu_2+y_2)-(\mu_1+y_1)y_2^2
  +\strut\nonumber\\&&\strut
  +\Lambda\left((1-\mu_1+\mu_2)y_1+(1+\mu_1-\mu_2)y_2+y_1y_2-\lambda\right)
  -\Lambda^2.
\end{eqnarray}
The three-body phase space is given by
\begin{equation}
dPS_3=(2\pi)^4\delta^{(4)}(p_1+p_2+p_3-q)\prod_{i=1}^3\frac{d^4p_i}{(2\pi)^4}
  (2\pi)\delta(p_i^2-m_i^2)\theta(p_i^0).
\end{equation}
which in the quark frame simplifies to
\begin{equation}
dPS_3=dPS_2\times\frac{q^2}{(4\pi)^2\sqrt{\lambda(1,\mu_1,\mu_2)}}dy_1dy_2,
\end{equation}
The phase limits in terms of $y_1$ and $y_2$ read
$y_{1-}(y_2)\le y_1\le y_{1+}(y_2)$ and
\begin{equation}
y_{20}:=\Lambda+2\sqrt{\Lambda\mu_2}\le y_2\le (1-\smu)^2-\mu_2=:y_{2-}.
\end{equation}
with
\begin{eqnarray}
  y_{1\pm}(y_2)&=&\frac1{2(\mu_2+y_2)}\Big(y_2\left(1-\mu_1-(\mu_2+y_2)\right)
  +\Lambda\left(1-\mu_1+(\mu_2+y_2)\right)+\strut\nonumber\\&&\qquad\strut
  \pm\sqrt{(y_2-\Lambda)^2-4\Lambda\mu_2}
  \sqrt{\lambda(1,\mu_1,\mu_2+y_2)}\Big).
\end{eqnarray}
The IR singularity resides in the phase space at $y_1=y_2=0$ and is
regularised by the parameter $\Lambda$. Details about the calculation of the
phase space integrals and the basic terms contained in the tree contributions
are found in Appendix~A.

\begin{figure}\begin{center}
  \epsfig{figure=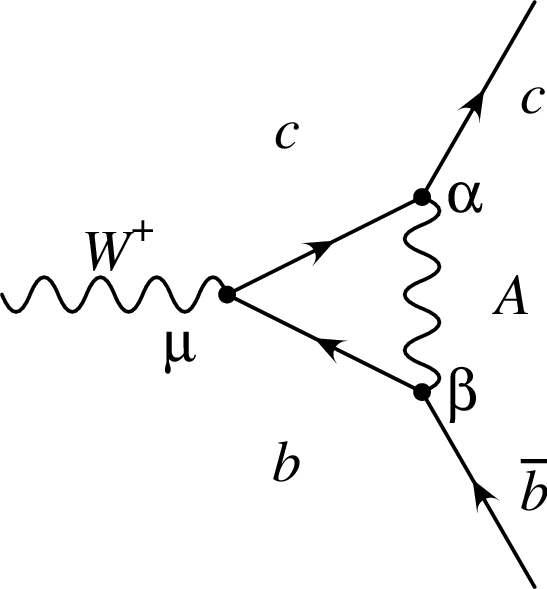, scale=0.3}\qquad
  \epsfig{figure=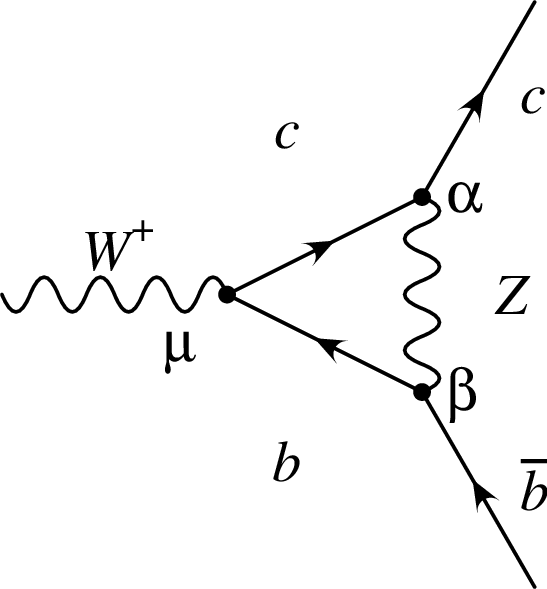,scale=0.3}\qquad
  \epsfig{figure=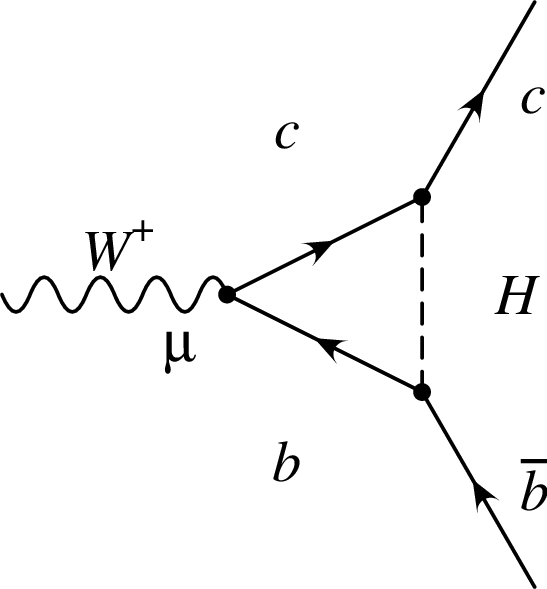, scale=0.3}\qquad
  \epsfig{figure=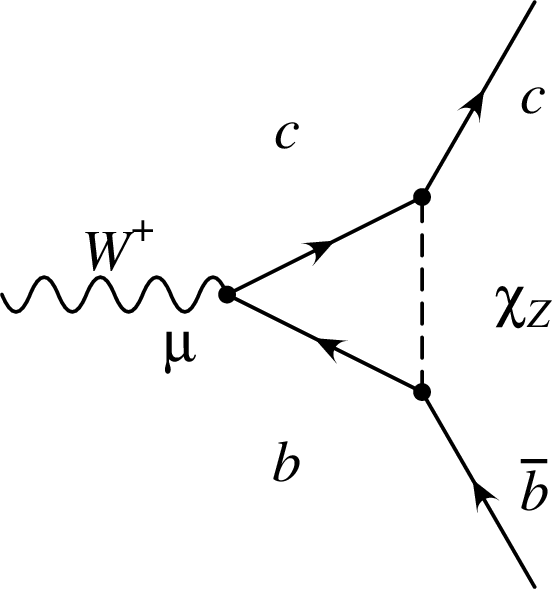, scale=0.3}\\[9pt]
  \epsfig{figure=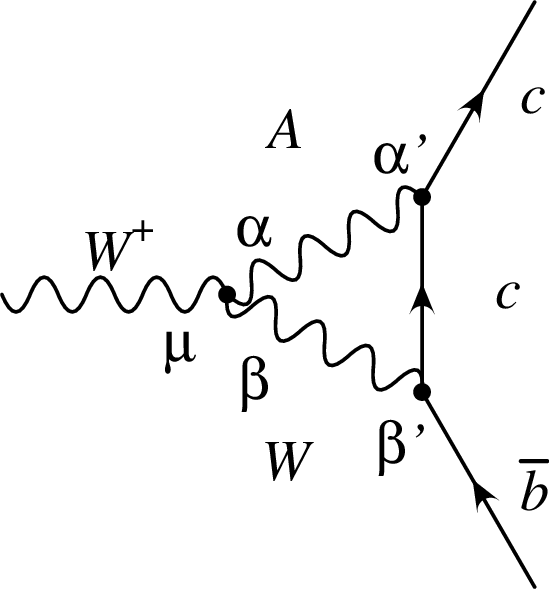, scale=0.3}\qquad
  \epsfig{figure=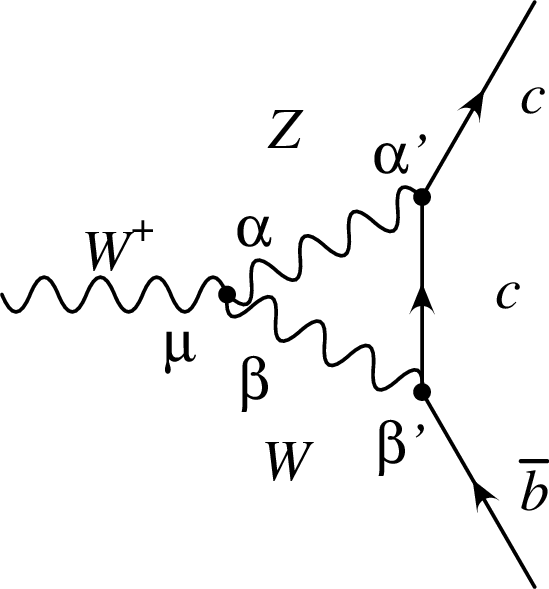, scale=0.3}\qquad
  \epsfig{figure=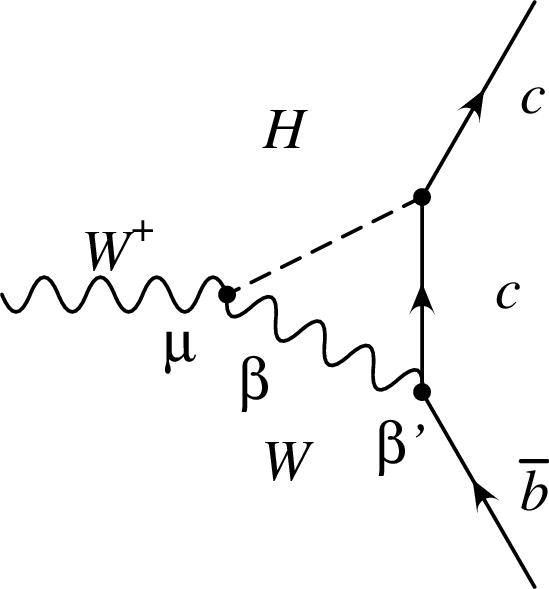, scale=0.3}\qquad
  \epsfig{figure=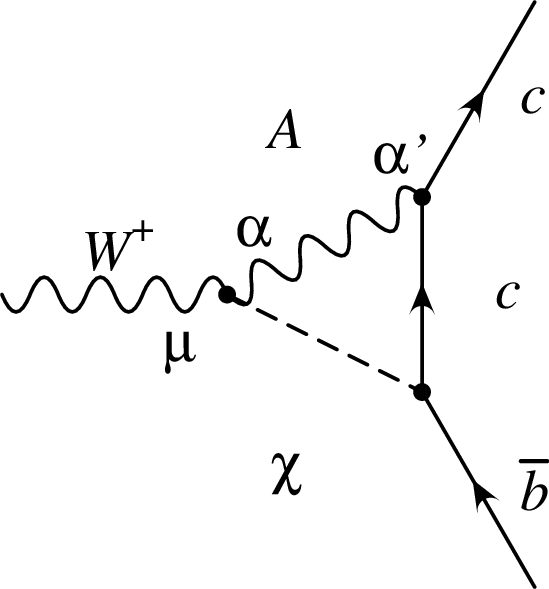, scale=0.3}\\[9pt]
  \epsfig{figure=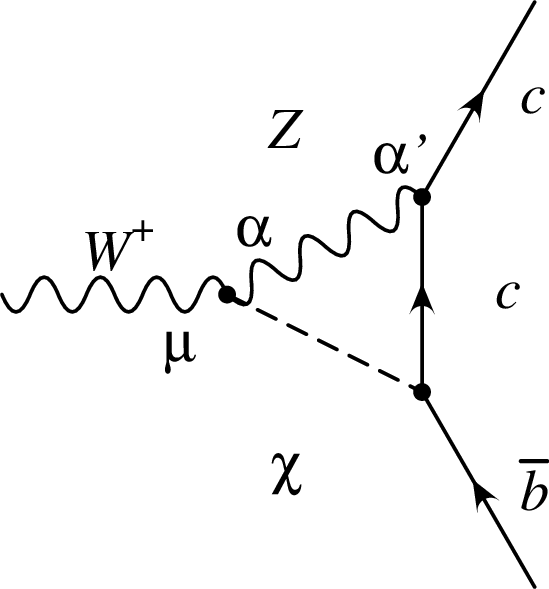, scale=0.3}\qquad
  \epsfig{figure=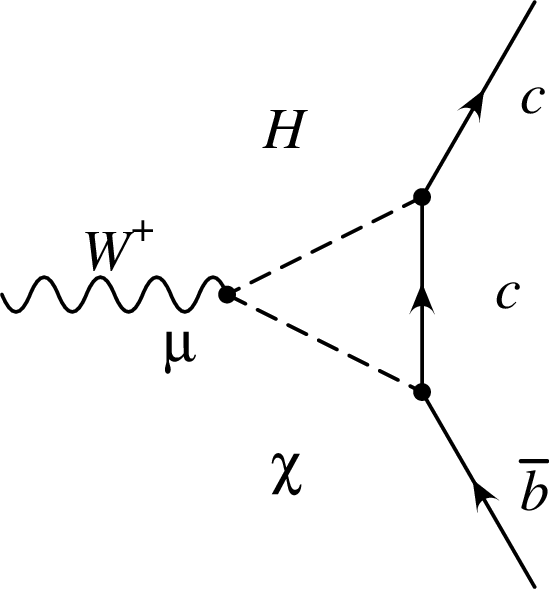, scale=0.3}\qquad
  \epsfig{figure=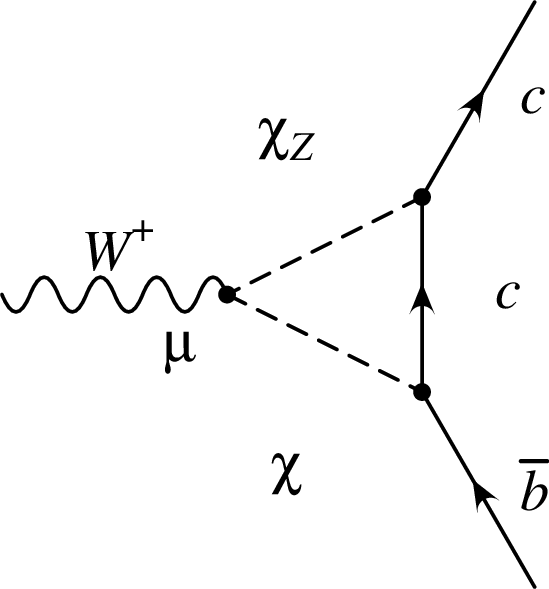, scale=0.3}\qquad
  \epsfig{figure=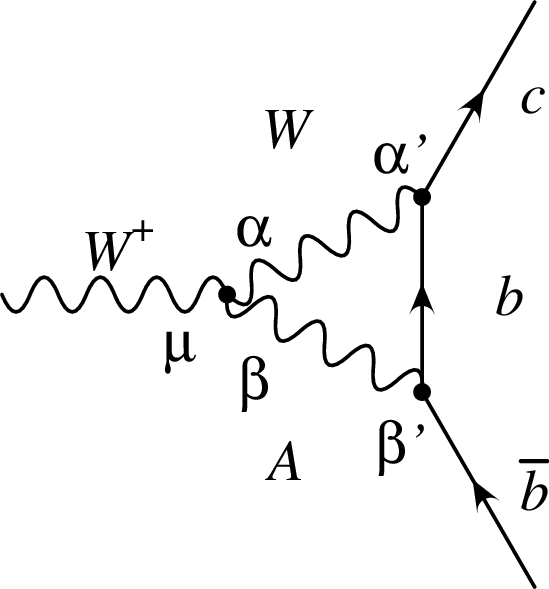, scale=0.3}\\[9pt]
  \epsfig{figure=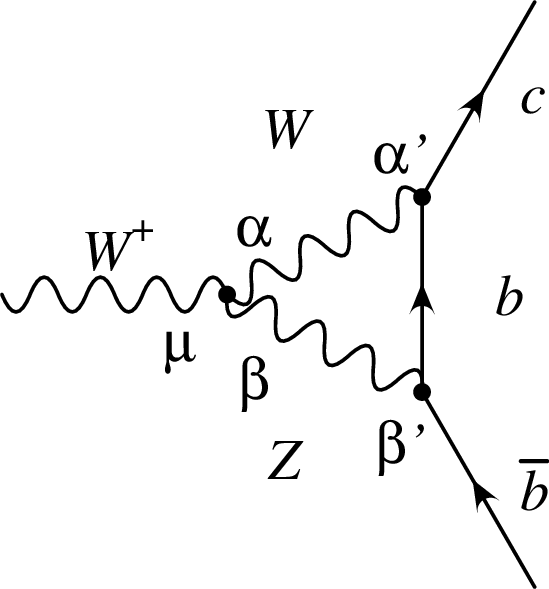, scale=0.3}\qquad
  \epsfig{figure=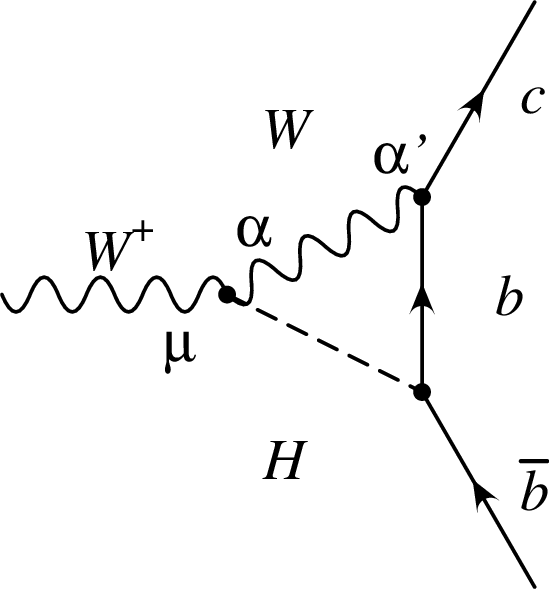, scale=0.3}\qquad
  \epsfig{figure=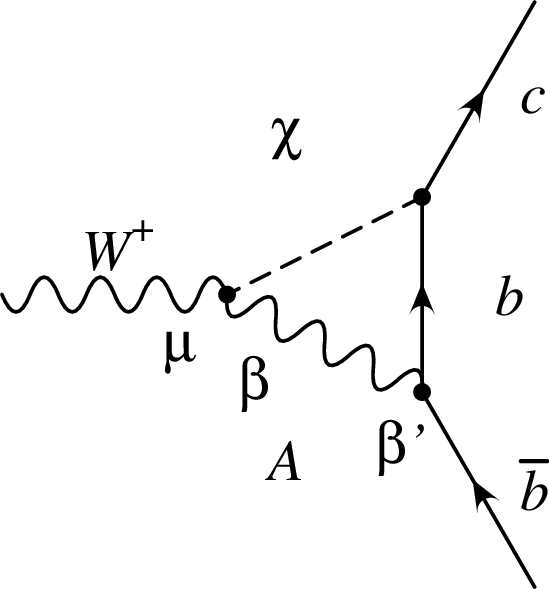, scale=0.3}\qquad
  \epsfig{figure=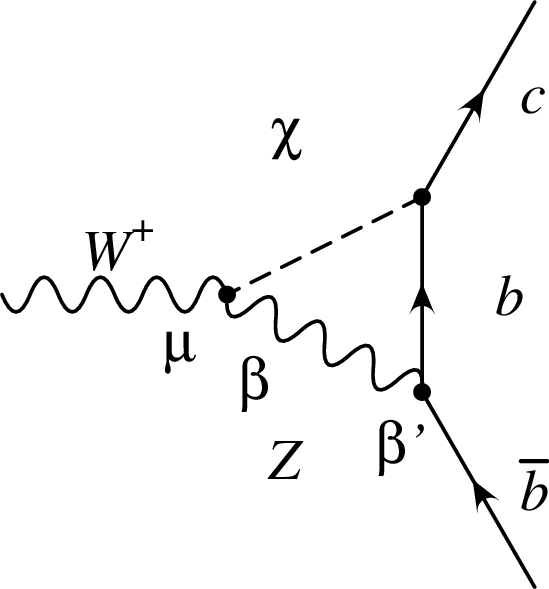, scale=0.3}\\[9pt]
  \epsfig{figure=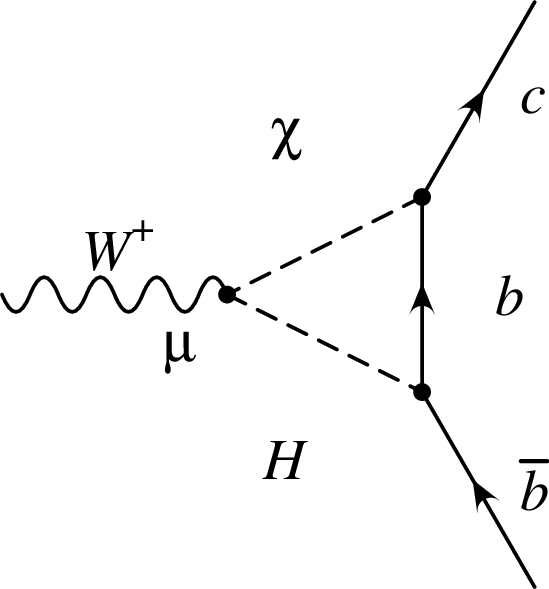, scale=0.3}\qquad
  \epsfig{figure=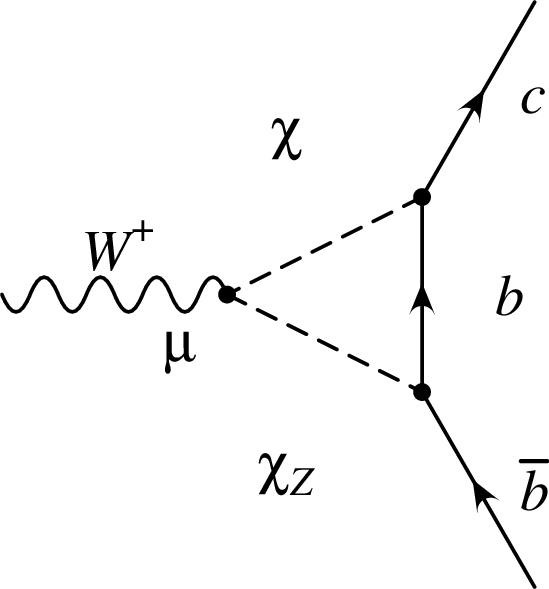, scale=0.3}\qquad
    \caption{\label{cur}Loop corrections}
\end{center}\end{figure}

\section{Loop contributions}
The tree contributions are not the only first order corrections. In addition,
we have to take into account a large number of loop corrections to the vertex
displayed in Fig.~\ref{cur}. The results we obtain from the loop corrections
are both infrared and ultraviolet singular. For the latter, we have to apply
the renormalisation procedure. The NLO result of the decay rate is contributed
by a matrix element ${\cal M}(q,\lambda)={\cal M}^\mu\eps_\mu(q,\lambda)$,
where $\eps(q,\lambda)$ is the polarisation vector of the $W$ boson, and
\begin{eqnarray}\label{Mcb1}
  {\cal M}^\mu&=&\frac{ie}{\sqrt2s_W}\Bigg\{\gamma^\mu\Lambda_-
  \Bigg[V_{cb}\left(1+\delta Z_e-\frac{\delta s_W}{s_W}+\delta Z_{WW}\right)
  +\strut\nonumber\\&&\strut\qquad\qquad\qquad
  +\delta V_{cb}+\sum_k(\delta Z_{ck}^{L*}V_{kb}+V_{ck}\delta Z_{kb}^L)
  \Bigg]+V_{cb}V^\mu_b\Bigg\},
\end{eqnarray}
where the sum runs over corresponding quark flavours. The result is
renormalised by the counter terms $\delta Z_e$ for the charge,
$\delta s_W/s_W$ for the sine of the Weinberg angle, and counter terms
related to the particles involved in the decay process. For different
renormalisation schemes, these counter terms are found in Appendix~C. The
result for the bare vertex correction $V^\mu_b$ can in principle be split up
into six contributions according to
\begin{eqnarray}
\lefteqn{V^\mu_b\ =\ V^0_-\bar u(p_1)\gamma^\mu\Lambda_-v(p_2)
  +V^0_+\bar u(p_1)\gamma^\mu\Lambda_+v(p_2)+\strut}\\&&\strut
  +V^1_-\bar u(p_1)p_1^\mu\Lambda_-v(p_2)
  +V^1_+\bar u(p_1)p_1^\mu\Lambda_+v(p_2)
  +V^2_-\bar u(p_1)p_2^\mu\Lambda_-v(p_2)
  +V^2_+\bar u(p_1)p_2^\mu\Lambda_+v(p_2)\nonumber
\end{eqnarray}
with $\Lambda_\pm=(1\pm\gamma_5)/2$. However, considering only on-shell $W$
boson decays, the renormalised loop corrections are divided up into four form
factors $V_-$, $V_+$, $V_1$ and $V_2$ only, defined by
\begin{eqnarray}\label{Vmp12def}
\frac\alpha{4\pi}V_-&=&\real(V^0_-+\delta Z_{\rm CKM}+\delta Z_e
  -\frac{\delta s_W}{s_W}+\delta Z_{WW}+\delta Z^L_{cc}+\delta Z^L_{bb}),
  \nonumber\\
\frac\alpha{4\pi}\mu_1\mu_2V_+q^2&=&m_1m_2\real V^0_+,\nonumber\\
\frac\alpha{4\pi}\mu_1V_1&=&m_1\real(V^1_--V^2_-),\nonumber\\
\frac\alpha{4\pi}\mu_2V_2&=&m_2\real(V^1_+-V^2_+),
\end{eqnarray}
where $\delta Z_{\rm CKM}$ is the counter term for the CKM matrix, as dealt
with in the following subsection. As will be discussed in Sec.~\ref{irdeal},
IR singularities will be subtracted from $V_-$ to obtain $V_-^*$. The form
factors $V_-^*$, $V_+$, $V_1$ and $V_2$ are cross-checked against the form
factors in Ref.~\cite{Denner:1991kt}. In doing so, we found agreement with
our result (cf.\ Appendix~B).
 
\subsection{Renormalisation of the CKM matrix\label{renver}}
As the CKM matrix element $V_{cb}$ dominating the decay $W^+\to c\bar b$ is
not close to unit, we have to take into account the mixing of quark states
and, related to this, the renormalisation of the mixing
matrix~\cite{Denner:1991kt,Balzereit:1998id,Gambino:1998ec,Barroso:2000is,%
Kniehl:2006bs,Kniehl:2006rc,Kniehl:2009kk,Kniehl:2009nz,Denner:2019vbn}. Using
\begin{eqnarray}
  \delta V_{cb}=\frac12\sum_k\left((\delta Z_{ck}^L-\delta Z_{ck}^{L*})
  V_{kb}-V_{ck}(\delta Z_{kb}^L-\delta Z_{kb}^{L*})\right),
\end{eqnarray}
one obtains
\begin{eqnarray}\label{Mcb2}
  {\cal M}^\mu&=&\frac{ie}{\sqrt2s_W}\Bigg\{\gamma^\mu\Lambda_-
  \Bigg[V_{cb}\left(1+\delta Z_e-\frac{\delta s_W}{s_W}+\delta Z_{WW}
  +\delta Z^L_{cc}+\delta Z^L_{bb}\right)
  +\strut\nonumber\\&&\strut\qquad\qquad
  +\real\delta Z_{cu}^LV_{ub}+\real\delta Z_{ct}^LV_{tb}
  +V_{cd}\real\delta Z_{db}^L+V_{cs}\real\delta Z_{sb}^L\Bigg]
  +V_{cb}V^\mu_b\Bigg\}.\qquad
\end{eqnarray}
Defining $\delta Z_{\rm CKM}:=\left(\real\delta Z_{cu}^LV_{ub}
+\real\delta Z_{ct}^LV_{tb}+V_{cd}\real\delta Z_{db}^L+V_{cs}
\real\delta Z_{sb}^L\right)/V_{cb}$, one ends up with $V_-$ as given in
Eq.~(\ref{Vmp12def}).

In the $\alpha(0)$ scheme, the UV singular part within $V^\mu_b$ is given by
$V^\mu_s=\gamma^\mu\Lambda_-V^0_s$ with
\begin{equation}
V^0_s=\frac{e^2}{4m_W^2s_W^2\eps}\left[m_c^2+m_b^2+12m_W^2
  -m_Z^2\left(1-2(Q_c-Q_b+2Q_cQ_b)s_W^2\right)\right]
\end{equation}
with space-time dimension $D=4-2\eps$, and $Q_c=2/3$ and $Q_b=-1/3$ for the
electric charges of the quarks, while
$\delta Z_e^s+\delta Z_{WW}^s-\delta s_W^s/s_W=-2e^2/(s_W^2\eps)$
and~\cite{Bohm:2001}
\begin{eqnarray}
  \delta Z^{Ls}_{bb}&=&-\frac{e^2}{8m_W^2s_W^2\eps}\left[m_b^2+\sum_k|V_{kb}|^2
  (m_k^2+2m_W^2)+(1+4Q_cQ_bs_W^2)m_Z^2\right],\nonumber\\
  \delta Z^{Ls}_{cc}&=&-\frac{e^2}{8m_W^2s_W^2\eps}\left[m_c^2+\sum_k|V_{ck}|^2
  (m_k^2+2m_W^2)+(1+4Q_cQ_bs_W^2)m_Z^2\right],\nonumber\\
  \delta Z^{Ls}_{ij}&=&\frac{e^2}{4m_W^2s_W^2\eps}\sum_kV_{ik}V_{kj}
  \frac{m_i^2m_j^2-2m_i^2m_k^2-m_j^2m_k^2+2m_j^2m_W^2}{m_i^2-m_j^2}.
\end{eqnarray}
Note that according to Eq.~(4.5.27) in Ref.~\cite{Bohm:2001}, one has
$\delta Z_{ij}^{Ls*}=\delta Z_{ij}^{Ls}|_{m_i^2\leftrightarrow m_j^2}$.
Therefore,
\begin{equation}
  \real\delta Z_{ij}^{Ls}
  =\frac{-e^2}{8m_W^2s_W^2\eps}\sum_kV_{ik}V_{kj}(m_k^2+2m_W^2),\qquad
  V_{kj}=V_{jk}^*.
\end{equation}
Without the mixed contributions (and without the general factor $V_{cb}$), one
first obtains
\begin{eqnarray}
\lefteqn{V^0_s+\delta Z_e^s-\frac{\delta s_W^s}{s_W}
    +\delta Z_{WW}^s+\delta Z^{Ls}_{cc}+\delta Z^{Ls}_{bb}\ =}\nonumber\\
  &=&\frac{e^2}{8m_W^2s_W^2\eps}\left[m_c^2+m_b^2+4m_W^2
  -\sum_k|V_{ck}|^2(m_k^2+2m_W^2)-\sum_k|V_{kb}|^2(m_k^2+2m_W^2)\right].\quad
\end{eqnarray}
On the other hand, the mixed part gives
\begin{eqnarray}
\lefteqn{\real\delta Z_{cu}^{Ls}V_{ub}+\real\delta Z_{ct}^{Ls}V_{tb}
    +V_{cd}\real\delta Z_{db}^{Ls}+V_{cs}\real\delta Z_{sb}^{Ls}\ =}\\
  &=&\frac{-e^2}{8m_W^2s_W^2\eps}\sum_k\left(V_{ck}V_{ku}V_{ub}
  +V_{ck}V_{kt}V_{tb}+V_{cd}V_{dk}V_{kb}+V_{cs}V_{sk}V_{kb}\right)
  (m_k^2+2m_W^2)\ =\nonumber\\
  &=&\frac{-e^2V_{cb}}{8m_W^2s_W^2\eps}\left[m_b^2+2m_W^2-\sum_k|V_{ck}|^2
  (m_k^2+2m_W^2)+m_c^2+2m_W^2-\sum_k|V_{kb}|^2(m_k^2+2m_W^2)\right].\kern-8pt
  \nonumber
\end{eqnarray}
This cancels exactly against the previous contribution, gaining an UV finite
form factor.

\section{Helicity bilinears}
The helicity bilinears are obtained by contracting the hadron tensor with the
polarisation vectors $\eps(q,\lambda)$ which for the Born term diagram is
given in Eq.~(\ref{M02}). First we use the helicity basis in the $W$ frame,
i.e., the rest frame of the decaying $W$ boson with the $z$ axis as the
initial direction of flight of the $W$ boson, given by
\begin{equation}\label{polvec}
\eps(q,\pm)=\frac1{\sqrt2}(0;\mp1,-i,0),\quad
  \eps(q,0)=(0;0,0,1),\quad\eps(q,t)=(1;0,0,0).
\end{equation}
Note that $\lambda=\pm,0$ describe the different magnetic quantum numbers for
the total angular momentum quantum number $j=1$ while $\lambda=t$ indicates
the magnetic quantum number zero for $j=0$. This time-like component does not
appear for on-shell $W$ bosons and will be skipped in the following. At Born
term level, up to a general factor $e^2q^2|V_{cb}|^2/s_W^2$ the helicity
bilinears are given by
\begin{equation}\label{helborn}
  q^2H^{\lambda_1\lambda_2}=\left[p_1^\mu p_2^\nu+p_1^\nu p_2^\mu
  -p_1p_2g^{\mu\nu}+ip_{1\kappa}p_{2\lambda}\epsilon^{\kappa\lambda\mu\nu}
  \right]\eps_\mu(q,\lambda_1)\eps_\nu^*(q,\lambda_2).
\end{equation}
This results in the $W$ frame read
\begin{eqnarray}\label{helexpl}
  H^{tt}(\theta)&=&\frac12\left(\mu_1+\mu_2-(\mu_1-\mu_2)^2\right),
  \nonumber\\
  H^{t0}(\theta)&=&H^{0t}(\theta)\ =\ \frac12(\mu_1-\mu_2)\sla\cos\theta,
  \nonumber\\
  H^{t\pm}(\theta)&=&H^{\pm t}(\theta)
    \ =\ \mp\frac1{2\sqrt2}(\mu_1-\mu_2)\sla\sin\theta,\nonumber\\
  H^{00}(\theta)&=&\frac12\left(1-\mu_1-\mu_2-\lambda\cos^2\theta\right),
  \nonumber\\
  H^{0\pm}(\theta)&=&H^{\pm 0}(\theta)
  \ =\ -\frac1{2\sqrt2}\sla(1\mp\sla\cos\theta)\sin\theta,\nonumber\\
  H^{\pm\pm}(\theta)&=&\frac14(1+\mu_1-\mu_2\mp\sla\cos\theta)
  (1-\mu_1+\mu_2\mp\sla\cos\theta),\nonumber\\
  H^{\pm\mp}(\theta)&=&\frac14\lambda\sin^2\theta.
\end{eqnarray}
Changing to the quark frame as the frame of reference for the angular
distribution~(\ref{Wtheta}), one has to take $\theta=0$. Only the the helicity
bilinears $H^{00}=H^{00}(\theta=0)$, $H^{\pm\pm}=H^{\pm\pm}(\theta=0)$
necessary for the angular distribution and $H^{\pm\mp}=H^{\pm\mp}(\theta=0)$
(for completeness) will be considered in the following.

\subsection{Helicity bilinears from the tree corrections}
The hadron tensor for the NLO tree correction to be contracted with the
polarisation vectors is obtained by replacing $|{\cal M}_0|^2$ in
Eq.~(\ref{M02}) by $|{\cal M}_1+{\cal M}_2+{\cal M}_3|^2$, where the matrix
elements ${\cal M}_i$ ($i=1,2,3$) correspond to the three Feynman diagrams in
Figure~\ref{wlagu}. We have calculate the helicity bilinears both for Feynman
and unitary gauge and found agreement. Up to a general factor
$e^2q^2|V_{cb}|^2/s_W^2\times\alpha/(4\pi\sla)$, the NLO tree contributions
to the helicity bilinears read
\begin{eqnarray}\label{treeH00}
\lefteqn{H^{00}({\it tree\/})\ =\
  \Frac12\Big[Q_c^2+Q_b^2+Q_W^2\Big]\sla(-\mu_1+\mu_1^2-\mu_2-2\mu_1\mu_2
  +\mu_2^2)\ell_\zeta+\strut}\nonumber\\&&\strut\kern-24pt
  +\Frac12\Big[(1-\mu_2)Q_c^2-\mu_1(Q_b^2-Q_W^2)\Big]
  (-\mu_1+\mu_1^2-\mu_2-2\mu_1\mu_2+\mu_2^2)(t_\zeta^\ell-2t_z^{\ell+})
  +\strut\nonumber\\&&\strut\kern-24pt
  -\Frac12\Big[Q_c^2-Q_b^2+(\mu_1-\mu_2)Q_W^2\Big]
  (-\mu_1+\mu_1^2-\mu_2-2\mu_1\mu_2+\mu_2^2)(t_{\zeta W}^\ell+2t_{zW}^{\ell+})
  +\strut\nonumber\\&&\strut\kern-24pt
  -2\mu_1\Big[(1+5\mu_1-\mu_2)Q_c^2+2\mu_1(Q_b^2-Q_W^2)\Big]
  (t_z^\ell+t_z^{-\ell}-t_z^{+\ell})+\strut\nonumber\\&&\strut\kern-24pt
  -\smu\Big[(1-10\mu_1-3\mu_1^2-2\mu_2+2\mu_1\mu_2+\mu_2^2)Q_c^2
  -2\mu_1(1+\mu_1-\mu_2)(Q_b^2-Q_W^2)\Big](t_z^{-\ell}+t_z^{+\ell})
  +\strut\kern-20pt\nonumber\\&&\strut\kern-24pt
  +2\Big[(1+\mu_1-\mu_2)(Q_c^2-Q_b^2)-(3+2\mu_1+\mu_1^2-4\mu_2-2\mu_1\mu_2
  +\mu_2^2)Q_W^2\Big](t_{zW}^\ell+t_{zW}^{-\ell}-t_{zW}^{+\ell})
  +\strut\kern-9pt\nonumber\\&&\strut\kern-24pt
  -\Frac1{\smu}(1+\mu_1-\mu_2)\Big[(1+\mu_1-\mu_2)(Q_c^2-Q_b^2)
  -2(1+2\mu_1-\mu_2)Q_W^2\Big](t_{zW}^{-\ell}+t_{zW}^{+\ell})
  +\strut\nonumber\\&&\strut\kern-24pt
  +\Frac14\Big[(4-25\mu_1-12\mu_1^2-9\mu_2+26\mu_1\mu_2-2\mu_1^2\mu_2
  +6\mu_2^2-3\mu_1\mu_2^2-\mu_2^3)Q_c^2+\strut\nonumber\\&&\strut
  -(4+6\mu_1+6\mu_1^2-\mu_1^3-6\mu_2-2\mu_1\mu_2+9\mu_1^2\mu_2-4\mu_1\mu_2^2
  +2\mu_2^3)Q_b^2+\strut\nonumber\\&&\strut
  -2(4+9\mu_1+3\mu_1^2+\mu_1^3-7\mu_2-7\mu_1\mu_2-2\mu_1^2\mu_2+2\mu_2^2
  -2\mu_1\mu_2^2+\mu_2^3)Q_W^2\Big]\ell_1+\strut\nonumber\\&&\strut\kern-24pt
  -\Frac14\Big[(4-23\mu_1-16\mu_1^2+2\mu_1^3-7\mu_2+12\mu_1\mu_2-6\mu_1^2\mu_2
  +4\mu_2^2+6\mu_1\mu_2^2-2\mu_2^3)Q_c^2+\strut\nonumber\\&&\strut
  -(4+5\mu_1+8\mu_1^2-2\mu_1^3-3\mu_2-4\mu_1\mu_2+6\mu_1^2\mu_2-4\mu_2^2
  -6\mu_1\mu_2^2+2\mu_2^3)Q_b^2+\strut\nonumber\\&&\strut
  -2(4+8\mu_1+5\mu_1^2-12\mu_2-12\mu_1\mu_2+7\mu_2^2)Q_W^2\Big]\ell_{1W}
  +\strut\nonumber\\&&\strut\kern-24pt
  +\Frac18\Big[(16-67\mu_1-15\mu_1^2-3\mu_2+38\mu_1\mu_2-11\mu_2^2)Q_c^2
  +\strut\nonumber\\&&\strut
  -(8+11\mu_1+11\mu_1^2-21\mu_2-38\mu_1\mu_2+15\mu_2^2)Q_b^2
  +\strut\nonumber\\&&\strut
  -2(24+\mu_1+9\mu_1^2-31\mu_2-14\mu_1\mu_2+9\mu_2^2)Q_W^2\Big]\sla,\\[12pt]
\lefteqn{H^{++}({\it tree\/})\ =\
  -\Frac12\Big[Q_c^2+Q_b^2+Q_W^2\Big]\sla\left((1-\mu_1-\mu_2-\sla)\ell_\zeta
  +2\sla\ell_+\right)+\strut}\nonumber\\&&\strut\kern-24pt
  -\Frac12\Big[(1-\mu_2)Q_c^2-\mu_1(Q_b^2-Q_W^2)\Big]\left((1-\mu_1-\mu_2)
  (t_\zeta^\ell-2t_z^{\ell+})-\sla(t_\zeta^\ell-2t_z^{\ell-})\right)
  +\strut\nonumber\\&&\strut\kern-24pt
  +\Frac12\Big[Q_c^2-Q_b^2+(\mu_1-\mu_2)Q_W^2\Big]\left((1-\mu_1-\mu_2)
  (t_{\zeta W}^\ell+2t_{zW}^{\ell+})-\sla(t_{\zeta W}^\ell+2t_{zW}^{\ell-})
  \right)+\strut\nonumber\\&&\strut\kern-24pt
  +\Big[(1-2\mu_1-2\mu_2-\mu_1\mu_2+\mu_2^2)Q_c^2-\mu_1(1+\mu_1-\mu_2)
  (Q_b^2-Q_W^2)\Big]t_z^\ell+\strut\nonumber\\&&\strut\kern-24pt
  +\mu_1\Big[(1+5\mu_1-\mu_2)Q_c^2+2\mu_1(Q_b^2-Q_W^2)\Big]
  (t_z^\ell+t_z^{-\ell}-t_z^{+\ell})+\strut\nonumber\\&&\strut\kern-24pt
  +\Frac12\smu\Big[(1-10\mu_1-3\mu_1^2-2\mu_2+2\mu_1\mu_2+\mu_2^2)Q_c^2
  -2\mu_1(1+\mu_1-\mu_2)(Q_b^2-Q_W^2)\Big](t_z^{-\ell}+t_z^{+\ell})
  +\strut\kern-19pt\nonumber\\&&\strut\kern-24pt
  -\Big[(1+\mu_1-\mu_2)(Q_c^2-Q_b^2)-(3+2\mu_1+\mu_1^2-4\mu_2-2\mu_1\mu_2
  +\mu_2^2)Q_W^2\Big](2t_{zW}^\ell+t_{zW}^{-\ell}-t_{zW}^{+\ell})
  +\strut\kern-19pt\nonumber\\&&\strut\kern-24pt
  +\Frac1{2\smu}(1+\mu_1-\mu_2)\Big[(1+\mu_1-\mu_2)(Q_c^2-Q_b^2)
  -2(1+2\mu_1-\mu_2)Q_W^2\Big](t_{zW}^{-\ell}+t_{zW}^{+\ell})
  +\strut\nonumber\\&&\strut\kern-24pt
  +\Frac12\Big[(1-\mu_1)(5-3\mu_1+4\mu_2)Q_c^2-(9-10\mu_1+\mu_1^2+6\mu_2
  -2\mu_1\mu_2)Q_b^2+\strut\nonumber\\&&\strut
  -2(1-\mu_1)(5-\mu_1+\mu_2)Q_W^2\Big]\ell_0+\strut\nonumber\\&&\strut\kern-24pt
  -\Frac14\Big[\left(3-8\mu_1-4\mu_1^2-6\mu_2+10\mu_1\mu_2+3\mu_2^2
  +(1+5\mu_1-\mu_2)\sla\right)Q_c^2+\strut\nonumber\\&&\strut
  -\mu_1(4+7\mu_1-4\mu_2)Q_b^2-2(1-\mu_2)(1+5\mu_1-\mu_2)Q_W^2\Big]\ell_1
  +\strut\nonumber\\&&\strut\kern-24pt
  +\Frac14\Big[(1-4\mu_1-6\mu_1^2-4\mu_2+6\mu_2^2+8\sla)Q_c^2
  -(1+2\mu_1+8\mu_1^2+6\mu_2-8\mu_1\mu_2+8\sla)Q_b^2
  +\strut\kern-6pt\nonumber\\&&\strut
  -2(2+3\mu_1+\mu_1^2-5\mu_2-6\mu_1\mu_2+5\mu_2^2+4\sla)Q_W^2\Big]\ell_{1W}
  +\strut\nonumber\\&&\strut\kern-24pt
  -\Frac14\Big[(17+7\mu_1-8\mu_2)Q_c^2-(13-3\mu_1+2\mu_2)Q_b^2
  -6(3+\mu_1-\mu_2)Q_W^2\Big]\lambda_-+\strut\nonumber\\&&\strut\kern-24pt
  +\Frac1{24}\Big[9(1+5\mu_1+\mu_2)Q_c^2+3(15+7\mu_1-29\mu_2)Q_b^2
  +\strut\nonumber\\&&\strut
  +4(29-10\mu_1-\mu_1^2-34\mu_2+2\mu_1\mu_2-\mu_2^2)Q_W^2\Big]\sla
  +\Frac18\Big[Q_c^2-9Q_b^2-12Q_W^2\Big]\lambda,\label{treeHpp}\\[12pt]
\lefteqn{H^{+-}({\it tree\/})\ =\ H^{-+}({\it tree\/})\ =\
  -\mu_1\Big[(1+5\mu_1-\mu_2)Q_c^2+2\mu_1(Q_b^2-Q_W^2)\Big]
  (t_z^\ell+t_z^{-\ell}-t_z^{+\ell})+\strut}\nonumber\\&&\strut\kern-24pt
  -\Frac12\smu\Big[(1-10\mu_1-3\mu_1^2-2\mu_2+2\mu_1\mu_2+\mu_2^2)Q_c^2
  -2\mu_1(1+\mu_1-\mu_2)(Q_b^2-Q_W^2)\Big](t_z^{-\ell}+t_z^{+\ell})
  +\strut\kern-14pt\nonumber\\&&\strut\kern-24pt
  +\Big[(1+\mu_1-\mu_2)(Q_c^2-Q_b^2)-(3+2\mu_1+\mu_1^2-4\mu_2-2\mu_1\mu_2
  +\mu_2^2)Q_W^2\Big](t_{zW}^\ell+t_{zW}^{-\ell}-t_{zW}^{+\ell})
  +\strut\nonumber\\&&\strut\kern-24pt
  -\Frac1{2\smu}(1+\mu_1-\mu_2)\Big[(1+\mu_1-\mu_2)(Q_c^2-Q_b^2)
  -2(1+2\mu_1-\mu_2)Q_W^2\Big](t_{zW}^{-\ell}+t_{zW}^{+\ell})
  +\strut\nonumber\\&&\strut\kern-24pt
  +\Frac12\Big[(1-6\mu_1-3\mu_1^2-2\mu_2+6\mu_1\mu_2+\mu_2^2)Q_c^2
  -(1+\mu_1+2\mu_1^2-2\mu_2-\mu_1\mu_2+\mu_2^2)Q_b^2+\strut\nonumber\\&&\strut
  -2(1+\mu_1-\mu_2)^2Q_W^2\Big]\ell_1+\strut\nonumber\\&&\strut\kern-24pt
  -\Frac12\Big[(1-6\mu_1-3\mu_1^2-2\mu_2+2\mu_1\mu_2+\mu_2^2)Q_c^2
  -(1+\mu_1+2\mu_1^2-\mu_2-2\mu_1\mu_2)Q_b^2+\strut\nonumber\\&&\strut
  -2(1+\mu_1-2\mu_2)(1+\mu_1-\mu_2)Q_W^2\Big]\ell_{1W}
  +\strut\nonumber\\&&\strut\kern-24pt
  +\Frac12\Big[2(1-5\mu_1-\mu_2)Q_c^2-(1+3\mu_1-\mu_2)Q_b^2
  -2(3+\mu_1-3\mu_2)Q_W^2\Big]\sla,\label{treeHpm}\\[12pt]
\lefteqn{H^{--}({\it tree\/})\ =\
  -\Frac12\Big[Q_c^2+Q_b^2+Q_W^2\Big]\sla\left((1-\mu_1-\mu_2+\sla)\ell_\zeta
  -2\sla\ell_+\right)+\strut}\nonumber\\&&\strut\kern-24pt
  -\Frac12\Big[(1-\mu_2)Q_c^2-\mu_1(Q_b^2-Q_W^2)\Big]\left((1-\mu_1-\mu_2)
  (t_\zeta^\ell-2t_z^{\ell+})+\sla(t_\zeta^\ell-2t_z^{\ell-})\right)
  +\strut\nonumber\\&&\strut\kern-24pt
  +\Frac12\Big[Q_c^2-Q_b^2+(\mu_1-\mu_2)Q_W^2\Big]\left((1-\mu_1-\mu_2)
  (t_{\zeta W}^\ell+2t_{zW}^{\ell+})+\sla(t_{\zeta W}^\ell+2t_{zW}^{\ell-})
  \right)+\strut\nonumber\\&&\strut\kern-24pt
  -\Big[(1-2\mu_1-2\mu_2-\mu_1\mu_2+\mu_2^2)Q_c^2-\mu_1(1+\mu_1-\mu_2)
  (Q_b^2-Q_W^2)\Big]t_z^\ell+\strut\nonumber\\&&\strut\kern-24pt
  +\mu_1\Big[(1+5\mu_1-\mu_2)Q_c^2+2\mu_1(Q_b^2-Q_W^2)\Big]
  (t_z^\ell+t_z^{-\ell}-t_z^{+\ell})+\strut\nonumber\\&&\strut\kern-24pt
  +\Frac12\smu\Big[(1-10\mu_1-3\mu_1^2-2\mu_2+2\mu_1\mu_2+\mu_2^2)Q_c^2
  -2\mu_1(1+\mu_1-\mu_2)(Q_b^2-Q_W^2)\Big](t_z^{-\ell}+t_z^{+\ell})
  +\strut\kern-19pt\nonumber\\&&\strut\kern-24pt
  -\Big[(1+\mu_1-\mu_2)(Q_c^2-Q_b^2)-(3+2\mu_1+\mu_1^2-4\mu_2-2\mu_1\mu_2
  +\mu_2^2)Q_W^2\Big](t_{zW}^{-\ell}-t_{zW}^{+\ell})
  +\strut\nonumber\\&&\strut\kern-24pt
  +\Frac1{2\smu}(1+\mu_1-\mu_2)\Big[(1+\mu_1-\mu_2)(Q_c^2-Q_b^2)
  -2(1+2\mu_1-\mu_2)Q_W^2\Big](t_{zW}^{-\ell}+t_{zW}^{+\ell})
  +\strut\nonumber\\&&\strut\kern-24pt
  -\Frac12\Big[(1-\mu_1)(5-3\mu_1+4\mu_2)Q_c^2-(9-10\mu_1+\mu_1^2+6\mu_2
  -2\mu_1\mu_2)Q_b^2+\strut\nonumber\\&&\strut
  -2(1-\mu_1)(5-\mu_1+\mu_2)Q_W^2\Big]\ell_0+\strut\nonumber\\&&\strut\kern-24pt
  -\Frac14\Big[\left(3-8\mu_1-4\mu_1^2-6\mu_2+10\mu_1\mu_2+3\mu_2^2
  -(1+5\mu_1-\mu_2)\sla\right)Q_c^2+\strut\nonumber\\&&\strut
  -\mu_1(4+7\mu_1-4\mu_2)Q_b^2-2(1-\mu_2)(1+5\mu_1-\mu_2)Q_W^2\Big]\ell_1
  +\strut\nonumber\\&&\strut\kern-24pt
  +\Frac14\Big[(1-4\mu_1-6\mu_1^2-4\mu_2+6\mu_2^2-8\sla)Q_c^2
  -(1+2\mu_1+8\mu_1^2+6\mu_2-8\mu_1\mu_2-8\sla)Q_b^2
  +\strut\kern-6pt\nonumber\\&&\strut
  -2(2+3\mu_1+\mu_1^2-5\mu_2-6\mu_1\mu_2+5\mu_2^2-4\sla)Q_W^2\Big]\ell_{1W}
  +\strut\nonumber\\&&\strut\kern-24pt
  +\Frac14\Big[(17+7\mu_1-8\mu_2)Q_c^2-(13-3\mu_1+2\mu_2)Q_b^2
  -6(3+\mu_1-\mu_2)Q_W^2\Big]\lambda_-+\strut\nonumber\\&&\strut\kern-24pt  
  +\Frac1{24}\Big[9(1+5\mu_1+\mu_2)Q_c^2+3(15+7\mu_1-29\mu_2)Q_b^2
  +\strut\nonumber\\&&\strut
  +4(29-10\mu_1-\mu_1^2-34\mu_2+2\mu_1\mu_2-\mu_2^2)Q_W^2\Big]\sla
  -\Frac18\Big[Q_c^2-9Q_b^2-12Q_W^2\Big]\lambda\label{treeHmm}
\end{eqnarray}
with $\lambda_\pm:=(1\pm\smu)^2-\mu_2$ and
$\lambda_-\lambda_+=\lambda:=\lambda(1,\mu_1,\mu_2)$. The dilogarithmic terms
contained in these expressions are given by
\begin{eqnarray}
t_\zeta^\ell&=&\int_{\zeta_-}^1\left(\frac1\zeta-\frac2{1+\zeta}\right)
  \ln\pfrac{\zeta+\zeta_0}{1+\zeta_0\zeta}d\zeta,\qquad
  \zeta_0=\frac{1-\mu_1-\mu_2-\sla}{1-\mu_1-\mu_2+\sla},\nonumber\\
t_{\zeta W}^\ell&=&\int_{\zeta_-}^1\left(\frac1\zeta-\frac2{1+\zeta}\right)
  \ln\pfrac{\zeta+\zeta_{0W}}{1+\zeta_{0W}\zeta}d\zeta,\qquad
  \zeta_{0W}=\frac{1-\mu_1+\mu_2-\sla}{1-\mu_1+\mu_2+\sla},\nonumber\\
t_z^{\ell-}&=&\int_{z_-}^1\left[\left(\frac1{z-z_-}-\frac1{z-z_+}\right)
  L_z(z)-\frac{L_z(z_-)}{z-z_+}\right]dz,\nonumber\\
t_z^{\ell+}&=&\int_{z_-}^1\left[\left(\frac1{z-z_-}+\frac1{z-z_+}
  -\frac1z\right)L_z(z)-\frac{L_z(z_-)}{z-z_+}\right]dz,\nonumber\\
t_{zW}^{\ell-}&=&\int_{z_-}^1\left[\left(\frac1{z-z_-}-\frac1{z-z_+}\right)
  L_{zW}(z)-\frac{L_{zW}(z_-)}{z-z_+}\right]dz,\nonumber\\
t_{zW}^{\ell+}&=&\int_{z_-}^1\left[\left(\frac1{z-z_-}+\frac1{z-z_+}
  -\frac1z\right)L_{zW}(z)-\frac{L_{zW}(z_-)}{z-z_+}\right]dz,\nonumber\\
t_z^{\pm\ell}&=&\int_{z_-}^1\frac{L_z(z)}{1\pm z}dz,\qquad
t_z^\ell\ =\ \int_{z_-}^1\frac{L_z(z)}zdz,\nonumber\\
t_{zW}^{\pm\ell}&=&\int_{z_-}^1\frac{L_{zW}(z)}{1\pm z}dz,\qquad
t_{zW}^\ell\ =\ \int_{z_-}^1\frac{L_{zW}(z)}zdz,
\end{eqnarray}
with
\begin{equation}
L_z(z)=\ln\pfrac{1-z\smu}{(z-\smu)z},\qquad
L_{zW}(z)=\ln\pfrac{z-\smu}{z(1-z\smu)}
\end{equation}
and
\begin{equation}
\zeta_-=\frac{\Lambda\mu_2}{((1-\smu)^2-\mu_2)^2},\qquad
z_\pm=\frac1{2\smu}(1+\mu_1-\mu_2\pm\sla),\qquad
\sla=\sqrt{\lambda(1,\mu_1,\mu_2)}.
\end{equation}
The logarithmic terms $\ell_\zeta$, $\ell_0$, $\ell_+$, $\ell_1$ and
$\ell_{1W}$ are listed in Eq.~(\ref{ells}).

\subsection{Helicity bilinears from the loop corrections}
Again up to a general factor $e^2q^2|V_{cb}|^2/s_W^2\times\alpha/(4\pi\sla)$,
the NLO loop contributions read
\begin{eqnarray}\label{Hloop}
H^{00}({\it loop\/})&=&-\frac12(-\mu_1+\mu_1^2-\mu_2-2\mu_1\mu_2+\mu_2^2)
  \sla V_-+\strut\nonumber\\&&\strut
  +\mu_1\mu_2\sla V_+-\frac14\lambda\sla(\mu_1V_1+\mu_2V_2),\nonumber\\
H^{++}({\it loop\/})&=&\frac12(1-\mu_1-\mu_2-\sla)\sla V_-
  +\mu_1\mu_2\sla V_+,\nonumber\\
H^{+-}({\it loop\/})&=&H^{-+}({\it loop\/})\ =\ 0,\nonumber\\
H^{--}({\it loop\/})&=&\frac12(1-\mu_1-\mu_2+\sla)\sla V_-+\mu_1\mu_2\sla V_+.
\end{eqnarray}

\section{Full results for the helicity bilinears}
Having used Feynman gauge up to now, it is worth mentioning here that we have
performed the calculation of the helicity bilinears for the tree corrections
also in unitary gauge. Unitary gauge used for the $W$ propagator in
Fig.~\ref{wlagu}.3 allows to drop contributions from the charged Goldstone
boson which is not indicated but implicitly assumed in this diagram. We found
that the results for the helicity bilinears $H^{00}({\it tree\/})$,
$H^{\pm\pm}({\it tree\/})$ and $H^{\pm\mp}({\it tree\/})$ turn out to be
exactly the same. Both tree and loop contributions, however, contain IR
singularities which have to cancel according to the Lee--Nauenberg theorem.

\subsection{Counter terms for the IR singularities\label{irdeal}}
In order to deal with the IR singularities in a consistent way, we need a
convenient method to extract them. For the first order tree contributions
we have seen that the IR singular parts are contained in $\ell_\zeta$,
$t_\zeta^\ell$ and $t_{\zeta W}^\ell$. While $\ell_\zeta$ itself, as defined
in Eq.~(\ref{ells}), can be used as counter term, for the other two
expressions in the first order tree contributions we use
\begin{equation}
  t_\zeta^\ell=t_\zeta^{\ell*}+\ell_1\ell_\zeta,\qquad
  t_{\zeta W}^\ell=t_{\zeta W}^{\ell*}+\ell_{1W}\ell_\zeta,
\end{equation}
where the starred quantities are found in Appendix~A. On the other hand, the
form factors contain IR singularities only in the three-point functions with
the photon on one of the internal lines. The scalar three-point functions are
given by
\begin{equation}
C(p_1^2,p_2^2,p_3^2;m_1,m_2,m_3)
  =\frac{i}{(4\pi)^2}C_f(p_1^2,p_2^2,p_3^2;m_1,m_2,m_3)
\end{equation}
with~\cite{Denner:1991kt}
\begin{equation}
C_f(p_1^2,p_2^2,p_3^2;m_1,m_2,m_3)=\frac{-1}{\sqrt{\lambda(p_1^2,p_2^2,p_3^2)}}
  \sum_{i=1}^3\sum_\pm\left[\Li_2\pfrac{1-y_{i0}}{y_{i\pm}\pm i\epsilon-y_{i0}}
  -\Li_2\pfrac{-y_{i0}}{y_{i\pm}\pm i\epsilon-y_{i0}}\right],\kern-25pt
\end{equation}
where
\begin{eqnarray}
y_{10}&=&\frac{(p_2^2+m_2^2-m_3^2)\left(\slap
  +p_1^2-p_2^2-p_3^2\right)+2p_2^2(p_3^2-m_1^2+m_2^2)}{2p_2^2\slap},\nonumber\\
y_{20}&=&\frac{(p_3^2+m_1^2-m_2^2)\left(\slap
  -p_1^2+p_2^2-p_3^2\right)+2p_3^2(p_1^2-m_3^2+m_1^2)}{2p_3^2\slap},\nonumber\\
y_{30}&=&\frac{(p_1^2+m_3^2-m_1^2)\left(\slap
  -p_1^2-p_2^2+p_3^2\right)+2p_1^2(p_2^2-m_2^2+m_3^2)}{2p_1^2\slap}
\end{eqnarray}
and
\begin{eqnarray}
y_{1\pm}&=&\frac{p_2^2+m_2^2-m_3^2
  \pm\sqrt{\lambda(p_2^2,m_2^2,m_3^2)}}{2p_2^2},\nonumber\\
y_{2\pm}&=&\frac{p_3^2+m_1^2-m_2^2
  \pm\sqrt{\lambda(p_3^2,m_1^2,m_2^2)}}{2p_3^2},\nonumber\\
y_{3\pm}&=&\frac{p_1^2+m_3^2-m_1^2
  \pm\sqrt{\lambda(p_1^2,m_3^2,m_1^2)}}{2p_1^2}
\end{eqnarray}
Therefore, we again define starred quantities by
\begin{eqnarray}
  C_f(m_1^2,m_2^2,m_W^2;m_A,m_W,m_1)&=&C_f^*(m_1^2,m_2^2,m_W^2;m_A,m_W,m_1)
  +\frac{\ell_1-\ell_{1W}}{2m_W^2\sla}\ell_\zeta,\nonumber\\
  C_f(m_1^2,m_2^2,m_W^2;m_W,m_A,m_2)&=&C_f^*(m_1^2,m_2^2,m_W^2;m_W,m_A,m_2)
  +\frac{\ell_{1W}}{2m_W^2\sla}\ell_\zeta,\nonumber\\
  C_f(m_1^2,m_2^2,m_W^2;m_1,m_2,m_A)&=&C_f^*(m_1^2,m_2^2,m_W^2;m_1,m_2,m_A)
  -\frac{\ell_1+2\pi i}{2m_W^2\sla}\ell_\zeta\qquad
\end{eqnarray}
where
\begin{eqnarray}
\lefteqn{C_f^*(m_1^2,m_2^2,m_W^2;m_A,m_W,m_1)\ =}\nonumber\\
  &=&-\frac1{m_W^2\sla}\Bigg[\Li_2\pfrac{1-\mu_1+\mu_2-\sla}{1
    -\mu_1+\mu_2+\sla}-\Li_2\pfrac{1-\mu_1-\mu_2-\sla}{1-\mu_1-\mu_2+\sla}
  +\strut\nonumber\\&&\strut
  -\frac14\ell_1^2+\frac14\ell_{1W}^2
  -\frac12\ln\pfrac{\mu_1}\lambda(\ell_1-\ell_{1W})\Bigg],\nonumber\\
\lefteqn{C_f^*(m_1^2,m_2^2,m_W^2;m_W,m_A,m_2)\ =}\nonumber\\
  &=&-\frac1{m_W^2\sla}\Bigg[\Li_2\pfrac{1+\mu_1-\mu_2-\sla}{1
    +\mu_1-\mu_2+\sla}-\Li_2\pfrac{1-\mu_1-\mu_2-\sla}{1-\mu_1-\mu_2+\sla}
  +\strut\nonumber\\&&\strut
  -\frac14\ell_1^2+\frac14(\ell_1-\ell_{1W})^2
  -\frac12\ln\pfrac{\mu_2}\lambda\ell_{1W}\Bigg],\nonumber\\
\lefteqn{C_f^*(m_1^2,m_2^2,m_W^2;m_1,m_2,m_A)\ =}\nonumber\\
  &=&-\frac1{m_W^2\sla}\Bigg[\Li_2\pfrac{1-\mu_1+\mu_2-\sla}{1
    -\mu_1+\mu_2+\sla}+\Li_2\pfrac{1+\mu_1-\mu_2-\sla}{1+\mu_1-\mu_2+\sla}
  +\frac{2\pi^2}3+\strut\nonumber\\&&\strut
  +\frac14(\ell_1-\ell_{1W})^2+\frac14\ell_{1W}^2
  +\frac12\ln\pfrac{\mu_1\mu_2}\lambda(\ell_1+2\pi i)\Bigg],
\end{eqnarray}
and inserted into the form factor $V_-$, one has
\begin{eqnarray}
\sla V_-&=&\sla V_-^*+(Q_c^2+Q_b^2+Q_W^2)\sla\ell_\zeta
  +(1+\mu_1-\mu_2)Q_cQ_W(\ell_1-\ell_{1W})\ell_\zeta+\strut\nonumber\\&&\strut
  -(1-\mu_1+\mu_2)Q_bQ_W\ell_{1W}\ell_\zeta
  +(1-\mu_1-\mu_2)Q_cQ_b(\ell_1+2\pi i)\ell_\zeta,
\end{eqnarray}
where the starred form factor $V_-^*$ contains the starred three-point
functions while the other form factors remain unchanged. The term
$(Q_c^2+Q_b^2+Q_W^2)\sla\ell_\zeta$ originates from the counter terms of the
renormalisation. Using the starred instead of the usual form factors, one has
\begin{eqnarray}
H^{00}({\it loop\/})&=&H^{00*}({\it loop\/})
  -\frac12(-\mu_1+\mu_1^2-\mu_2-2\mu_1\mu_2+\mu_2^2)
  \Big[(Q_c^2+Q_b^2+Q_W^2)\sla+\strut\nonumber\\&&\strut\kern-36pt
  +\left((1-\mu_2)Q_c^2-\mu_1(Q_b^2-Q_W^2)\right)\ell_1
  -\left(Q_c^2-Q_b^2+(\mu_1-\mu_2)Q_W^2\right)\ell_{1W}\Big]\ell_\zeta.\quad
\end{eqnarray}

\subsection{Analytic results for the $O(\alpha)$ helicity bilinears}
Adding up first order tree and loop corrections for all five helicity
bilinears under review, the IR singularities cancel and we obtain
\begin{eqnarray}\label{full}
\lefteqn{H^{00}(\alpha)\ =\ -\Frac12(-\mu_1+\mu_1^2-\mu_2-2\mu_1\mu_2
  +\mu_2^2)\sla V_-^*+\mu_1\mu_2\sla V_+-\Frac12\lambda\sla(\mu_1V_1+\mu_2V_2)
    +\strut}\nonumber\\&&\strut\kern-12pt
  +\Frac12\Big[(1-\mu_2)Q_c^2-\mu_1(Q_b^2-Q_W^2)\Big]
  (-\mu_1+\mu_1^2-\mu_2-2\mu_1\mu_2+\mu_2^2)(t_\zeta^{\ell*}-2t_z^{\ell+})
  +\strut\nonumber\\&&\strut\kern-12pt
  -\Frac12\Big[Q_c^2-Q_b^2+(\mu_1-\mu_2)Q_W^2\Big](-\mu_1+\mu_1^2-\mu_2
  -2\mu_1\mu_2+\mu_2^2)(t_{\zeta W}^{\ell*}+2t_{zW}^{\ell+})
  +\strut\nonumber\\&&\strut\kern-12pt
  -2\mu_1\Big[(1+5\mu_1-\mu_2)Q_c^2+2\mu_1(Q_b^2-Q_W^2)\Big]
  (t_z^{-\ell}-t_z^{+\ell}+t_z^\ell)+\strut\nonumber\\&&\strut\kern-12pt
  -\smu\Big[(1-10\mu_1-3\mu_1^2-2\mu_2+2\mu_1\mu_2+\mu_2^2)Q_c^2
  +\strut\nonumber\\&&\strut
  -2\mu_1(1+\mu_1-\mu_2)(Q_b^2-Q_W^2)\Big](t_z^{-\ell}+t_z^{+\ell})
  +\strut\nonumber\\&&\strut\kern-12pt
  +2\Big[(1+\mu_1-\mu_2)(Q_c^2-Q_b^2)+\strut\nonumber\\&&\strut
  -(3+2\mu_1+\mu_1^2-4\mu_2-2\mu_1\mu_2+\mu_2^2)Q_W^2\Big]
  (t_{zW}^{-\ell}-t_{zW}^{+\ell}+t_{zW}^\ell)
  +\strut\nonumber\\&&\strut\kern-12pt
  -\frac{1+\mu_1-\mu_2}{\smu}\Big[(1+\mu_1-\mu_2)(Q_c^2-Q_b^2)
  -2(1+2\mu_1-\mu_2)Q_W^2\Big](t_{zW}^{-\ell}+t_{zW}^{+\ell})
  +\strut\nonumber\\&&\strut\kern-12pt
  +\Frac14\Big[\left((4-\mu_2)\lambda-\mu_1(17+16\mu_1-32\mu_2+\mu_1\mu_2
  +5\mu_2^2)\right)Q_c^2+\strut\nonumber\\&&\strut
  -\left((4-\mu_1+2\mu_2)\lambda+\mu_1(15+8\mu_2+5\mu_1\mu_2+\mu_2^2)\right)
  Q_b^2+\strut\nonumber\\&&\strut
  -2\left((4+\mu_1+\mu_2)\lambda+\mu_1(16+\mu_1+5\mu_2-\mu_1\mu_2-\mu_2^2)
  \right)Q_W^2\Big]\ell_1+\strut\nonumber\\&&\strut\kern-12pt
  -\Frac14\Big[\left(2(2+\mu_1-\mu_2)\lambda-17\mu_1-16\mu_1^2+3\mu_2
  +20\mu_1\mu_2-4\mu_2^2\right)Q_c^2+\strut\nonumber\\&&\strut
  -\left(2(2-\mu_1+\mu_2)\lambda+15\mu_1+3\mu_2+4\mu_1\mu_2-4\mu_2^2\right)
  Q_b^2+\strut\nonumber\\&&\strut
  -2(4+8\mu_1+5\mu_1^2-12\mu_2-12\mu_1\mu_2+7\mu_2^2)Q_W^2\Big]\ell_{1W}
  +\strut\nonumber\\&&\strut\kern-12pt
  +\Frac18\Big[(16-67\mu_1-15\mu_1^2-3\mu_2+38\mu_1\mu_2-11\mu_2^2)Q_c^2
  +\strut\nonumber\\&&\strut
  -(8+11\mu_1+11\mu_1^2-21\mu_2-38\mu_1\mu_2+15\mu_2^2)Q_b^2
  +\strut\nonumber\\&&\strut
  -2(24+\mu_1+9\mu_1^2-31\mu_2-14\mu_1\mu_2+9\mu_2^2)Q_W^2\Big]\sla,
  \nonumber\\[12pt]
\lefteqn{H^{++}(\alpha)\ =\ \Frac12(1-\mu_1-\mu_2-\sla)\sla V_-^*
  +\mu_1\mu_2\sla V_+-(Q_c^2+Q_b^2+Q_W^2)\lambda\ell_+
  +\strut}\nonumber\\&&\strut\kern-12pt
  -\Frac12\Big[(1-\mu_2)Q_c^2-\mu_1(Q_b^2-Q_W^2)\Big]
  \left((1-\mu_1-\mu_2)(t_\zeta^{\ell*}-2t_z^{\ell+})
  -\sla(t_\zeta^{\ell*}-2t_z^{\ell-})\right)
  +\strut\nonumber\\&&\strut\kern-12pt
  +\Frac12\Big[Q_c^2-Q_b^2+(\mu_1-\mu_2)Q_W^2\Big]
  \left((1-\mu_1-\mu_2)(t_{\zeta W}^{\ell*}+2t_{zW}^{\ell+})
  -\sla(t_{\zeta W}^{\ell*}+2t_{zW}^{\ell-})\right)
  +\strut\nonumber\\&&\strut\kern-12pt
  +\Big[(1-2\mu_1-2\mu_2-\mu_1\mu_2+\mu_2^2)Q_c^2
  -\mu_1(1+\mu_1-\mu_2)(Q_b^2-Q_W^2)\Big]t_z^\ell
  +\strut\nonumber\\&&\strut\kern-12pt
  +\mu_1\Big[(1+5\mu_1-\mu_2)Q_c^2+2\mu_1(Q_b^2-Q_W^2)\Big]
  (t_z^{-\ell}-t_z^{+\ell}+t_z^\ell)+\strut\nonumber\\&&\strut\kern-12pt
  +\Frac12\smu\Big[(1-10\mu_1-3\mu_1^2-2\mu_2+2\mu_1\mu_2+\mu_2^2)Q_c^2
  +\strut\nonumber\\&&\strut
  -2\mu_1(1+\mu_1-\mu_2)(Q_b^2-Q_W^2)\Big](t_z^{-\ell}+t_z^{+\ell})
  +\strut\nonumber\\&&\strut\kern-12pt
  -\Big[(1+\mu_1-\mu_2)(Q_c^2-Q_b^2)+\strut\nonumber\\&&\strut
  -(3+2\mu_1+\mu_1^2-4\mu_2-2\mu_1\mu_2+\mu_2^2)Q_W^2\Big]
  (t_{zW}^{-\ell}-t_{zW}^{+\ell}+2t_{zW}^\ell)
  +\strut\nonumber\\&&\strut\kern-12pt
  +\frac{1+\mu_1-\mu_2}{2\smu}\Big[(1+\mu_1-\mu_2)(Q_c^2-Q_b^2)
  -2(1+2\mu_1-\mu_2)Q_W^2\Big](t_{zW}^{-\ell}+t_{zW}^{+\ell})
  +\strut\nonumber\\&&\strut\kern-12pt
  +\Frac12\Big[(1-\mu_1)(5-3\mu_1+4\mu_2)Q_c^2
  -(9-10\mu_1+\mu_1^2+6\mu_2-2\mu_1\mu_2)Q_b^2+\strut\nonumber\\&&\strut
  -2(1-\mu_1)(5-\mu_1+\mu_2)Q_W^2\Big]\ell_0+\strut\nonumber\\&&\strut\kern-12pt
  -\Frac14\Big[\left(3-8\mu_1-4\mu_1^2-6\mu_2+10\mu_1\mu_2+3\mu_2^2
  +(1+5\mu_1-\mu_2)\sla\right)Q_c^2+\strut\nonumber\\&&\strut
  -\mu_1(4+7\mu_1-4\mu_2)Q_b^2-2(1+5\mu_1-\mu_2)(1-\mu_2)Q_W^2\Big]\ell_1
  +\strut\nonumber\\&&\strut\kern-12pt
  +\Frac14\Big[(1-4\mu_1-6\mu_1^2-4\mu_2+6\mu_2^2+8\sla)Q_c^2
  +\strut\nonumber\\&&\strut
  -(1+2\mu_1+8\mu_1^2+6\mu_2-8\mu_1\mu_2+8\sla)Q_b^2+\strut\nonumber\\&&\strut
  -2(2+3\mu_1+\mu_1^2-5\mu_2-6\mu_1\mu_2+5\mu_2^2+4\sla)Q_W^2\Big]\ell_{1W}
  +\strut\nonumber\\&&\strut\kern-12pt
  -\Frac14\Big[(17+7\mu_1-8\mu_2)Q_c^2-(13-3\mu_1+2\mu_2)Q_b^2
  -6(3+\mu_1-\mu_2)Q_W^2\Big]\lambda_-+\strut\nonumber\\&&\strut\kern-12pt
  +\Frac1{24}\Big[3(3+15\mu_1+3\mu_2+\sla)Q_c^2
  +3(15+7\mu_1-29\mu_2-9\sla)Q_b^2+\strut\nonumber\\&&\strut
  +4(29-10\mu_1-\mu_1^2-34\mu_2+2\mu_1\mu_2-\mu_2^2-9\sla)Q_W^2\Big]\sla,
  \nonumber\\[12pt]
\lefteqn{H^{+-}(\alpha)\ =\ H^{-+}(\alpha)\ =\
  -\mu_1\Big[(1+5\mu_1-\mu_2)Q_c^2+2\mu_1(Q_b^2-Q_W^2)\Big](t_z^{-\ell}
  -t_z^{+\ell}+t_z^\ell)+\strut}\nonumber\\&&\strut\kern-12pt
  -\Frac12\smu\Big[(1-10\mu_1-3\mu_1^2-2\mu_2+2\mu_1\mu_2+\mu_2^2)Q_c^2
  +\strut\nonumber\\&&\strut
  -2\mu_1(1+\mu_1-\mu_2)(Q_b^2-Q_W^2)\Big](t_z^{-\ell}+t_z^{+\ell})
  +\strut\nonumber\\&&\strut\kern-12pt
  +\Big[(1+\mu_1-\mu_2)(Q_c^2-Q_b^2)+\strut\nonumber\\&&\strut
  -(3+2\mu_1+\mu_1^2-4\mu_2-2\mu_1\mu_2+\mu_2^2)Q_W^2\Big]
  (t_{zW}^{-\ell}-t_{zW}^{+\ell}+t_{zW}^\ell)
  +\strut\nonumber\\&&\strut\kern-12pt
  -\frac{1+\mu_1-\mu_2}{2\smu}\Big[(1+\mu_1-\mu_2)(Q_c^2-Q_b^2)
  -2(1+2\mu_1-\mu_2)Q_W^2\Big](t_{zW}^{-\ell}+t_{zW}^{+\ell})
  +\strut\nonumber\\&&\strut\kern-12pt
  +\Frac12\Big[(1-6\mu_1-3\mu_1^2-2\mu_2+6\mu_1\mu_2+\mu_2^2)Q_c^2
  +\strut\nonumber\\&&\strut
  -(1+\mu_1+2\mu_1^2-2\mu_2-\mu_1\mu_2+\mu_2^2)Q_b^2
  -2(1+\mu_1-\mu_2)^2Q_W^2\Big]\ell_1+\strut\nonumber\\&&\strut\kern-12pt
  -\Frac12\Big[(1-6\mu_1-3\mu_1^2-2\mu_2+2\mu_1\mu_2+\mu_2^2)Q_c^2
  +\strut\nonumber\\&&\strut
  -(1+\mu_1+2\mu_1^2-\mu_2-2\mu_1\mu_2)Q_b^2
  -2(1+\mu_1-2\mu_2)(1+\mu_1-\mu_2)Q_W^2\Big]\ell_{1W}
  +\strut\nonumber\\&&\strut\kern-12pt
  +\Frac12\Big[2(1-5\mu_1-\mu_2)Q_c^2-(1+3\mu_1-\mu_2)Q_b^2
  -2(3+\mu_1-3\mu_2)Q_W^2\Big]\sla,\nonumber\\[12pt]
\lefteqn{H^{--}(\alpha)\ =\ \Frac12(1-\mu_1-\mu_2+\sla)\sla V_-^*
  +\mu_1\mu_2\sla V_++(Q_c^2+Q_b^2+Q_W^2)\lambda\ell_+
  +\strut}\nonumber\\&&\strut\kern-12pt
  -\Frac12\Big[(1-\mu_2)Q_c^2-\mu_1(Q_b^2-Q_W^2)\Big]
  \left((1-\mu_1-\mu_2)(t_\zeta^{\ell*}-2t_z^{\ell+})
  +\sla(t_\zeta^{\ell*}-2t_z^{\ell-})\right)
  +\strut\nonumber\\&&\strut\kern-12pt
  +\Frac12\Big[Q_c^2-Q_b^2+(\mu_1-\mu_2)Q_W^2\Big]
  \left((1-\mu_1-\mu_2)(t_{\zeta W}^{\ell*}+2t_{zW}^{\ell+})
  +\sla(t_{\zeta W}^{\ell*}+2t_{zW}^{\ell-})\right)
  +\strut\nonumber\\&&\strut\kern-12pt
  -\Big[(1-2\mu_1-2\mu_2-\mu_1\mu_2+\mu_2^2)Q_c^2
  -\mu_1(1+\mu_1-\mu_2)(Q_b^2-Q_W^2)\Big]t_z^\ell
  +\strut\nonumber\\&&\strut\kern-12pt
  +\mu_1\Big[(1+5\mu_1-\mu_2)Q_c^2+2\mu_1(Q_b^2-Q_W^2)\Big]
  (t_z^{-\ell}-t_z^{+\ell}+t_z^\ell)+\strut\nonumber\\&&\strut\kern-12pt
  +\Frac12\smu\Big[(1-10\mu_1-3\mu_1^2-2\mu_2+2\mu_1\mu_2+\mu_2^2)Q_c^2
  +\strut\nonumber\\&&\strut
  -2\mu_1(1+\mu_1-\mu_2)(Q_b^2-Q_W^2)\Big](t_z^{-\ell}+t_z^{+\ell})
  +\strut\nonumber\\&&\strut\kern-12pt
  -\Big[(1+\mu_1-\mu_2)(Q_c^2-Q_b^2)+\strut\nonumber\\&&\strut
  -(3+2\mu_1+\mu_1^2-4\mu_2-2\mu_1\mu_2+\mu_2^2)Q_W^2\Big]
  (t_{zW}^{-\ell}-t_{zW}^{+\ell})+\strut\nonumber\\&&\strut\kern-12pt
  +\frac{1+\mu_1-\mu_2}{2\smu}\Big[(1+\mu_1-\mu_2)(Q_c^2-Q_b^2)
  -2(1+2\mu_1-\mu_2)Q_W^2\Big](t_{zW}^{-\ell}+t_{zW}^{+\ell})
  +\strut\nonumber\\&&\strut\kern-12pt
  -\Frac12\Big[(1-\mu_1)(5-3\mu_1+4\mu_2)Q_c^2
  -(9-10\mu_1+\mu_1^2+6\mu_2-2\mu_1\mu_2)Q_b^2+\strut\nonumber\\&&\strut
  -2(1-\mu_1)(5-\mu_1+\mu_2)Q_W^2\Big]\ell_0+\strut\nonumber\\&&\strut\kern-12pt
  -\Frac14\Big[\left(3-8\mu_1-4\mu_1^2-6\mu_2+10\mu_1\mu_2+3\mu_2^2
  -(1+5\mu_1-\mu_2)\sla\right)Q_c^2+\strut\nonumber\\&&\strut
  -\mu_1(4+7\mu_1-4\mu_2)Q_b^2-2(1+5\mu_1-\mu_2)(1-\mu_2)Q_W^2\Big]\ell_1
  +\strut\nonumber\\&&\strut\kern-12pt
  +\Frac14\Big[(1-4\mu_1-6\mu_1^2-4\mu_2+6\mu_2^2-8\sla)Q_c^2
  +\strut\nonumber\\&&\strut
  -(1+2\mu_1+8\mu_1^2+6\mu_2-8\mu_1\mu_2-8\sla)Q_b^2+\strut\nonumber\\&&\strut
  -2(2+3\mu_1+\mu_1^2-5\mu_2-6\mu_1\mu_2+5\mu_2^2-4\sla)Q_W^2\Big]\ell_{1W}
  +\strut\nonumber\\&&\strut\kern-12pt
  +\Frac14\Big[(17+7\mu_1-8\mu_2)Q_c^2-(13-3\mu_1+2\mu_2)Q_b^2
  -6(3+\mu_1-\mu_2)Q_W^2\Big]\lambda_-+\strut\nonumber\\&&\strut\kern-12pt
  +\Frac1{24}\Big[3(3+15\mu_1+3\mu_2-\sla)Q_c^2
  +3(15+7\mu_1-29\mu_2+9\sla)Q_b^2+\strut\nonumber\\&&\strut
  +4(29-10\mu_1-\mu_1^2-34\mu_2+2\mu_1\mu_2-\mu_2^2+9\sla)Q_W^2\Big]\sla.
\end{eqnarray}
Analytical expressions for the form factors $V_-^*$, $V_+$, $V_1$ and $V_2$
are found in Appendix~B.

\section{Discussion}
In general, electroweak corrections are assumed to be small. However, the
size of the correction depends strongly on the subtraction scheme used. Using
the $\alpha(0)$ scheme as the canonical choice from the perturbation series is
not an option for the decay of a massive $W$ boson, as the scale of the problem
is far from being zero. Instead, one might consider the $\alpha(m_W^2)$ scheme
with the renormalisation scale fixed to the mass of the $W$ boson. As argued
in Ref.~\cite{Denner:2019vbn}, however, the best choice is the $G_\mu$ scheme
(traditionally known as the $G_F$ scheme related to the Fermi constant as the
best measured quantity in the process). This scheme and the consequences of
its application is discussed in Appendix~C.1. Frankly speaking, using the
latter two schemes, large logarithms from light fermion loops are removed from
the result which then is indeed a small correction to the Born term result.

\begin{figure}\begin{center}
\epsfig{figure=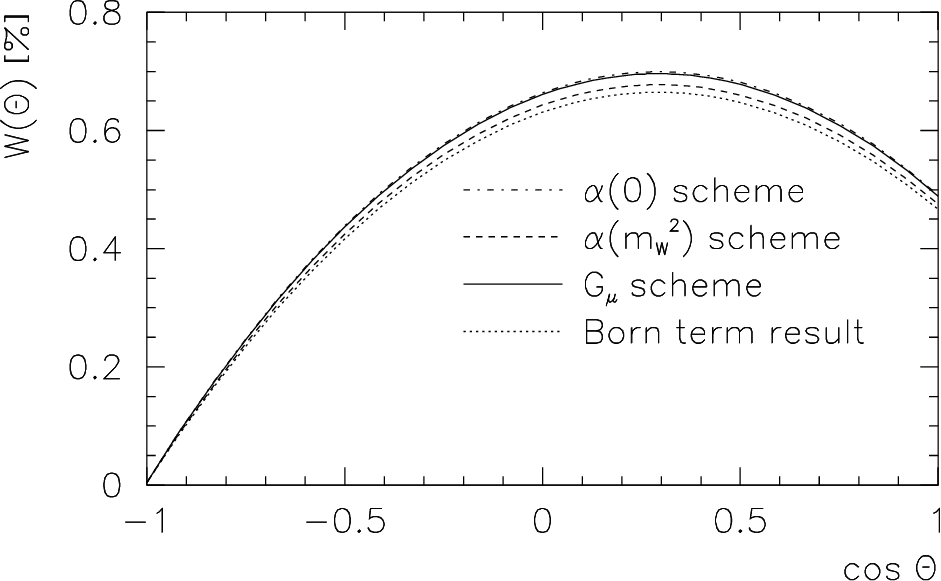, scale=0.8}
\caption{\label{distrib}Angular distribution in dependence on the subtraction
scheme employed}
\end{center}\end{figure}

The effect of the different schemes is displayed in the angular distributions
shown in Fig.~\ref{distrib}. The numerical analysis is done by employing the
values $m_W=80.3962(133)\GeV$, $m_Z=91.1880(20)\GeV$, $m_H=125.20(11)\GeV$,
$m_c=1.2730(46)\GeV$, $m_b=4.183(7)\GeV$, $m_t=172.57(29)\GeV$, together with
values for the lighter leptons taken from
Ref.~\cite{ParticleDataGroup:2024cfk}. For the light quarks we use the values
$m_u\approx m_d=0.06\GeV$ and $m_s=0.17\GeV$ obtained by fitting these masses
to the hadronic vacuum polarisation (cf.\ Appendix~E).
As it turns out, in the $\alpha(m_W^2)$ scheme the electroweak correction is
of the order of $1.9\%$ while for the other schemes the corrections are
larger, $4.7\%$ and $5.2\%$ for the $G_\mu$ and $\alpha(0)$ schemes,
respectively.\footnote{Note that different from
Refs.~\cite{Groote:2012xr,Groote:2013hc,Groote:2013xt} we have chosen the
positive $z$ axis in the quark frame as the direction of motion of the quark.
A different convention means an interchange $\theta\leftrightarrow-\theta$,
i.e., the diagram for the angular distribution will be mirrored at the
vertical axis.} In the second column of Table~\ref{helischeme} we give the
helicity bilinears for the different schemes.
\begin{table}\begin{center}
\caption{\label{helischeme}helicity bilinears for the Born term contribution
and the electroweak radiative corrections. Shown are numerical values for the
exact fermion masses, and for the fermion masses except for the top quark mass
set to zero, i.e., in the collinear limit.}
\vspace{12pt}
\begin{tabular}{|lc||c|c|}\hline
scheme,&bilinear&$m_f$&$m_f=0$, $f\ne t$\\\hline
LO&$H^{00}$&$0.00148$&$0$\\
  &$H^{++}$&$7\times 10^{-7}$&$0$\\
  &$H^{+-}$&$0$&$0$\\
  &$H^{--}$&$0.99704$&$1$\\\hline
$\alpha(0)$&$H^{00}$&$+0.00019$&$+0.00206$\\
           &$H^{++}$&$+0.00003$&$-0.00094$\\
           &$H^{+-}$&$+0.00005$&$+0.00103$\\
           &$H^{--}$&$+0.05195$&$+0.07333$\\\hline
$\alpha(m_W^2)$&$H^{00}$&$+0.00015$&$+0.00222$\\
               &$H^{++}$&$+0.00003$&$-0.00101$\\
               &$H^{+-}$&$+0.00006$&$+0.00111$\\
               &$H^{--}$&$+0.01911$&$+0.04208$\\\hline
$G_\mu$&$H^{00}$&$+0.00018$&$+0.00214$\\
       &$H^{++}$&$+0.00003$&$-0.00097$\\
       &$H^{+-}$&$+0.00006$&$+0.00107$\\
       &$H^{--}$&$+0.04701$&$+0.06916$\\\hline
\end{tabular}\end{center}
\end{table}

\begin{figure}\begin{center}
\epsfig{figure=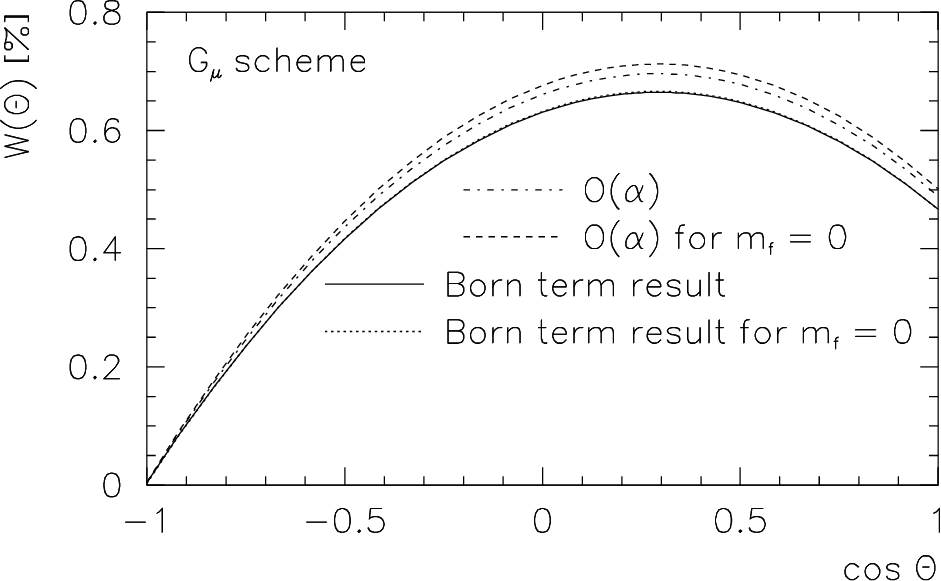, scale=0.8}
\caption{\label{distribg}Mass dependence of the angular distribution in the
  $G_\mu$ scheme}
\end{center}\end{figure}

On the other hand, we can investigate the dependence on the masses. For this,
the collinear limit $m_c,m_b\to 0$ can be taken, as the collinear
singularities between tree and loop contributions were not only implicitly
cancelled but explicitly truncated on both sides in our approach. The last
column shows the results in the collinear limit where all fermion masses
except for the top quark mass are set to zero. The collinear limit is used
also for the observables in Table~\ref{obsscheme}. Finally, in
Figure~\ref{distribg} we have compared the angular distribution for the LO
Born term result and for the NLO result in the $G_\mu$ scheme with the results
in the collinear limit in order to study mass effects.
While mass corrections for the Born term results are difficult to discern and
are of the order $-0.2\%$, mass effects in the $G_\mu$ scheme amount to
$-2.3\%$, with the mass effects for the $\alpha(m_W^2)$ and $\alpha(0)$
schemes given by $-2.4\%$ and $-2.2\%$, respectively. At the same time, we see
that in the collinear limit the electroweak correction is larger and amounts
to $7.0\%$, $4.2\%$ and $7.4\%$ for the $G_\mu$, $\alpha(m_W^2)$ and
$\alpha(0)$ schemes, respectively. This is also echoed in Tab.~\ref{obsscheme}.
Vice versa, it seems that taking exact fermion masses, the differences due to
the choice of the scheme diminish.

Up to a general factor of $e^2q^2|V_{cb}|^2/s_W^2$, the Born term results in
the collinear limit are given by $H^{00}({\it Born\/})=H^{++}({\it Born\/})
=H^{+-}({\it Born\/})=H^{-+}({\it Born\/})=0$, while the only non-vanishing
amplitude is $H^{--}({\it Born\/})=1$. On the other hand, up to an additional
general factor of $\alpha/(4\pi)$, the analytic expressions for the NLO EW
corrections to the helicity bilinears in the collinear limit for the $G_\mu$
scheme read
\begin{eqnarray}
\lefteqn{H^{00}(\alpha)\ =\ 2Q_c^2-Q_b^2-6Q_W^2+\frac{2\pi^2}3Q_cQ_W
  -\frac{4\pi^2}3Q_bQ_W,}\nonumber\\
\lefteqn{H^{++}(\alpha)\ =\ -\frac{15}4Q_c^2+4Q_b^2+\frac{47}6Q_W^2
  -\frac{\pi^2}2Q_cQ_W+\frac{3\pi^2}2Q_bQ_W,}\nonumber\\
\lefteqn{H^{+-}(\alpha)\ =\ H^{-+}(\alpha)\ =\ Q_c^2-\frac12Q_b^2-3Q_W^2
  +\frac{\pi^2}3Q_cQ_W-\frac{2\pi^2}3Q_bQ_W,}\nonumber\\
\lefteqn{H^{--}(\alpha)\ =\ \frac52Q_c^2-\frac94Q_b^2+\frac{11}6Q_W^2
  -\frac{5\pi^2}6Q_cQ_W+\frac{\pi^2}2Q_bQ_W+\strut}\nonumber\\&&\strut
  -2g_c^-g_b^-\pfrac{m_W^2+m_Z^2}{m_W^2}^2
  \left[\Li_2\left(-\frac{m_W^2}{m_Z^2}\right)
  +\ln\left(1+\frac{m_W^2}{m_Z^2}\right)\ln\left(-\frac{m_W^2}{m_Z^2}\right)
  \right]+\strut\nonumber\\&&\strut
  +2g_c^-\frac{m_W^2+2m_Z^2}{m_Zm_Ws_W}
  \Bigg[\Li_2\pfrac2{1-\sqrt{1-4m_W^2/m_Z^2}}
  +\Li_2\pfrac2{1+\sqrt{1-4m_W^2/m_Z^2}}+\strut\nonumber\\&&\strut\qquad
  -\Li_2\pfrac{2(1-m_W^2/m_Z^2)}{1-\sqrt{1-4m_W^2/m_Z^2}}
  -\Li_2\pfrac{2(1-m_W^2/m_Z^2)}{1+\sqrt{1-4m_W^2/m_Z^2}}\Bigg]
  +\strut\nonumber\\&&\strut
  -2g_b^-\frac{m_W^2+2m_Z^2}{m_Zm_Ws_W}
  \left[\Li_2\pfrac{2m_W^2/m_Z^2}{1-\sqrt{1-4m_W^2/m_Z^2}}
  +\Li_2\pfrac{2m_W^2/m_Z^2}{1+\sqrt{1-4m_W^2/m_Z^2}}\right]
  +\strut\nonumber\\&&\strut
  +\frac{m_Z^2}{2m_W^4}\left(1-2Q_cs_W^2+Q_bs_W^2-4Q_cQ_bs_W^2
  \right)\left[(2m_W^2+m_Z^2)\left(\ln\pfrac{m_Z^2}{m_W^2}+1\right)\right]
  +\strut\nonumber\\&&\strut
  +\frac{4m_W^2-m_Z^2}{2m_W^2s_W^2}+\frac1{4s_W^2}
  +\frac{g_c^{-2}+g_b^{-2}}2\left(\ln\pfrac{m_Z^2}{m_W^2}+\frac12\right)
  +\strut\nonumber\\&&\strut
  -\frac{m_t^2|V_{tb}|^2}{16(m_t^2-m_W^2)^2m_W^2s_W^2}
  \left[3(m_t^4-m_W^4)-2m_t^2(m_t^2+2m_W^2)\ln\pfrac{m_t^2}{m_W^2}\right]
  +\strut\nonumber\\&&\strut
  +\frac{48m_W^{10}+20m_W^8m_Z^2-320m_W^6m_Z^4+249m_W^4m_Z^6-30m_W^2m_Z^8
  -3m_Z^{10}}{48m_Z^4m_W^6s_W^4}\ln\pfrac{m_Z^2}{m_W^2}
  +\strut\nonumber\\&&\strut
  +\frac{m_t^2(5m_t^4-9m_W^4)}{12m_W^6s_W^2}\ln\pfrac{m_t^2}{m_W^2}
  -\frac{(m_t^2-m_W^2)(5m_t^4+5m_t^2m_W^2-4m_W^4)}{12m_W^6s_W^2}
  \ln\left(\frac{m_t^2}{m_W^2}-1\right)+\strut\kern-3pt\nonumber\\&&\strut
  -\frac{3m_H^8-18m_H^6m_W^2+51m_H^4m_W^4-64m_H^2m_W^6+12m_W^8}{48(m_H^2-m_W^2)
  m_W^6s_W^2}\ln\pfrac{m_H^2}{m_W^2}+\strut\nonumber\\&&\strut
  +\frac{28m_W^6+52m_W^4m_Z^2-13m_W^2m_Z^4-m_Z^6}{4m_Z^2m_W^6s_W^2}
  \sqrt{4m_W^2m_Z^2-m_Z^4}\arctan\left(\sqrt{\frac{2m_W-m_Z}{2m_W+m_Z}}\right)
  +\strut\nonumber\\&&\strut
  +\frac{m_H^6-7m_H^4m_W^2+20m_H^2m_W^4-28m_W^6}{4(4m_W^2-m_H^2)m_W^6s_W^2}
  \sqrt{4m_H^2m_W^2-m_H^4}\arctan\left(\sqrt{\frac{2m_W-m_H}{2m_W+m_H}}\right)
  +\strut\nonumber\\&&\strut
  +\frac{3m_H^4-10m_t^4-10m_H^2m_W^2-8m_t^2m_W^2-184m_W^4+38m_W^2m_Z^2
  +3m_Z^4}{24m_W^4s_W^2}+\strut\nonumber\\&&\strut
  +\frac{3V_{tb}^2m_t^2}{16m_W^2s_W^2}
  \left[\frac{m_t^2+m_W^2}{m_t^2-m_W^2}-\frac{2m_t^2(m_t^2+2m_W^2)}{3
  (m_t^2-m_W^2)^2}\ln\pfrac{m_t^2}{m_W^2}\right].
\end{eqnarray}
Note that only $H^{--}(\alpha)$ depends on the loop corrections and the
corrections due to the renormalisation factors and, therefore, on the chosen
subtraction scheme.

\begin{figure}\begin{center}
\epsfig{figure=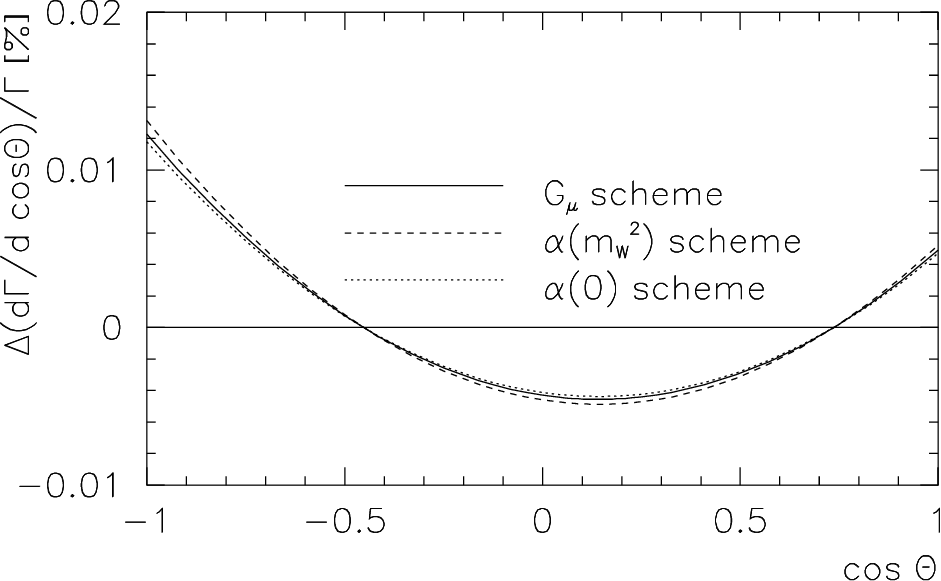, scale=0.8}
\caption{\label{distrin}Differences between the normalised differential decay
rate $\hat\Gamma=(d\Gamma/d\cos\theta)/\Gamma$ to first order for the different
schemes and the Born result.}
\end{center}\end{figure}

The considerations done so far are instructive to compare the effect of first
order radiative corrections and final state mass effects with the Born term
result in the collinear limit. Note, however, that neither $W(\theta)$ nor the
helicity bilinears are observables. Observables are based on the differential
decay rate given by
\begin{eqnarray}\label{Gamma}
\frac{d\Gamma}{d\cos\theta}&=&\Gamma_0^0\sum_{m=0,\pm1}\rho_{mm}H^{mm}(\theta)
  \ =\ \Gamma_0^0\sum_{m,m'=0,\pm1}\rho_{mm}\,d^1_{mm'}(\theta)
  \,d^1_{mm'}(\theta)\,\,H^{m'm'}\ =\nonumber\\
  &=&\Gamma_0^0\Bigg[\frac38(1+\cos^2\theta)\,\Big((\rho_{++}+\rho_{--})\,
  (H^{++}+H^{--})+2\rho_{00}H^{00}\Big)+\strut\nonumber\\&&\strut
  +\frac34\cos\theta\,\Big((\rho_{++}-\rho_{--})\,
  (H^{++}-H^{--})\Big)+\strut\nonumber\\&&\strut
  +\frac34\sin^2\theta\,\Big((\rho_{++}+\rho_{--})H^{00}
  +\rho_{00}(H^{++}+H^{--}-H^{00})\Big)\Bigg],
\end{eqnarray}
where
\begin{equation}
\Gamma_0^0=\frac{G_\mu m_W^3}{6\pi\sqrt2}N_c|V_{ij}|^2
  =\frac{\alpha m_W}{12s_W^2}N_c|V_{ij}|^2
\end{equation}
is the Born term rate~(\ref{Gamma0}) in the collinear limit into quark
flavours $i$ and $j$. In Figure~\ref{distrin} we show the differences between
the normalised decay rates $\hat\Gamma=(d\Gamma/d\cos\theta)/\Gamma$ to first
order for the different schemes and the Born term result for the same quantity.

Three observables characterising the angular distribution are considered.
These observables are the maximal curvature of the angular distribution,
called the convexity parameter, given by
\begin{equation}
c_f=\frac{d^2\hat\Gamma}{d(\cos\theta)^2}
  =\frac34(\rho_{++}-2\rho_{00}+\rho_{--})
  \pfrac{H^{++}-2H^{00}+H^{--}}{H^{++}+H^{00}+H^{--}},
\end{equation}
and the forward--backward asymmetry of the decay distribution defined by
\begin{equation}
A_{FB}=\frac{\hat\Gamma(F)-\hat\Gamma(B)}{\hat\Gamma(F)+\hat\Gamma(B)}
  =\frac34(\rho_{++}-\rho_{--})\pfrac{H^{++}-H^{--}}{H^{++}+H^{00}+H^{--}}
\end{equation}
with $\hat\Gamma(F)=\hat\Gamma(0\le\theta\le\pi/2)$ and
$\hat\Gamma(B)=\hat\Gamma(\pi/2\le\theta\le\pi)$, and the position of the
extremum
\begin{equation}
\cos\theta\Big|_{\rm extr}=-\frac{A_{FB}}{c_f}=
  -\frac{(\rho_{++}-\rho_{--})}{(\rho_{++}-2\rho_{00}
  +\rho_{--})}\pfrac{H^{++}-H^{--}}{H^{++}-2H^{00}+H^{--}}.
\end{equation}
The values for these observables are found in the first lines in the scheme
blocks of Table~\ref{obsscheme}.
\begin{table}\begin{center}
\caption{\label{obsscheme}Numerical values for the three observables
$c_f$, $A_{\rm FB}$ and $\cos\theta_{\rm extr}$ to leading order and for
three next-to-leading order schemes with exact fermion masses and in the
collinear limit}
\begin{tabular}{|lc||c|c|c|}\hline
scheme,&case&$c_f$&$A_{\rm FB}$&$\cos\theta_{\rm extr}$\\\hline
             LO&$m_f\ne 0$&$-0.791269$&$+0.231288$&$+0.292300$\\
                  &$m_f=0$&$-0.795975$&$+0.231975$&$+0.291435$\\\hline
    $\alpha(0)$&$m_f\ne 0$&$-0.832342$&$+0.243333$&$+0.292347$\\
                  &$m_f=0$&$-0.850489$&$+0.249258$&$+0.293076$\\\hline
$\alpha(m_W^2)$&$m_f\ne 0$&$-0.806268$&$+0.235714$&$+0.292352$\\
                  &$m_f=0$&$-0.825215$&$+0.242002$&$+0.293260$\\\hline
        $G_\mu$&$m_f\ne 0$&$-0.828419$&$+0.242187$&$+0.292349$\\
                  &$m_f=0$&$-0.847012$&$+0.248296$&$+0.293143$\\\hline
\end{tabular}\end{center}
\end{table}
The percentage values for the radiatively corrected to the
Born term normalised decay rate are smaller, for $\alpha(0)$, $\alpha(m_W^2)$
and $G_\mu$ schemes given by $+0.006\%$, $-0.007\%$ and $-0.006\%$,
respectively, while the corresponding values in the collinear limit are an
order of magnitude higher, $-0.09\%$, $-0.07\%$ and $-0.10\%$, respectively.
The mass correction in the Born term case amount of $-0.08\%$, while the mass
corrections at first order are $-0.0003\%$, $+0.003\%$, and $+0.003\%$, thus
unvisible.

\section{Conclusions}
In this publication we have given analytic results for the NLO electroweak
corrections to the decay process $W^+(\uparrow)\to c\bar b$ of a polarised
$W$ boson into quarks, taking into account all fermion and boson masses
contributing to this process. We have estimated the size of the corrections
and shown the dependence on the subtraction scheme of the renormalisation
procedure. Finally, we have performed the collinear limit in order to estimate
the influence of mass effects to this decay process.

\subsection*{Acknowledgments}
SG acknowledges useful and fuitful discussions with Ansgar Denner. The
research was supported by the European Regional Development Fund under Grant
No.~TK133, and by the Estonian Research Council under Grant No.~PRG356.


\begin{appendix}

\section{Analytical expressions for the tree corrections}
\setcounter{equation}{0}\def\theequation{A\arabic{equation}}
The integrals necessary for the tree corrections are of the form
\begin{equation}
I(n_1,n_2;\Lambda)=\int_{y_{20}}^{y_{2-}}dy_2\int_{y_{1-}}^{y_{1+}}dy_1
  \frac{y_1^{n_1}y_2^{n_2}}{(y_1+y_2)^2}.
\end{equation}
While the integrals are regular for $n_1+n_2>0$ and can be calculated taking
$\Lambda=0$, the integrals at the border line $n_1+n_2=0$ are IR singular.
After performing the integation over $y_1$, the integration over $y_2$ cannot
be calculated analytically for a general parameter $\Lambda$ in a closed form.
Instead, we split the integral up into a divergent part $D(n_1,n_2;\Lambda)$
and a convergent part $C(n_1,n_2)$, where the divergent part is given by
$I(n_1,n_2;\Lambda)$ with the integrand approximated for small values of
$y_2$, while the convergent part is given by the limit $\Lambda\to 0$ of the
difference $I(n_1,n_2;\Lambda)-D(n_1,n_2;\Lambda)$ that is finite by
construction. In technical terms, for the calculation of the divergent part
in terms of a couple of elementary $\zeta$-integrals, we use the substitution
$y_2=\Lambda+\sqrt{\Lambda\mu_2}(1+\zeta)/\sqrt\zeta$ to perform the integral
over $\zeta$ between $\zeta_-=\Lambda\mu_2/((1-\smu)^2-\mu_2)^2$ to $1$. After
having performed this substitution, all occurrences of $\Lambda$ in the
(simplified) integrand can be neglected and the integral can be performed. The
simplified integrand is then transformed back to $y_2$ and subtracted from the
full integrand and integrated in the limit $\Lambda\to 0$ by using the
substitution $y_2=1+\mu_1-\mu_2-\smu(z+1/z)$ to obtain the convergent part
again in terms of a couple of $z$-integrals. The corresponding limits are
given by $z_-=(1+\mu_1-\mu_2-\sqrt{\lambda(1,\mu_1,\mu_2)})/(2\smu)$ and $1$.
A large portion of the $\zeta$- and $z$-integrals can be expressed in terms of
rational functions and logarithms, containing
\begin{eqnarray}\label{ells}
\ell_\zeta&=&\ln\pfrac{\lambda^2}{\Lambda\mu_1\mu_2},\qquad
  \ell_0\ =\ \ln\pfrac{1-\smu}{\sqrt{\mu_2}},\qquad
  \ell_+\ =\ \ln\pfrac{(1+\smu)^2-\mu_2}\smu,\nonumber\\
\ell_1&=&\ln\pfrac{1-\mu_1-\mu_2-\sla}{1-\mu_1-\mu_2+\sla},\qquad
  \ell_{1W}\ =\ \ln\pfrac{1-\mu_1+\mu_2-\sla}{1-\mu_1+\mu_2+\sla}.
\end{eqnarray}
Exceptional are integrals containing a logarithm together with $\zeta$ or $z$
to the power of $-1$. These integrals contain dilogarithms and are kept as
closed form terms $t_\zeta$ and $t_z$, the analytic expressions for these
found in the following. The $\zeta$-terms
\begin{eqnarray}
t_\zeta^{\ell*}&=&\Li_2\left(-\frac{1-\mu_1-\mu_2-\sla}{1-\mu_1-\mu_2+\sla}
  \right)-\Li_2\left(-\frac{1-\mu_1-\mu_2+\sla}{1-\mu_1-\mu_2-\sla}\right)
  +\strut\nonumber\\&&\strut
  -2\Li_2\pfrac\sla{1-\mu_1-\mu_2+\sla}
  +2\Li_2\pfrac{-\sla}{1-\mu_1-\mu_2-\sla}+\strut\nonumber\\&&\strut
  +2\Li_2\pfrac{2\sla}{1-\mu_1-\mu_2+\sla}
  -2\Li_2\pfrac{-2\sla}{1-\mu_1-\mu_2-\sla}+\strut\nonumber\\&&\strut
  +\ell_1\ln\Bigg(\frac{4\mu_1\left((1-\smu)^2-\mu_2\right)^2}{\lambda^2}
  \Bigg),\nonumber\\[7pt]
t_{\zeta W}^{\ell*}&=&\Li_2\left(-\frac{1-\mu_1+\mu_2-\sla}{1-\mu_1+\mu_2+\sla}
  \right)-\Li_2\left(-\frac{1-\mu_1+\mu_2+\sla}{1-\mu_1+\mu_2-\sla}\right)
  +\strut\nonumber\\&&\strut
  -2\Li_2\pfrac\sla{1-\mu_1+\mu_2+\sla}
  +2\Li_2\pfrac{-\sla}{1-\mu_1+\mu_2-\sla}+\strut\nonumber\\&&\strut
  +2\Li_2\pfrac{2\sla}{1-\mu_1+\mu_2+\sla}
  -2\Li_2\pfrac{-2\sla}{1-\mu_1+\mu_2-\sla}+\strut\nonumber\\&&\strut
  +\ell_{1W}\ln\Bigg(\frac{4\mu_1\left((1-\smu)^2-\mu_2\right)^2}{\lambda^2}
  \Bigg)
\end{eqnarray}
are IR subtracted, cf.\ the discussion in Sec.~\ref{irdeal}. The remaining
$z$-terms read
\begin{eqnarray}
\lefteqn{t_z^{\ell-}\ =\ \Li_2\left(-\frac{(1-\smu)^2-\mu_2+\sla}{2\smu}\right)
  -\Li_2\left(-\frac{(1-\smu)^2-\mu_2-\sla}{2\smu}\right)
  +\strut}\nonumber\\&&\strut
  +\Li_2\pfrac{2\sla}{1+\mu_1-\mu_2+\sla}
  +\Li_2\pfrac{2\sla}{1-\mu_1-\mu_2+\sla}
  -\Li_2\pfrac{-2\sla}{1-\mu_1+\mu_2-\sla}+\strut\nonumber\\&&\strut
  +\Li_2\pfrac{(1-\smu)^2-\mu_2-\sla}{1-\mu_1-\mu_2-\sla}
  -\Li_2\pfrac{(1-\smu)^2-\mu_2+\sla}{1-\mu_1-\mu_2+\sla}
  +\strut\nonumber\\&&\strut
  +\Li_2\left(-\frac{(1-\smu)^2-\mu_2+\sla}{1-\mu_1+\mu_2-\sla}\right)
  -\Li_2\left(-\frac{(1-\smu)^2-\mu_2-\sla}{1-\mu_1+\mu_2+\sla}\right)
  +\strut\nonumber\\&&\strut
  +\ell_1\ln\pfrac{(1+\smu)^2-\mu_2-\sla}{(1+\smu)^2-\mu_2+\sla},\\[12pt]
\lefteqn{t_z^{\ell+}\ =\ \Li_2\left(-\frac{(1-\smu)^2-\mu_2-\sla}{2\smu}\right)
  +\Li_2\left(-\frac{(1-\smu)^2-\mu_2+\sla}{2\smu}\right)+2\Li_2(\smu)
  +\strut}\nonumber\\&&\strut
  +\Li_2\pfrac{(1-\smu)^2-\mu_2-\sla}{1-\mu_1-\mu_2-\sla}
  +\Li_2\pfrac{(1-\smu)^2-\mu_2+\sla}{1-\mu_1-\mu_2+\sla}
  +\strut\nonumber\\&&\strut
  -\Li_2\left(-\frac{(1-\smu)^2-\mu_2-\sla}{1-\mu_1+\mu_2+\sla}\right)
  -\Li_2\left(-\frac{(1-\smu)^2-\mu_2+\sla}{1-\mu_1+\mu_2-\sla}\right)
  +\strut\nonumber\\&&\strut
  -\Li_2\pfrac{2\sla}{1-\mu_1-\mu_2+\sla}
  +\Li_2\pfrac{-2\sla}{1-\mu_1+\mu_2-\sla}
  -\Li_2\pfrac{2\sla}{1+\mu_1-\mu_2+\sla}+\strut\nonumber\\&&\strut
  -\Li_2\pfrac{1+\mu_1-\mu_2-\sla}2
  -\Li_2\pfrac{1+\mu_1-\mu_2+\sla}2+\nonumber\\&&\strut
  -\ell_1\ln\pfrac{(1+\smu)^2-\mu_2}{\smu}
  -\ln^2\pfrac{(1+\smu)^2-\mu_2-\sla}{(1+\smu)^2-\mu_2+\sla},\nonumber\\\\
\lefteqn{t_z^{-\ell}\ =\ \Li_2\left(-\frac{(1-\smu)^2-\mu_2-\sla}{2\smu}\right)
  -\Li_2\pfrac{(1-\smu)^2-\mu_2-\sla}{2(1-\smu)}+\strut}\nonumber\\&&\strut
  +\Li_2\left(-\frac{(1-\smu)^2-\mu_2-\sla}{2(1-\smu)\smu}\right),\\[12pt]
\lefteqn{t_z^{+\ell}\ =\ -\Li_2\pfrac{1-\mu_1+\mu_2+\sla}{2(1+\smu)}
  +\Li_2\left(-\frac{1-\mu_1-\mu_2-\sla}{2\smu(1+\smu)}\right)
  +\Li_2\left(-\frac{1+\mu_1-\mu_2-\sla}{2\smu}\right)
  +\strut}\nonumber\\&&\strut
  -\Li_2\left(-\frac{1-\smu}{1+\smu}\right)
  +\Li_2\pfrac{1-\smu}{1+\smu}-\Li_2(-1)+\strut\nonumber\\&&\strut
  +\frac12\ell_1\ln\pfrac{(1+\smu)^2-\mu_2}{\smu(1+\smu)}
  -\frac12\ell_{1W}\ln\pfrac{1+\smu}{\smu}+\strut\nonumber\\&&\strut
  +\frac14\ln\pfrac{1-\mu_1-\mu_2-\sla}{1-\mu_1-\mu_2+\sla}
  \ln\pfrac{1+\mu_1-\mu_2-\sla}{1+\mu_1-\mu_2+\sla}
  +\ln(\smu)\ln\pfrac{1-\smu}{\sqrt{\mu_2}},\\[12pt]
\lefteqn{t_z^\ell\ =\ \Li_2\pfrac{1+\mu_1-\mu_2+\sla}2
  +\Li_2\pfrac{1+\mu_1-\mu_2-\sla}2-2\Li_2(\smu)+\strut}\nonumber\\&&\strut
  +\frac14\ln^2\pfrac{1+\mu_1-\mu_2-\sla}{1+\mu_1-\mu_2+\sla},\\[12pt]
\lefteqn{t_{zW}^{\ell-}
  \ =\ \Li_2\pfrac{(1-\smu)^2-\mu_2+\sla}{1-\mu_1-\mu_2+\sla}
  -\Li_2\pfrac{(1-\smu)^2-\mu_2-\sla}{1-\mu_1-\mu_2-\sla}
  +\strut}\nonumber\\&&\strut
  -\Li_2\pfrac{(1-\smu)^2-\mu_2+\sla}{1+\mu_1-\mu_2+\sla}
  +\Li_2\pfrac{(1-\smu)^2-\mu_2-\sla}{1+\mu_1-\mu_2-\sla}
  +\strut\nonumber\\&&\strut
  +\Li_2\left(-\frac{(1-\smu)^2-\mu_2-\sla}{1-\mu_1+\mu_2+\sla}\right)
  -\Li_2\left(-\frac{(1-\smu)^2-\mu_2+\sla}{1-\mu_1+\mu_2-\sla}\right)
  +\strut\nonumber\\&&\strut
  -\Li_2\pfrac{2\sla}{1-\mu_1-\mu_2+\sla}
  +\Li_2\pfrac{-2\sla}{1-\mu_1+\mu_2-\sla}
  +\Li_2\pfrac{2\sla}{1+\mu_1-\mu_2+\sla}+\strut\nonumber\\&&\strut
  -\ell_{1W}\ln\pfrac{(1+\smu)^2-\mu_2-\sla}{(1+\smu)^2-\mu_2+\sla},\\[12pt]
\lefteqn{t_{zW}^{\ell+}
  \ =\ \Li_2\pfrac{(1-\smu)^2-\mu_2-\sla}{1+\mu_1-\mu_2-\sla}
  +\Li_2\pfrac{(1-\smu)^2-\mu_2+\sla}{1+\mu_1-\mu_2+\sla}-2\Li_2(\smu)
  +\strut}\nonumber\\&&\strut
  -\Li_2\pfrac{(1-\smu)^2-\mu_2-\sla}{1-\mu_1-\mu_2-\sla}
  -\Li_2\pfrac{(1-\smu)^2-\mu_2+\sla}{1-\mu_1-\mu_2+\sla}
  +\strut\nonumber\\&&\strut
  +\Li_2\left(-\frac{(1-\smu)^2-\mu_2-\sla}{1-\mu_1+\mu_2+\sla}\right)
  +\Li_2\left(-\frac{(1-\smu)^2-\mu_2+\sla}{1-\mu_1+\mu_2-\sla}\right)
  +\strut\nonumber\\&&\strut
  +\Li_2\pfrac{2\sla}{1-\mu_1-\mu_2+\sla}
  -\Li_2\pfrac{-2\sla}{1-\mu_1+\mu_2-\sla}
  -\Li_2\pfrac{2\sla}{1+\mu_1-\mu_2+\sla}+\strut\nonumber\\&&\strut
  +\Li_2\pfrac{1+\mu_1-\mu_2-\sla}2
  +\Li_2\pfrac{1+\mu_1-\mu_2+\sla}2
  +\ell_{1W}\ln\pfrac{(1+\smu)^2-\mu_2}{\smu},\nonumber\\\\
\lefteqn{t_{zW}^{-\ell}
  \ =\ \Li_2\left(-\frac{(1-\smu)^2-\mu_2-\sla}{2\smu}\right)
  +\Li_2\pfrac{(1-\smu)^2-\mu_2-\sla}{2(1-\smu)}+\strut}\nonumber\\&&\strut
  -\Li_2\left(-\frac{(1-\smu)^2-\mu_2-\sla}{2\smu(1-\smu)}\right),\\[12pt]
\lefteqn{t_{zW}^{+\ell}
  \ =\ \Li_2\left(-\frac{1+\mu_1-\mu_2-\sla}{2\smu}\right)
  +\Li_2\pfrac{1-\mu_1+\mu_2+\sla}{2(1+\smu)}
  -\Li_2\left(-\frac{1-\mu_1-\mu_2-\sla}{2\smu(1+\smu)}\right)
  +\strut}\nonumber\\&&\strut
  +\Li_2\left(-\frac{1-\smu}{1+\smu}\right)-\Li_2\pfrac{1-\smu}{1+\smu}
  -\Li_2(-1)+\ln(1+\smu)\ln\pfrac{1-\mu_1-\mu_2-\sla}{2\smu(1-\smu)}
  +\strut\kern-21pt\nonumber\\&&\strut
  -\ln\left(1+\frac1{\smu}\right)\ln\pfrac{1-\mu_1+\mu_2+\sla}{2(1-\smu)}
  -\ell_{1W}\ln\pfrac{(1+\smu)^2-\mu_2-\sla}{2\smu},\\[12pt]
\lefteqn{t_{zW}^\ell\ =\ 2\Li_2(\smu)-\Li_2\pfrac{1+\mu_1-\mu_2+\sla}2
  -\Li_2\pfrac{1+\mu_1-\mu_2-\sla}2.}
\end{eqnarray}

\newpage

\section{Form factors for the vertex correction}
\setcounter{equation}{0}\def\theequation{B\arabic{equation}}
After IR subtraction explained in Sec.~\ref{irdeal}, the renormalised form
factors for the vertex correction defined in Eq.~(\ref{Vmp12def}) and used in
Eqs.~(\ref{full}) read
\begin{eqnarray}\label{V0m}
\lefteqn{V_-^*\ =\ \frac{m_c^2+m_b^2+2m_W^2-2(2Q_c-1)(2Q_b+1)m_Z^2s_W^2}{4m_W^2
  s_W^2}+\strut}\nonumber\\&&\strut\kern-12pt
  +\delta_{\rm CKM}^f+\delta Z_e^f-\frac{\delta s_W^f}{s_W}
  +\delta Z_{WW}^f+\delta Z^{Lf}_{cc}+\delta Z^{Lf}_{bb}
+\strut\nonumber\\&&\strut\kern-12pt
  +2(m_c^2-m_b^2+m_W^2)Q_cC_f^*(m_c^2,m_b^2,m_W^2;m_A,m_W,m_c)
+\strut\nonumber\\&&\strut\kern-12pt
  +2(m_c^2-m_b^2-m_W^2)Q_bC_f^*(m_c^2,m_b^2,m_W^2;m_W,m_A,m_b)
+\strut\nonumber\\&&\strut\kern-12pt
  +2(m_c^2+m_b^2-m_W^2)Q_cQ_bC_f^*(m_c^2,m_b^2,m_W^2;m_c,m_b,m_A)
+\strut\nonumber\\&&\strut\kern-12pt
  -\frac{C_f(m_c^2,m_b^2,m_W^2;m_Z,m_W,m_c)}{4\lambda'm_W^2s_W^2}
  \Big(m_c^2\left(m_b^2(3m_Z^2-4m_W^2)m_Z^2+(4m_W^2+m_Z^2)\lambda'\right)
  +\strut\nonumber\\&&\strut
  +2\Big(m_b^2(m_c^2-m_b^2+3m_W^2)m_Z^4+2(m_c^2-m_b^2+m_W^2)m_W^2\lambda'
  +\strut\nonumber\\&&\strut\qquad
  +2\left(m_b^2(m_c^2-m_b^2+m_W^2)+2\lambda'\right)m_W^2m_Z^2\Big)
  (2Q_cs_W^2-1)\Big)+\strut\nonumber\\&&\strut\kern-12pt
  -\frac{C_f(m_c^2,m_b^2,m_W^2;m_W,m_Z,m_b)}{4\lambda'm_W^2s_W^2}
  \Big(m_b^2\left(m_c^2(3m_Z^2-4m_W^2)m_Z^2+(4m_W^2+m_Z^2)\lambda'\right)
  +\strut\nonumber\\&&\strut
  -2\Big(m_c^2(m_b^2-m_c^2+3m_W^2)m_Z^4+2(m_b^2-m_c^2+m_W^2)m_W^2\lambda'
  +\strut\nonumber\\&&\strut\qquad
  +2\left(m_c^2(m_b^2-m_c^2+m_W^2)+2\lambda'\right)m_W^2m_Z^2\Big)
  (2Q_bs_W^2+1)\Big)+\strut\nonumber\\&&\strut\kern-12pt
  +\frac{m_Z^2C_f(m_c^2,m_b^2,m_W^2;m_c,m_b,m_Z)}{2\lambda'm_W^2s_W^2}
  \times\nonumber\\&&\strut\times(m_c^2+m_b^2-m_W^2-m_Z^2)
  \Big(m_W^2m_Z^2+\lambda'\Big)(2Q_cs_W^2-1)(2Q_bs_W^2+1)
+\strut\nonumber\\&&\strut\kern-12pt
  +\frac{m_c^2C_f(m_c^2,m_b^2,m_W^2;m_H,m_W,m_c)}{4\lambda'm_W^2s_W^2}
  (m_H^2-4m_W^2)\Big(m_b^2m_H^2+\lambda'\Big)
+\strut\nonumber\\&&\strut\kern-12pt
  +\frac{m_b^2C_f(m_c^2,m_b^2,m_W^2;m_W,m_H,m_b)}{4\lambda'm_W^2s_W^2}
  (m_H^2-4m_W^2)\Big(m_c^2m_H^2+\lambda'\Big)
+\strut\nonumber\\&&\strut\kern-12pt
  -\frac{C_f(m_c^2,m_b^2,m_W^2;m_c,m_b,m_H)}{\lambda'm_W^2s_W^2}m_c^2m_b^2
  \Big(m_W^2m_H^2+\lambda'\Big)+\strut\nonumber\\&&\strut\kern-12pt
  +\frac2{\lambda'}B_f(m_c^2;m_c,m_A)Q_c(Q_b+1)
  \Big(m_c^2(m_b^2-m_c^2+m_W^2)+\lambda'\Big)
+\strut\nonumber\\&&\strut\kern-12pt
  +\frac2{\lambda'}B_f(m_b^2;m_b,m_A)Q_b(Q_c-1)
  \Big(m_b^2(m_c^2-m_b^2+m_W^2)+\lambda'\Big)
+\strut\nonumber\\&&\strut\kern-12pt
  +\frac{2m_W^2}{\lambda'}(m_c^2+m_b^2-m_W^2)(Q_c-Q_b)B_f(m_W^2;m_W,m_A)
+\strut\nonumber\\&&\strut\kern-12pt
  -\frac{B_f(m_c^2;m_c,m_Z)}{8\lambda'm_W^2s_W^2}
  \Big(m_c^2(m_c^2+m_b^2-m_W^2)(4m_W^2-3m_Z^2)+\strut\nonumber\\&&\strut
  +4\left(2m_c^2(m_b^2-m_c^2+m_W^2)m_W^2+2m_W^2\lambda'
  -(m_c^2-m_W^2)(m_c^2-m_b^2+m_W^2)m_Z^2\right)(2Q_cs_W^2-1)
  +\strut\kern-33pt\nonumber\\&&\strut
  -2\left(2m_c^2(m_b^2-m_c^2+m_W^2)+(m_c^2-m_b^2+m_W^2)m_Z^2+2\lambda'\right)
  m_Z^2(2Q_cs_W^2-1)(2Q_bs_W^2+1)\Big)
+\strut\kern-21pt\nonumber\\&&\strut\kern-12pt
  -\frac{B_f(m_b^2;m_b,m_Z)}{8\lambda'm_W^2s_W^2}
  \Big(m_b^2(m_c^2+m_b^2-m_W^2)(4m_W^2-3m_Z^2)+\strut\nonumber\\&&\strut
  -4\Big(2m_W^2m_b^2(m_c^2-m_b^2+m_W^2)+2m_W^2\lambda'
  -(m_b^2-m_W^2)(m_b^2-m_c^2+m_W^2)m_Z^2\Big)(2Q_bs_W^2+1)
  +\strut\kern-29pt\nonumber\\&&\strut
  -2\left(2m_b^2(m_c^2-m_b^2+m_W^2)+(m_b^2-m_c^2+m_W^2)m_Z^2+2\lambda'\right)
  m_Z^2(2Q_cs_W^2-1)(2Q_bs_W^2+1)\Big)
+\strut\kern-21pt\nonumber\\&&\strut\kern-12pt
  +\frac{B_f(m_W^2;m_W,m_Z)}{8\lambda'm_W^2s_W^2}
  \Big(\left((m_c^2+m_b^2-m_W^2)m_W^2+\lambda'\right)(4m_W^2-3m_Z^2)
  +\strut\nonumber\\&&\strut
  -4\left(2(m_c^2+m_b^2-m_W^2)m_W^4+2(m_c^2-m_W^2)m_W^2m_Z^2+m_Z^2\lambda'
  \right)(2Q_cs_W^2-1)+\strut\nonumber\\&&\strut
  +4\left(2(m_c^2+m_b^2-m_W^2)m_W^4+2(m_b^2-m_W^2)m_W^2m_Z^2+m_Z^2\lambda'
  \right)(2Q_bs_W^2+1)\Big)
+\strut\nonumber\\&&\strut\kern-12pt
  -\frac{m_c^2B_f(m_c^2;m_c,m_H)}{8\lambda'm_W^2s_W^2}
  \Big((m_c^2+m_b^2-m_W^2)m_H^2+4m_c^2(m_b^2-m_c^2+m_W^2)+4\lambda'\Big)
+\strut\nonumber\\&&\strut\kern-12pt
  -\frac{m_b^2B_f(m_b^2;m_b,m_H)}{8\lambda'm_W^2s_W^2}
  \Big((m_c^2+m_b^2-m_W^2)m_H^2+4m_b^2(m_c^2-m_b^2+m_W^2)+4\lambda'\Big)
+\strut\nonumber\\&&\strut\kern-12pt
  +\frac{B_f(m_W^2;m_W,m_H)}{8\lambda'm_W^2s_W^2}(m_H^2-4m_W^2)
  \Big((m_c^2+m_b^2-m_W^2)m_W^2+\lambda'\Big)
+\strut\nonumber\\&&\strut\kern-12pt
  +\frac{B_f(m_c^2;m_b,m_W)}{4\lambda'm_W^2s_W^2}\Big(m_c^2m_b^2(m_H^2-3m_Z^2)
  +(m_b^2+6m_W^2)\lambda'+\strut\nonumber\\&&\strut
  +2m_c^2(3m_b^2-3m_c^2+5m_W^2)m_W^2
  +2m_c^2(m_b^2-m_c^2+3m_W^2)m_Z^2(2Q_b+1)s_W^2\Big)
+\strut\nonumber\\&&\strut\kern-12pt
  +\frac{B_f(m_b^2;m_W,m_c)}{4\lambda'm_W^2s_W^2}\Big(m_c^2m_b^2(m_H^2-3m_Z^2)
  +(m_c^2+6m_W^2)\lambda'+\strut\nonumber\\&&\strut
  +2m_b^2(3m_c^2-3m_b^2+5m_W^2)m_W^2
  -2m_b^2(m_c^2-m_b^2+3m_W^2)m_Z^2(2Q_c-1)s_W^2\Big)
+\strut\nonumber\\&&\strut\kern-12pt
  +\frac{B_f(m_W^2;m_c,m_b)}{4\lambda'm_W^2s_W^2}
  \Big(\left(\lambda'-4m_c^2m_b^2-2(m_c^2+m_b^2-2m_W^2)m_W^2\right)m_W^2
  +\strut\\&&\strut+4m_W^4m_Z^2(1+Q_b-Q_c)s_W^2
  -\left(\lambda'-2(m_c^2+m_b^2-m_Z^2)m_W^2\right)m_Z^2(2Q_c-1)(2Q_b+1)s_W^2
  \Big),\kern-9pt\nonumber\\
\lefteqn{V_+\ =\ -\frac{C_f(m_c^2,m_b^2,m_W^2;m_Z,m_W,m_c)}{4\lambda's_W^2}
  \Big((4m_W^2+m_Z^2)\lambda'-2(m_c^2+m_b^2-m_W^2)m_W^2m_Z^2
  +\strut}\nonumber\\&&\strut
  +(2m_c^2+m_b^2-2m_W^2)m_Z^4-2(m_b^2-m_c^2+m_W^2)(2m_W^2+m_Z^2)m_Z^2
  (2Q_cs_W^2-1)\Big)+\strut\nonumber\\&&\strut\kern-12pt
  -\frac{C_f(m_c^2,m_b^2,m_W^2;m_W,m_Z,m_b)}{4\lambda's_W^2}
  \Big((4m_W^2+m_Z^2)\lambda'-2(m_c^2+m_b^2-m_W^2)m_W^2m_Z^2
  +\strut\nonumber\\&&\strut
  +(2m_b^2+m_c^2-2m_W^2)m_Z^4+2(m_c^2-m_b^2+m_W^2)(2m_W^2+m_Z^2)m_Z^2
  (2Q_bs_W^2+1)\Big)+\strut\nonumber\\&&\strut\kern-12pt
  +\frac{m_Z^2C_f(m_c^2,m_b^2,m_W^2;m_c,m_b,m_Z)}{4\lambda's_W^2}
  \Big(m_W^2m_Z^2+2\lambda'+2\lambda'(2Q_cs_W^2-1)+\strut\nonumber\\&&\strut
  -2\lambda'(2Q_bs_W^2+1)+4m_W^2m_Z^2(2Q_cs_W^2-1)(2Q_bs_W^2+1)\Big)
+\strut\nonumber\\&&\strut\kern-12pt
  -\frac{m_H^2C_f(m_c^2,m_b^2,m_W^2;m_H,m_W,m_c)}{4\lambda's_W^2}
  \Big(m_b^2m_H^2+\lambda'+2(m_c^2+m_b^2-m_W^2)m_W^2\Big)
+\strut\nonumber\\&&\strut\kern-12pt
  -\frac{m_H^2C_f(m_c^2,m_b^2,m_W^2;m_W,m_H,m_b)}{4\lambda's_W^2}
  \Big(m_c^2m_H^2+\lambda'+2(m_c^2+m_b^2-m_W^2)m_W^2\Big)
+\strut\nonumber\\&&\strut\kern-12pt
  -\frac{m_H^2C_f(m_c^2,m_b^2,m_W^2;m_c,m_b,m_H)}{4\lambda's_W^2}
  \Big(m_W^2m_H^2+2\lambda'+2(m_c^2+m_b^2-m_W^2)m_W^2\Big)
+\strut\nonumber\\&&\strut\kern-12pt
  -\frac{2m_W^2}{\lambda'}(m_c^2-m_b^2+m_W^2)Q_c(Q_b+1)B_f(m_c^2;m_c,m_A)
+\strut\nonumber\\&&\strut\kern-12pt
  -\frac{2m_W^2}{\lambda'}(m_b^2-m_c^2+m_W^2)Q_b(Q_c-1)B_f(m_b^2;m_b,m_A)
  +\frac{4m_W^4}{\lambda'}(Q_c-Q_b)B_f(m_W^2;m_W,m_A)
  +\strut\kern-15pt\nonumber\\&&\strut\kern-12pt
  -\frac{B_f(m_c^2;m_c,m_Z)}{4\lambda's_W^2}\Big(m_c^2(4m_W^2-3m_Z^2)
  -2(m_c^2-m_b^2+m_W^2)(2m_W^2+m_Z^2)(2Q_cs_W^2-1)+\strut\nonumber\\&&\strut
  +2(m_c^2-m_b^2+m_W^2)m_Z^2(2Q_cs_W^2-1)(2Q_bs_W^2+1)\Big)
+\strut\nonumber\\&&\strut\kern-12pt
  -\frac{B_f(m_b^2;m_b,m_Z)}{4\lambda's_W^2}\Big(m_b^2(4m_W^2-3m_Z^2)
  +2(m_b^2-m_c^2+m_W^2)(2m_W^2+m_Z^2)(2Q_bs_W^2+1)+\strut\nonumber\\&&\strut
  +2(m_b^2-m_c^2+m_W^2)m_Z^2(2Q_cs_W^2-1)(2Q_bs_W^2+1)\Big)
+\strut\nonumber\\&&\strut\kern-12pt
  +\frac{m_W^2B_f(m_W^2;m_W,m_Z)}{4\lambda's_W^2}
  \Big(4m_W^2-5m_Z^2+\strut\nonumber\\&&\strut
  -4(2m_W^2+m_Z^2)(2Q_cs_W^2-1)+4(2m_W^2+m_Z^2)(2Q_bs_W^2+1)\Big)
+\strut\nonumber\\&&\strut\kern-12pt
  +\frac{m_c^2B_f(m_c^2;m_c,m_H)}{4\lambda's_W^2}
  \Big(m_H^2+2(m_c^2-m_b^2+m_W^2)\Big)
+\strut\nonumber\\&&\strut\kern-12pt
  +\frac{m_b^2B_f(m_b^2;m_b,m_H)}{4\lambda's_W^2}
  \Big(m_H^2+2(m_b^2-m_c^2+m_W^2)\Big)
+\strut\nonumber\\&&\strut\kern-12pt
  +\frac{m_W^2B_f(m_W^2;m_W,m_H)}{4\lambda's_W^2}\Big(m_H^2-4m_W^2\Big)
+\strut\nonumber\\&&\strut\kern-12pt
  -\frac{B_f(m_c^2;m_b,m_W)}{4\lambda's_W^2}
  \left(m_c^2(m_H^2+3m_Z^2)+4(m_c^2-m_b^2+m_W^2)(Q_bm_Z^2s_W^2+m_W^2)\right)
+\strut\nonumber\\&&\strut\kern-12pt
  -\frac{B_f(m_b^2;m_W,m_c)}{4\lambda's_W^2}
  \left(m_b^2(m_H^2+3m_Z^2)-4(m_b^2-m_c^2+m_W^2)(Q_cm_Z^2s_W^2-m_W^2)\right)
+\strut\nonumber\\&&\strut\kern-12pt
  -\frac{B_f(m_W^2;m_c,m_b)}{4\lambda's_W^2}
  \Big(m_W^2(m_H^2-m_Z^2)+2\lambda'+2(m_c^2+m_b^2+m_W^2)m_W^2
  +\strut\nonumber\\&&\strut
  -4m_Z^2m_W^2s_W^2(2Q_c-1)(2Q_b+1)\Big),\\
\lefteqn{V_1\ =\ \frac{C_f(m_c^2,m_b^2,m_W^2;m_Z,m_W,m_c)}{2\lambda^{\prime2}
  s_W^2}\times\strut}\nonumber\\&&\strut
  \Big\{\lambda'(\lambda'+2m_b^2m_Z^2)(4m_W^2+m_Z^2)
  +(7\lambda'+3m_b^2m_Z^2)(m_W^2-m_c^2-m_b^2)m_W^2m_Z^2
  +\strut\nonumber\\&&\strut
  +2\Big[\lambda'\left(\lambda'-2m_b^2(2m_W^2-m_Z^2)\right)
  -2(\lambda'+3m_b^2m_Z^2)(m_W^2-m_c^2+m_b^2)m_W^2\Big]m_Z^2(2Q_cs_W^2-1)\Big\}
+\strut\kern-28pt\nonumber\\&&\strut\kern-12pt
  +\frac{m_Z^2C_f(m_c^2,m_b^2,m_W^2;m_W,m_Z,m_b)}{2\lambda^{\prime2}s_W^2}
  \times\strut\nonumber\\&&\strut
  \Big\{m_b^2\left(\lambda'(2m_W^2+m_Z^2)+6m_c^2m_W^2m_Z^2\right)
  +2\Big[\lambda^{\prime2}-4\lambda'(m_c^2-m_W^2)m_W^2
  +\strut\nonumber\\&&\strut\qquad\qquad
  +2\left(\lambda'(m_c^2+m_W^2)+3m_c^2(m_W^2-m_c^2+m_b^2)m_W^2\right)m_Z^2
  \Big](2Q_bs_W^2+1)\Big\}
+\strut\nonumber\\&&\strut\kern-12pt
  +\frac{m_Z^2C_f(m_c^2,m_b^2,m_W^2,m_c,m_b,m_Z)}{2\lambda^{\prime2}s_W^2}
  \times\strut\nonumber\\&&\strut
  \Big\{m_b^2(m_W^2-m_b^2+m_c^2)(\lambda'+3m_W^2m_Z^2)
  +2\lambda'(\lambda'+2m_W^2m_Z^2)(2Q_bs_W^2+1)+\strut\nonumber\\&&\strut
  -2(m_W^2-m_c^2+m_b^2)(2\lambda'+3m_W^2m_Z^2)
  m_Z^2(2Q_cs_W^2-1)(2Q_bs_W^2+1)\Big\}
+\strut\nonumber\\&&\strut\kern-12pt
  +\frac{m_H^2C_f(m_c^2,m_b^2,m_W^2,m_H,m_W,m_c)}{2\lambda^{\prime2}s_W^2}
  \times\strut\nonumber\\&&\strut\times
  \Big\{\lambda'\left(\lambda'+4m_c^2(m_W^2-m_c^2+m_b^2)
  -3(m_W^2-m_c^2-m_b^2)m_W^2\right)+\strut\nonumber\\&&\strut
  -m_b^2\left(\lambda'-3(m_c^2-m_b^2)(m_W^2-m_c^2-m_b^2)\right)m_H^2\Big\}
+\strut\nonumber\\&&\strut\kern-12pt
  -\frac{m_b^2m_H^2C_f(m_c^2,m_b^2,m_W^2;m_W,m_H,m_b)}{2\lambda^{\prime2}s_W^2}
  \Big\{2\lambda'(m_W^2-2m_b^2+2m_c^2)-\left(\lambda'-6m_c^2(m_c^2-m_b^2)\right)
  m_H^2\Big\}
+\strut\kern-40pt\nonumber\\&&\strut\kern-12pt
  -\frac{3m_b^2m_H^2C_f(m_c^2,m_b^2,m_W^2;m_c,m_b,m_H)}{2\lambda^{\prime2}s_W^2}
  (m_W^2-m_b^2+m_c^2)\left(\lambda'+m_W^2m_H^2\right)
+\strut\nonumber\\&&\strut\kern-12pt
  +\frac{4m_W^2}{\lambda'}Q_c(Q_b+1)(m_W^2-m_c^2-m_b^2)B_f(m_c^2;m_c,m_A)
  +\frac{8m_W^2}{\lambda'}(Q_c-1)Q_bm_b^2B_f(m_b^2;m_b,m_A)
+\strut\kern-26pt\nonumber\\&&\strut\kern-12pt
  -\frac{4m_W^2}{\lambda'}(Q_c-Q_b)(m_W^2-m_c^2+m_b^2)B_f(m_W^2;m_W,m_A)
+\strut\nonumber\\&&\strut\kern-12pt
  +\frac{B_f(m_c^2;m_c,m_Z)}{4\lambda^{\prime2}m_c^2s_W^2}
  \Big\{\lambda'm_c^2\Big[m_c^2(8m_W^2+m_Z^2)+8(m_W^2-m_b^2+m_c^2)m_W^2
  +3(m_W^2-m_c^2-m_b^2)m_Z^2\Big]+\strut\kern-37pt\nonumber\\&&\strut
  -4\lambda'm_c^2(m_W^2-m_b^2+m_c^2)m_Z^2(2Q_bs_W^2+1)
  -4\Big[\lambda'(2m_c^2-m_Z^2)(m_W^2-m_c^2-m_b^2)m_W^2
  +\strut\nonumber\\&&\strut\qquad
  +\left(\lambda'm_c^2(m_c^2+m_b^2)
  +6m_c^2m_b^2(m_W^2-m_b^2+m_c^2)m_W^2\right)m_Z^2\Big](2Q_cs_W^2-1)
  +\strut\nonumber\\&&\strut
  +2\Big[2\lambda'm_c^2(m_W^2-m_c^2-m_b^2)+\strut\nonumber\\&&\strut\qquad
  -\left(\lambda'(m_W^2-m_b^2+2m_c^2)-6m_c^2(m_W^2-m_c^2-m_b^2)m_W^2\right)
  m_Z^2\Big]m_Z^2(2Q_cs_W^2-1)(2Q_bs_W^2+1)\Big\}
+\strut\kern-48pt\nonumber\\&&\strut\kern-12pt
  -\frac{B_f(m_b^2;m_b,m_Z)}{4\lambda^{\prime2}s_W^2}\Big\{\lambda'm_b^2m_Z^2
  +\strut\nonumber\\&&\strut
  -4\Big[4\lambda'm_b^2m_W^2
  -\left(\lambda'(2m_b^2-m_W^2)-6m_b^2(m_W^2-m_b^2+m_c^2)m_W^2\right)m_Z^2\Big]
  (2Q_bs_W^2+1)+\strut\nonumber\\&&\strut
  -2\Big[4\lambda'm_b^2+(\lambda'+12m_b^2m_W^2)m_Z^2\Big]
  m_Z^2(2Q_cs_W^2-1)(2Q_bs_W^2+1)\Big\}
+\strut\nonumber\\&&\strut\kern-12pt
  -\frac{B_f(m_W^2;m_W,m_Z)}{4\lambda^{\prime2}s_W^2}
  \Big\{\lambda'm_W^2(16m_W^2+m_Z^2)+4\lambda'(m_W^2-m_b^2+m_c^2)m_W^2
  +\strut\nonumber\\&&\strut
  +3\left(\lambda'(m_W^2-m_c^2+m_b^2)+4m_b^2(m_W^2-m_b^2+m_c^2)m_W^2\right)
  m_Z^2+\strut\nonumber\\&&\strut
  -4\Big[2\lambda'(m_W^2-m_c^2+m_b^2)m_W^2
  +\left(\lambda'(m_c^2-m_b^2)+12m_b^2m_W^4\right)m_Z^2\Big](2Q_cs_W^2-1)
  +\strut\nonumber\\&&\strut
  +4\Big[2\lambda'(m_W^2-m_c^2+m_b^2)m_W^2
  +\left(\lambda'(m_c^2-m_b^2)+6(m_W^2-m_c^2-m_b^2)m_W^4\right)m_Z^2\Big]
  (2Q_bs_W^2+1)\Big\}
+\strut\kern-40pt\nonumber\\&&\strut\kern-12pt
  +\frac{B_f(m_c^2;m_c,m_H)}{4\lambda^{\prime2}s_W^2}
  \Big\{\lambda'm_c^2m_H^2+\strut\nonumber\\&&\strut
  -\lambda'(4m_c^2+m_H^2)(m_W^2-m_c^2-m_b^2) 
  +12m_c^2m_b^2(m_W^2-m_b^2+m_c^2)m_H^2\Big\}
+\strut\nonumber\\&&\strut\kern-12pt
  -\frac{m_b^2B_f(m_b^2;m_b,m_H)}{4\lambda^{\prime2}s_W^2}\Big\{8\lambda'm_b^2
  -3\left(\lambda'+4m_b^2(m_W^2-m_b^2+m_c^2)\right)m_H^2\Big\}
+\strut\nonumber\\&&\strut\kern-12pt
  +\frac{B_f(m_W^2;m_W,m_H)}{4\lambda's_W^2}
  \Big\{4(m_W^2-m_c^2+m_b^2)m_W^2-(m_c^2-m_b^2)m_H^2\Big\}
+\strut\nonumber\\&&\strut\kern-12pt
  -\frac{B_f(m_c^2;m_b,m_W)}{2\lambda^{\prime2}m_c^2s_W^2}
  \Big\{\lambda^{\prime2}m_W^2+(\lambda^{\prime2}-12m_c^4m_W^4)
  (m_W^2-m_c^2+m_b^2)+\strut\nonumber\\&&\strut
  +\lambda'm_c^2\left(2m_b^2m_W^2-(5m_W^2-m_c^2)(m_W^2-m_b^2+m_c^2)\right)
  +\strut\nonumber\\&&\strut
  -m_c^2m_b^2(\lambda'+6m_c^2m_W^2)m_Z^2
  -m_c^2m_b^2\left(\lambda'-6m_c^2(m_c^2-m_b^2)\right)m_H^2
  +\strut\nonumber\\&&\strut
  -4m_c^2\Big[\lambda'(m_c^2+m_W^2)+3m_c^2(m_W^2-m_c^2+m_b^2)m_W^2\Big]
  m_Z^2(2Q_b+1)s_W^2\Big\}
+\strut\nonumber\\&&\strut\kern-12pt
  -\frac{B_f(m_b^2;m_W,m_c)}{2\lambda^{\prime2}s_W^2}
  \Big\{\lambda^{\prime2}-\lambda'(3m_W^2+m_b^2)(m_W^2-m_c^2+m_b^2)
  -12m_b^2(m_W^2-m_c^2+m_b^2)m_W^4+\strut\kern-23pt\nonumber\\&&\strut
  -m_b^2\left(2\lambda'+3(m_W^2-m_c^2-m_b^2)m_W^2\right)m_Z^2
  +m_b^2\left(\lambda'-3(m_c^2-m_b^2)(m_W^2-m_c^2-m_b^2)\right)m_H^2
  +\strut\kern-23pt\nonumber\\&&\strut
  -4m_b^2\Big[\lambda'-3(m_W^2-m_c^2+m_b^2)m_W^2\Big]m_Z^2(2Q_c-1)s_W^2\Big\}
+\strut\nonumber\\&&\strut\kern-12pt
  +\frac{B_f(m_W^2;m_c,m_b)}{2\lambda^{\prime2}s_W^2}
  \Big\{\lambda'(\lambda'+4m_W^4)+\left(\lambda'(m_c^2+m_b^2)+6m_W^6\right)
  (m_W^2-m_c^2+m_b^2)+\strut\nonumber\\&&\strut
  +3m_b^2(m_W^2-m_b^2+m_c^2)m_W^2(m_Z^2-m_H^2)
  -6(m_W^2-m_c^2+m_b^2)m_W^4m_Z^2(2Q_c-1)s_W^2
  +\strut\nonumber\\&&\strut
  +2\left(2\lambda'+3(m_W^2-m_c^2+m_b^2)m_W^2\right)m_W^2m_Z^2(2Q_b+1)s_W^2
  +\strut\nonumber\\&&\strut
  -(m_W^2-m_c^2+m_b^2)(\lambda'+6m_W^2m_Z^2s_W^2)m_Z^2(2Q_c-1)(2Q_b+1)s_W^2
  \Big\}+\strut\nonumber\\&&\strut\kern-12pt
  +\frac{\ln(m_c/\bar\mu)-1}{\lambda's_W^2}\Big\{
  m_c^4-m_c^2m_b^2+4m_c^2m_W^2-m_b^2m_W^2+m_W^4
  +\strut\nonumber\\&&\strut
  +(m_W^2-m_c^2-m_b^2)m_Z^2(2Q_c-1)(2Q_b+1)s_W^2\Big\}
+\strut\nonumber\\&&\strut\kern-12pt
  -\frac{m_b^2(\ln(m_b/\bar\mu)-1)}{\lambda'm_c^2s_W^2}
  \Big\{m_b^2(m_W^2-m_c^2+m_b^2)-2m_W^4
  -2m_c^2m_Z^2(2Q_c-1)(2Q_b+1)s_W^2\Big\}
+\strut\nonumber\\&&\strut\kern-12pt
  -\frac{m_W^2(\ln(m_W/\bar\mu)-1)}{\lambda'm_c^2s_W^2}
  \Big\{2m_c^4-m_c^2m_b^2+2m_c^2m_W^2-m_b^4-m_b^2m_W^2+2m_W^4\Big\}
+\strut\nonumber\\&&\strut\kern-12pt
  -\frac{m_Z^2(2\ln(m_Z/\bar\mu)-1)}{4\lambda'm_c^2s_W^2}
  \Big\{m_c^2(m_W^2-m_c^2+m_b^2)-4(m_W^2-m_c^2-m_b^2)m_W^2(2Q_cs_W^2-1)
  +\strut\nonumber\\&&\strut
  +8m_c^2m_W^2(2Q_bs_W^2+1)
  +2(m_W^2-m_b^2+m_c^2)m_Z^2(2Q_cs_W^2-1)(2Q_bs_W^2+1)\Big\}
+\strut\nonumber\\&&\strut\kern-12pt
  -\frac{m_H^2(2\ln(m_H/\bar\mu)-1)}{4\lambda's_W^2}(m_W^2-m_c^2+m_b^2)
  -\frac{m_b^2+2m_W^2}{2m_c^2s_W^2},\\
\lefteqn{V_2\ =\ \frac{m_Z^2C_f(m_c^2,m_b^2,m_W^2,m_Z,m_W,m_c)}{2
  \lambda^{\prime2}s_W^2}\times\strut}\nonumber\\&&\strut
  \Big\{m_c^2\left(\lambda'(2m_W^2+m_Z^2)+6m_b^2m_W^2m_Z^2\right)
  -2\Big[\lambda^{\prime2}+4\lambda'(m_W^2-m_b^2)m_W^2
  +\strut\nonumber\\&&\strut\qquad
  +2\left(\lambda'(m_W^2+m_b^2)+3m_b^2(m_W^2-m_b^2+m_c^2)m_W^2\right)m_Z^2
  \Big](2Q_cs_W^2-1)\Big\}
+\strut\nonumber\\&&\strut\kern-12pt
  +\frac{C_f(m_c^2,m_b^2,m_W^2,m_W,m_Z,m_b)}{2\lambda^{\prime2}s_W^2}
  \Big\{\lambda^{\prime2}(4m_W^2+m_Z^2)+8\lambda'm_c^2m_W^2m_Z^2
  +\strut\nonumber\\&&\strut
  +7\lambda'(m_W^2-m_c^2-m_b^2)m_W^2m_Z^2
  +m_c^2\left(2\lambda'+3(m_W^2-m_c^2-m_b^2)m_W^2\right)m_Z^4
  +\strut\nonumber\\&&\strut
  -2\Big[\lambda^{\prime2}-2\lambda'm_c^2(4m_W^2-m_Z^2)
  -2\lambda'(m_W^2-m_c^2-m_b^2)m_W^2+\strut\nonumber\\&&\strut\qquad
  -6m_c^2(m_W^2-m_b^2+m_c^2)m_W^2m_Z^2\Big]m_Z^2(2Q_bs_W^2+1)\Big\}
+\strut\nonumber\\&&\strut\kern-12pt
  +\frac{m_Z^2C_f(m_c^2,m_b^2,m_W^2;m_c,m_b,m_Z)}{2\lambda^{\prime2}s_W^2}
  \times\strut\nonumber\\&&\strut
  \Big\{m_c^2(m_W^2-m_c^2+m_b^2)(\lambda'+3m_W^2m_Z^2)
  -2\lambda'(\lambda'+2m_W^2m_Z^2)(2Q_cs_W^2-1)+\strut\nonumber\\&&\strut
  -2(m_W^2-m_b^2+m_c^2)(2\lambda'+3m_W^2m_Z^2)
  m_Z^2(2Q_bs_W^2+1)(2Q_cs_W^2-1)\Big\}
+\strut\nonumber\\&&\strut\kern-12pt
  -\frac{m_c^2m_H^2C_f(m_c^2,m_b^2,m_W^2;m_H,m_W,m_c)}{2\lambda^{\prime2}s_W^2}
  \times\strut\nonumber\\&&\strut
  \Big\{2\lambda'(m_W^2-2m_c^2+2m_b^2)
  -\left(\lambda'+6m_b^2(m_c^2-m_b^2)\right)m_H^2\Big\}
+\strut\nonumber\\&&\strut\kern-12pt
  +\frac{m_H^2C_f(m_c^2,m_b^2,m_W^2;m_W,m_H,m_b)}{2\lambda^{\prime2}s_W^2}
  \times\strut\nonumber\\&&\strut
  \Big\{\lambda'\Big[\lambda'
  +4m_b^2(m_W^2-m_b^2+m_c^2)-3(m_W^2-m_c^2-m_b^2)m_W^2\Big]
  +\strut\nonumber\\&&\strut
  -m_c^2\left(\lambda'+3(m_c^2-m_b^2)(m_W^2-m_c^2-m_b^2)\right)m_H^2\Big\}
+\strut\nonumber\\&&\strut\kern-12pt
  -\frac{3m_c^2m_H^2C_f(m_c^2,m_b^2,m_W^2;m_c,m_b,m_H)}{2\lambda^{\prime2}
  s_W^2}(m_W^2-m_c^2+m_b^2)(\lambda'+m_W^2m_H^2)
+\strut\nonumber\\&&\strut\kern-12pt
  +\frac{8m_W^2}{\lambda'}Q_c(Q_b+1)m_c^2B_f(m_c^2;m_c,m_A)
  +\frac{4m_W^2}{\lambda'}(Q_c-1)Q_b(m_W^2-m_c^2-m_b^2)B_f(m_b^2;m_b,m_A)
+\strut\kern-26pt\nonumber\\&&\strut\kern-12pt
  -\frac{4m_W^2}{\lambda'}(Q_c-Q_b)(m_W^2-m_b^2+m_c^2)B_f(m_W^2;m_W,m_A)
+\strut\nonumber\\&&\strut\kern-12pt
  -\frac{B_f(m_c^2;m_c,m_Z)}{4\lambda^{\prime2}s_W^2}
  \Big\{\lambda'm_c^2m_Z^2+\strut\nonumber\\&&\strut
  +4\Big[4\lambda'm_c^2m_W^2
  +\left(\lambda'(m_W^2-2m_c^2)+6m_c^2(m_W^2-m_c^2+m_b^2)m_W^2\right)m_Z^2
  \Big](2Q_cs_W^2-1)+\strut\nonumber\\&&\strut
  -2\left(4\lambda'm_c^2+(\lambda'+12m_c^2m_W^2)m_Z^2\right)
  m_Z^2(2Q_bs_W^2+1)(2Q_cs_W^2-1)\Big\}
+\strut\nonumber\\&&\strut\kern-12pt
  +\frac{B_f(m_b^2;m_b,m_Z)}{4\lambda^{\prime2}m_b^2s_W^2}
  \Big\{\lambda'm_b^2\Big[m_b^2(8m_W^2+m_Z^2)+8(m_W^2-m_c^2+m_b^2)m_W^2
  +3(m_W^2-m_c^2-m_b^2)m_Z^2\Big]+\strut\kern-39pt\nonumber\\&&\strut
  -4\Big[\lambda'(m_Z^2-2m_b^2)(m_W^2-m_c^2-m_b^2)m_W^2
  +\strut\nonumber\\&&\strut\qquad
  -\left(\lambda'm_b^2(m_c^2+m_b^2)
  +6m_c^2m_b^2(m_W^2-m_c^2+m_b^2)m_W^2\right)m_Z^2\Big](2Q_bs_W^2+1)
  +\strut\nonumber\\&&\strut
  +4\lambda'm_b^2(m_W^2-m_c^2+m_b^2)m_Z^2(2Q_cs_W^2-1)
  +2\Big[2m_b^2(m_W^2-m_c^2-m_b^2)(\lambda'+3m_W^2m_Z^2)
  +\strut\nonumber\\&&\strut\qquad
  -\lambda'(m_W^2-m_c^2+2m_b^2)m_Z^2\Big]m_Z^2(2Q_bs_W^2+1)(2Q_cs_W^2-1)\Big\}
+\strut\nonumber\\&&\strut\kern-12pt
  -\frac{B_f(m_W^2;m_W,m_Z)}{4\lambda^{\prime2}s_W^2}
  \Big\{\lambda'm_W^2(16m_W^2+m_Z^2)
  +4\lambda'(m_W^2-m_c^2+m_b^2)m_W^2+\strut\nonumber\\&&\strut\qquad
  +3\left(\lambda'(m_W^2-m_b^2+m_c^2)
  +4m_c^2(m_W^2-m_c^2+m_b^2)m_W^2\right)m_Z^2+\strut\nonumber\\&&\strut
  +4\Big[2\lambda'(m_W^2-m_b^2+m_c^2)m_W^2
  -\left(\lambda'(m_c^2-m_b^2)-12m_c^2m_W^4\right)m_Z^2\Big](2Q_bs_W^2+1)
  +\strut\nonumber\\&&\strut
  -4\Big[2\lambda'(m_W^2-m_b^2+m_c^2)m_W^2
  -\left(\lambda'(m_c^2-m_b^2)-6(m_W^2-m_c^2-m_b^2)m_W^4\right)m_Z^2\Big]
  (2Q_cs_W^2-1)\Big\}
+\strut\kern-39pt\nonumber\\&&\strut\kern-12pt
  -\frac{m_c^2B_f(m_c^2;m_c,m_H)}{4\lambda^{\prime2}s_W^2}
  \Big\{\lambda'(8m_c^2-3m_H^2)-12m_c^2(m_W^2-m_c^2+m_b^2)m_H^2\Big\}
+\strut\nonumber\\&&\strut\kern-12pt
  -\frac{B_f(m_b^2;m_b,m_H)}{4\lambda^{\prime2}s_W^2}
  \Big\{\lambda'(4m_b^2+m_H^2)(m_W^2-m_c^2-m_b^2)
  -m_b^2\left(\lambda'+12m_c^2(m_W^2-m_c^2+m_b^2)\right)m_H^2\Big\}
+\strut\kern-39pt\nonumber\\&&\strut\kern-12pt
  +\frac{B_f(m_W^2;m_W,m_H)}{4\lambda's_W^2}
  \Big\{4(m_W^2-m_b^2+m_c^2)m_W^2+(m_c^2-m_b^2)m_H^2\Big\}
+\strut\nonumber\\&&\strut\kern-12pt
  -\frac{B_f(m_c^2;m_b,m_W)}{2\lambda^{\prime2}s_W^2}
  \Big\{\lambda^{\prime2}
  -\lambda'(3m_W^2+m_c^2)(m_W^2-m_b^2+m_c^2)-12m_c^2(m_W^2-m_b^2+m_c^2)m_W^4
  +\strut\kern-10pt\nonumber\\&&\strut
  -m_c^2\Big[2\lambda'+3(m_W^2-m_c^2-m_b^2)m_W^2\Big]m_Z^2
  +m_c^2\left(\lambda'+3(m_c^2-m_b^2)(m_W^2-m_c^2-m_b^2)\right)m_H^2
  +\strut\kern-11pt\nonumber\\&&\strut
  +4m_c^2\left(\lambda'-3(m_W^2-m_b^2+m_c^2)m_W^2\right)m_Z^2
  (2Q_b+1)s_W^2\Big\}
+\strut\nonumber\\&&\strut\kern-12pt
  -\frac{B_f(m_b^2;m_W,m_c)}{2\lambda^{\prime2}m_b^2s_W^2}
  \Big\{\lambda^{\prime2}m_W^2+(\lambda^{\prime2}-12m_b^4m_W^4)
  (m_W^2-m_b^2+m_c^2)+\strut\nonumber\\&&\strut
  +\lambda'm_b^2\left(2m_c^2m_W^2-(5m_W^2-m_b^2)(m_W^2-m_c^2+m_b^2)\right)
  +\strut\nonumber\\&&\strut
  -m_c^2m_b^2(\lambda'+6m_b^2m_W^2)m_Z^2
  -m_c^2m_b^2\left(\lambda'+6m_b^2(m_c^2-m_b^2)\right)m_H^2
  +\strut\nonumber\\&&\strut
  +4m_b^2\Big[\lambda'(m_W^2+m_b^2)+3m_b^2(m_W^2-m_b^2+m_c^2)m_W^2\Big]
  m_Z^2(2Q_c-1)s_W^2\Big\}
+\strut\nonumber\\&&\strut\kern-12pt
  +\frac{B_f(m_W^2;m_c,m_b)}{2\lambda^{\prime2}s_W^2}
  \Big\{\lambda'(\lambda'+4m_W^4)
  +\left(\lambda'(m_c^2+m_b^2)+6m_W^6\right)(m_W^2-m_b^2+m_c^2)
  +\strut\nonumber\\&&\strut
  +3m_c^2(m_W^2-m_c^2+m_b^2)m_W^2(m_Z^2-m_H^2)
  +6(m_W^2-m_b^2+m_c^2)m_W^4m_Z^2(2Q_b+1)s_W^2
  +\strut\nonumber\\&&\strut
  -2\left(2\lambda'+3(m_W^2-m_b^2+m_c^2)m_W^2\right)m_W^2m_Z^2(2Q_c-1)s_W^2
  +\strut\nonumber\\&&\strut
  -(m_W^2-m_b^2+m_c^2)(\lambda'+6m_W^2m_Z^2s_W^2)m_Z^2(2Q_b+1)(2Q_c-1)s_W^2
  \Big\}
+\strut\nonumber\\&&\strut\kern-12pt
  +\frac{m_c^2(\ln(m_c/\bar\mu)-1)}{\lambda'm_b^2s_W^2}
  \Big\{2m_W^4-m_c^2(m_W^2-m_b^2+m_c^2)+2m_b^2m_Z^2(2Q_b+1)(2Q_c-1)s_W^2\Big\}
+\strut\nonumber\\&&\strut\kern-12pt
  +\frac{\ln(m_b/\bar\mu)-1}{\lambda's_W^2}\Big\{2m_b^2m_W^2
  +\strut\nonumber\\&&\strut
  +(m_W^2+m_b^2)(m_W^2-m_c^2+m_b^2)
  +(m_W^2-m_c^2-m_b^2)m_Z^2(2Q_b+1)(2Q_c-1)s_W^2\Big\}
+\strut\nonumber\\&&\strut\kern-12pt
  +\frac{m_W^2(\ln(m_W/\bar\mu)-1)}{\lambda'm_b^2s_W^2}
  \Big\{\lambda'-3m_b^2(m_W^2-m_c^2+m_b^2)-3(m_W^2-m_c^2-m_b^2)m_W^2\Big\}
+\strut\nonumber\\&&\strut\kern-12pt
  -\frac{m_Z^2(2\ln(m_Z/\bar\mu)-1)}{4\lambda'm_b^2s_W^2}
  \Big\{m_b^2(m_W^2-m_b^2+m_c^2)+\strut\nonumber\\&&\strut
  +4(m_W^2-m_c^2-m_b^2)m_W^2(2Q_bs_W^2+1)-8m_b^2m_W^2(2Q_cs_W^2-1)
  +\strut\nonumber\\&&\strut
  +2(m_W^2-m_c^2+m_b^2)m_Z^2(2Q_bs_W^2+1)(2Q_cs_W^2-1)\Big\}
+\strut\nonumber\\&&\strut\kern-12pt
  -\frac{m_H^2(2\ln(m_H/\bar\mu)-1)}{4\lambda's_W^2}(m_W^2-m_b^2+m_c^2)
  -\frac{2m_W^2+m_c^2}{2m_b^2s_W^2},
\end{eqnarray}
where $\lambda'=\lambda(m_W^2,m_c^2,m_b^2)$. $\bar\mu$ is the $\overline{\rm
MS}$ renormalisation scale which for the decay process considered here is set
to be the mass of the $W$ boson. The IR subtracted main form factor $V_-^*$
contains also the UV finite parts of the counter terms which are listed in
Appendix~C. MATHEMATICA output formulas for these lengthy expressions can be
obtained from the authors on request, though they can be generated also by
e.g.\ {\tt FeynArts} or {\tt FormCalc}.

\section{Renormalisation counter terms}
\setcounter{equation}{0}\def\theequation{C\arabic{equation}}
In the electroweak theory with its multiple parameters depending on each
other, it is essential to decide which of the parameters are independent. In
the so-called $\alpha(0)$ scheme, the coupling and the masses of the particles
are used as such. However, depending on the energy of the process, more and
more fermion loops from the photon and $Z$ boson vacuum polarisations can be
resummed to the coupling, so that $\alpha(0)$ is changed to e.g.\
$\alpha(m_W^2)$. This is shown in Sec.~8.2.1 of Ref.~\cite{Denner:1991kt}.
The renormalisation counter terms are defined by
\begin{eqnarray}
\delta Z_e&=&\frac12\Pi_{AA}^T(m_A^2)-\frac{s_W}{c_W}
  \frac{\Sigma_{AZ}^T(m_A^2)}{m_Z^2}-\frac12\Delta r,\qquad
  \Pi_{AA}^T(m_A^2)=\frac{\partial\Sigma_{AA}^T(k^2)}{\partial k^2}
  \Big|_{k^2=m_A^2}\nonumber\\
\frac{\delta s_W}{s_W}&=&-\frac{c_W^2}{2s_W^2}
  \left(\frac{\Sigma_{WW}^T(m_W^2)}{m_W^2}-\frac{\Sigma_{ZZ}^T(m_Z^2)}{m_Z^2}
  \right),\nonumber\\
\delta Z_{WW}&=&-\frac12\frac{\partial\Sigma_{WW}^T(k^2)}{\partial k^2}
  \Big|_{k^2=m_W^2},\nonumber\\
\delta Z_{ii}^L&=&-\frac12\Sigma_{ii}^{bL}(m_i^2)
  -m_i\left(m_i\Sigma_{ii}^{bL\prime}(m_i^2)+\Sigma_{ii}^{br\prime}(m_i^2)
  \right),\nonumber\\
\delta Z_{ij}^L&=&\frac1{m_i^2-m_j^2}\Big(m_j^2\Sigma_{ij}^{bL}(m_j^2)
  +m_im_j\Sigma_{ij}^{bR}(m_j^2)+m_i\Sigma_{ij}^{bl}(m_j^2)
  +m_j\Sigma_{ij}^{br}(m_j^2)\Big),\qquad
\end{eqnarray}
where the choice of $\Delta r$ that is absorbed into the coupling
$\alpha=\alpha(0)/(1-\Delta r)$ determines the scheme. While $\Delta r=0$
for the $\alpha(0)$ scheme, in the $\alpha(m_W^2)$ scheme $\Delta r$ contains
the light fermion loops.

In the $\alpha(0)$ scheme where $\Delta r=0$, the UV finite parts of the UV
counter terms that are used for the IR subtracted vertex correction $V_-^*$ in
Eq.~(\ref{V0m}) read
\begin{eqnarray}
\lefteqn{\delta Z_e^f\ =\ -\frac13\Bigg\{1-\frac{21}2\left(1-A_f(m_W)\right)
  +2\sum_fQ_f^2\left(1-A_f(m_f)\right)\Bigg\},}\nonumber\\
\lefteqn{\frac{\delta s_W^f}{s_W}\ =\ \frac{-1}{72m_W^2m_Z^6s_W^4}\Bigg\{
  4m_W^2(m_W^2-m_Z^2)(36m_W^4+24m_W^2m_Z^2+m_Z^4)+\strut}\nonumber\\&&\strut
  +3(4m_W^2-m_Z^2)(12m_W^4+20m_W^2m_Z^2+m_Z^4)
  \times\strut\nonumber\\&&\strut\qquad\times
  \left(m_W^2B_f(m_Z^2;m_W,m_W)-m_W^2A_f(m_W)
  -m_Z^2B_f(m_W^2;m_W,m_Z)+m_Z^2A_f(m_Z)\right)
  +\strut\kern-8pt\nonumber\\&&\strut
  +3m_Z^4\Big((12m_W^4-4m_W^2m_H^2+m_H^4)B_f(m_W^2;m_W,m_H)
  -m_W^2m_H^2(A_f(m_H)-A_f(m_W))\Big)+\strut\kern-11pt\nonumber\\&&\strut
  -3m_W^2m_Z^2\Big((12m_Z^4-4m_Z^2m_H^2+m_H^4)B_f(m_Z^2;m_Z,m_H)
  -m_Z^2m_H^2(A_f(m_H)-A_f(m_Z))\Big)+\strut\kern-14pt\nonumber\\[3pt]&&\strut
  +144m_W^4m_Z^2(m_W^2-m_Z^2)B_f(m_W^2,m_W,m_A)
  +6m_W^4m_Z^2(42m_W^2-m_Z^2)(A_f(m_W)-A_f(m_Z))
  +\kern-33pt\strut\nonumber\\[7pt]&&\strut
  +6m_W^2m_Z^2(m_W^2-m_Z^2)(18m_W^2-5m_Z^2)A_f(m_Z)
  +3(m_W^2-m_Z^2)m_Z^2m_H^4A_f(m_H)+\strut\nonumber\\[7pt]&&\strut\kern-12pt
  -m_Z^4\sum_i\Big[4m_W^2(m_W^2-3m_{\nu_i}^2-3m_{\ell_i}^2)
  +\strut\nonumber\\&&\strut
  +6\left((m_{\nu_i}^2-m_{\ell_i}^2)^2
  +(m_{\nu_i}^2+m_{\ell_i}^2-2m_W^2)m_W^2\right)
  B_f(m_W^2;m_{\nu_i},m_{\ell_i})+\strut\nonumber\\&&\strut
  -6(m_{\nu_i}^2-m_{\ell_i}^2)\left(m_{\nu_i}^2A_f(m_{\nu_i})
  -m_{\ell_i}^2A_f(m_{\ell_i})\right)+12m_W^2\left(m_{\nu_i}^2A_f(m_{\nu_i})
  +m_{\ell_i}^2A_f(m_{\ell_i})\right)\Big]+\strut\nonumber\\&&\strut\kern-12pt
  -m_Z^4\sum_{i,j}|V_{ij}|^2\Big[4m_W^2(m_W^2-3m_i^2-3m_j^2)
  +\strut\nonumber\\&&\strut
  +6\left((m_i^2-m_j^2)^2+(m_i^2+m_j^2-2m_W^2)m_W^2\right)
  B_f(m_W^2;m_i,m_j)+\strut\nonumber\\&&\strut
  -6(m_i^2-m_j^2)\left(m_i^2A_f(m_i)-m_j^2A_f(m_j)\right)
  +12m_W^2\left(m_i^2A_f(m_i)+m_j^2A_f(m_j)\right)\Big]
  +\strut\nonumber\\&&\strut\kern-12pt
  -8m_W^4(m_W^2-m_Z^2)\sum_f\Big[(g_f^{-2}+g_f^{+2})
  \left(m_Z^2-6m_f^2+6m_f^2A_f(m_f)\right)+\strut\nonumber\\&&\strut
  +3\left((g_f^{-2}+g_f^{+2}-6g_f^-g_f^+)m_f^2
  -(g_f^{-2}+g_f^{+2})m_Z^2\right)B_f(m_Z^2;m_f,m_f)\Big]\Bigg\},\nonumber\\
\lefteqn{\delta Z_{WW}^f\ =\ \frac1{72m_W^4m_Z^2s_W^2}\Bigg\{4m_W^4m_Z^2
    +\strut}\nonumber\\&&\strut
  +3(48m_W^6-16m_W^4m_Z^2+6m_W^2m_Z^4+m_Z^6)B_f(m_W^2;m_W,m_Z)
  +\strut\nonumber\\&&\strut
  -3m_Z^2(2m_W^2-m_H^2)m_H^2B_f(m_W^2;m_W,m_H)
  -144m_W^4(m_W^2-m_Z^2)B_f(m_W^2;m_W,m_A)+\strut\nonumber\\&&\strut
  +3m_W^2(4m_W^2-m_Z^2)(12m_W^4+20m_W^2m_Z^2+m_Z^4)B'_f(m_W^2;m_W,m_Z)
  +\strut\nonumber\\&&\strut
  -3m_W^2m_Z^2(12m_W^4-4m_W^2m_H^2+m_H^4)B'_f(m_W^2;m_W,m_H)
  +\strut\nonumber\\&&\strut
  +144m_W^6m_Z^2s_W^2B'_f(m_W^2;m_W,m_A)
  -3m_W^2m_Z^2(2m_W^2-m_Z^2-m_H^2)A_f(m_W)+\strut\nonumber\\&&\strut
  +3m_Z^2(m_W^2-m_Z^2)(8m_W^2+m_Z^2)A_f(m_Z)
  +3m_Z^2(m_W^2-m_H^2)m_H^2A_f(m_H)+\strut\nonumber\\&&\strut\kern-12pt
  -m_Z^2\sum_i
  \Big[3\Big(2\left((m_{\nu_i}^2-m_{\ell_i}^2)^2+2m_W^4\right)
  B_f(m_W^2;m_{\nu_i},m_{\ell_i})+\strut\nonumber\\&&\strut\qquad
  -2m_W^2\left((m_{\nu_i}^2-m_{\ell_i}^2)^2
  +(m_{\nu_i}^2+m_{\ell_i}^2-2m_W^2)m_W^2\right)
  B'_f(m_W^2;m_{\nu_i},m_{\ell_i})+\strut\nonumber\\&&\strut\qquad
  -2(m_{\nu_i}^2-m_{\ell_i}^2)(m_{\nu_i}^2A_f(m_{\nu_i})
  -m_{\ell_i}^2A_f(m_{\ell_i}))\Big)-4m_W^4\Big]
  +\strut\nonumber\\&&\strut\kern-12pt
  -m_Z^2\sum_{i,j}|V_{ij}|^2
  \Big[3\Big(2\left((m_i^2-m_j^2)^2+2m_W^4\right)B_f(m_W^2;m_i,m_j)
  +\strut\nonumber\\&&\strut\qquad
  -2m_W^2\left((m_i^2-m_j^2)^2+(m_i^2+m_j^2-2m_W^2)m_W^2\right)
  B'_f(m_W^2;m_i,m_j)+\strut\nonumber\\&&\strut\qquad
  -2(m_i^2-m_j^2)(m_i^2A_f(m_i)-m_j^2A_f(m_j))\Big)
  -4m_W^4\Big]\Big\},\nonumber\\
\lefteqn{\delta Z^{Lf}_{cc}\ =\ \frac{-1}{16m_c^2m_W^2s_W^2}\Bigg\{2m_c^4
  +2m_c^2\left(4(Q_c^2+g_c^{-2})m_W^2s_W^2+m_c^2\right)(A_f(m_c)-1)
  +\strut}\nonumber\\&&\strut
  +m_Z^2(8g_c^{-2}m_W^2s_W^2+m_c^2)\left(B_f(m_c^2;m_c,m_Z)-A_f(m_Z)\right)
  +\strut\nonumber\\&&\strut
  +m_c^2m_H^2\left(B_f(m_c^2;m_c,m_H)-A_f(m_H)\right)
  +\strut\nonumber\\&&\strut
  +2m_c^2\left(8g_c^{-2}(2m_c^2-m_Z^2)m_W^2s_W^2
  -32g_c^-g_c^+m_c^2m_W^2s_W^2-m_c^2m_Z^2\right)B'_f(m_c^2;m_c,m_Z)
  +\strut\nonumber\\&&\strut
  +2m_c^4(4m_c^2-m_H^2)B'_f(m_c^2;m_c,m_H)
  -32m_c^4m_W^2s_W^2Q_c^2B'_f(m_c^2;m_c,m_A)
  +\strut\nonumber\\&&\strut\kern-12pt
  +\sum_k|V_{ck}|^2\Big[-4m_c^2m_W^2+2(m_k^2+2m_W^2)
  \times\strut\nonumber\\&&\strut\qquad\times\left((m_c^2-m_k^2+m_W^2)
  B_f(m_c^2;m_k,m_W)+m_k^2A_f(m_k)-m_W^2A_f(m_W)\right)
  +\strut\nonumber\\&&\strut
  -4m_c^2(m_c^2m_k^2-m_k^4-2m_c^2m_W^2-m_k^2m_W^2+2m_W^4)B'_f(m_c^2;m_k,m_W)
  \Big]\Bigg\},\nonumber\\
\lefteqn{\delta Z^{Lf}_{bb}\ =\ \frac{-1}{16m_b^2m_W^2s_W^2}\Bigg\{2m_b^4
  +2m_b^2\left(4(Q_b^2+g_b^{-2})m_W^2s_W^2+m_b^2\right)(A_f(m_b)-1)
  +\strut}\nonumber\\&&\strut
  +m_Z^2(8g_b^{-2}m_W^2s_W^2+m_b^2)\left(B_f(m_b^2;m_b,m_Z)-A_f(m_Z)\right)
  +\strut\nonumber\\&&\strut
  +m_b^2m_H^2\left(B_f(m_b^2;m_b,m_H)-A_f(m_H)\right)
  +\strut\nonumber\\&&\strut
  +2m_b^2\left(8g_b^{-2}(2m_b^2-m_Z^2)m_W^2s_W^2
  -32g_b^-g_b^+m_b^2m_W^2s_W^2-m_b^2m_Z^2\right)B'_f(m_b^2;m_b,m_Z)
  +\strut\nonumber\\&&\strut
  +2m_b^4(4m_b^2-m_H^2)B'_f(m_b^2;m_b,m_H)
  -32m_b^4m_W^2s_W^2Q_b^2B'_f(m_b^2;m_b,m_A)
  +\strut\nonumber\\&&\strut\kern-12pt
  +\sum_k|V_{kb}|^2\Big[-4m_b^2m_W^2+2(m_k^2+2m_W^2)
  \times\strut\nonumber\\&&\strut\qquad\times\left((m_b^2-m_k^2+m_W^2)
  B_f(m_b^2;m_k,m_W)+m_k^2A_f(m_k)-m_W^2A_f(m_W)\right)
  +\strut\nonumber\\&&\strut
  -4m_b^2(m_b^2m_k^2-m_k^4-2m_b^2m_W^2-m_k^2m_W^2+2m_W^4)B'_f(m_b^2;m_k,m_W)
  \Big]\Bigg\},\nonumber\\
\lefteqn{\delta Z^{Lf}_{ij}\ =\ \frac{-1}{8m_W^2s_W^2}\sum_kV_{ik}V_{kj}
    \Bigg\{m_k^2A_f(m_k)-m_W^2A_f(m_W)-2m_W^2+\strut}\nonumber\\&&\strut
  +\frac{B_f(m_i^2;m_k,m_W)}{m_i^2-m_j^2}
  \left((m_i^2-m_k^2)(m_j^2-m_k^2)+(2m_i^2-m_j^2+m_k^2-2m_W^2)m_W^2\right)
  +\strut\nonumber\\&&\strut
  +\frac{B_f(m_j^2;m_k,m_W)}{m_j^2-m_i^2}
  \left((m_j^2-m_k^2)(m_i^2-m_k^2)+(2m_j^2-m_i^2+m_k^2-2m_W^2)m_W^2\right)
  \Bigg\}.
\end{eqnarray}
Here, $\delta Z_e$ and $\delta s_W/s_W$ are the counter terms for the electric
charge and the sine of the Weinberg angle. $\delta Z_{WW}$, $\delta Z^L_{cc}$
and $\delta Z^L_{bb}$ are counter terms for the renormalisation of the wave
functions of $W$ boson, left handed charm and bottom quarks, respectively.
For the CKM matrix we need flavour changing wave function counter terms
$\delta Z^L_{ij}$. We have checked all these results against the results given
in Ref.~\cite{Denner:1991kt} and found agreement.

In order to save the symmetry, the expressions presented here contain also
vanishing contributions from e.g.\ the neutrinos. However, in order to handle
two-point functions containing the neutrino masses appropriately, a couple of
limiting cases has to be calculated. Calculating $B_f(m_W^2;m_\nu,m_\ell)$ and
the derivative $B'_f(m_W^2;m_\nu,m_\ell)$, we start with the expansion of the
square root of K\"all\'en function for this particular mass configuration,
\begin{equation}
\sqrt{\lambda(m_W^2,m_\nu^2,m_\ell^2)}=m_W^2-m_\ell^2
  -\frac{p^2+m_\ell^2}{p^2-m_\ell^2}m_\nu^2+O(m_\nu^4).
\end{equation}
The finite part of the two-point function
$B(p^2;m_1,m_2)=i\bar\mu^{-2\eps}/(4\pi)^2(1/\eps+B_f(p^2;m_1,m_2))$
(for simplicity, $\bar\mu$ is taken to be the $\overline{\rm MS}$ scale) is
written as (cf.\ (B.1) in Ref.~\cite{Bohm:1986rj})
\begin{eqnarray}
\lefteqn{B_f(p^2;m_1,m_2)\ =\ 2-\frac12\left(\ln\pfrac{m_1^2}{\bar\mu^2}
  +\ln\pfrac{m_2^2}{\bar\mu^2}\right)
  -\frac{m_1^2-m_2^2}{2p^2}\ln\pfrac{m_1^2}{m_2^2}+\frac1{p^2}\times\strut}
  \nonumber\\&&\strut\kern-24pt
  \cases{\displaystyle-\sqrt{(m_1+m_2)^2-p^2}\sqrt{(m_1-m_2)^2-p^2}
  \ln\pfrac{\sqrt{(m_1+m_2)^2-p^2}-\sqrt{(m_1-m_2)^2-p^2}}
  {\sqrt{(m_1+m_2)^2-p^2}+\sqrt{(m_1-m_2)^2-p^2}}\cr
  \displaystyle-2\sqrt{(m_1+m_2)^2-p^2}\sqrt{p^2-(m_1-m_2)^2}
  \arctan\pfrac{\sqrt{p^2-(m_1-m_2)^2}}{\sqrt{(m_1+m_2)^2-p^2}}\cr
  \displaystyle\sqrt{p^2-(m_1+m_2)^2}\sqrt{p^2-(m_1-m_2)^2}\left(
  \ln\pfrac{\sqrt{p^2-(m_1+m_2)^2}-\sqrt{p^2-(m_1-m_2)^2}}
  {\sqrt{p^2-(m_1+m_2)^2}+\sqrt{p^2-(m_1-m_2)^2}}+i\pi\right)\cr}\kern-17pt
\end{eqnarray}
in the cases $p^2<(m_1-m_2)^2$, $(m_1-m_2)^2<p^2<(m_1+m_2)^2$, and
$p^2>(m_1+m_2)^2$, respectively. The derivative of this finite part with
respect to $p^2$ results in
\begin{eqnarray}
\lefteqn{B'_f(p^2;m_1,m_2)\ =\ \frac{m_1^2-m_2^2}{2p^4}\ln\pfrac{m_1^2}{m_2^2}
  -\frac1{p^2}+\frac1{p^4}\times\strut}\nonumber\\&&\strut\kern-24pt
  \cases{\displaystyle\frac{(m_1^2-m_2^2)^2-(m_1^2+m_2^2)p^2}{\sqrt{(m_1+m_2)^2
  -p^2}\sqrt{(m_1-m_2)^2-p^2}}\ln\pfrac{\sqrt{(m_1+m_2)^2-p^2}
  -\sqrt{(m_1-m_2)^2-p^2}}{\sqrt{(m_1+m_2)^2-p^2}+\sqrt{(m_1-m_2)^2-p^2}}\cr
  \displaystyle-2\frac{(m_1^2-m_2^2)^2-(m_1^2+m_2^2)p^2}{\sqrt{(m_1+m_2)^2
  -p^2}\sqrt{p^2-(m_1-m_2)^2}}\arctan\pfrac{\sqrt{p^2-(m_1-m_2)^2}}
  {\sqrt{(m_1+m_2)^2-p^2}}\cr
  \displaystyle-\frac{(m_1^2-m_2^2)^2-(m_1^2+m_2^2)p^2}{\sqrt{p^2-(m_1+m_2)^2}
  \sqrt{p^2-(m_1-m_2)^2}}\left(\ln\pfrac{\sqrt{p^2-(m_1+m_2)^2}
  -\sqrt{p^2-(m_1-m_2)^2}}{\sqrt{p^2-(m_1+m_2)^2}+\sqrt{p^2-(m_1-m_2)^2}}
  +i\pi\right)\cr}\kern-28pt
\end{eqnarray}
in the same cases. The power series expansions are given by
\begin{eqnarray}
B_f(m_W^2;m_\nu,m_\ell)&=&2-\frac{m_\ell^2}{m_W^2}\ln\pfrac{m_\ell^2}{\bar
  \mu^2}-\frac{m_W^2-m_\ell^2}{m_W^2}\left(\ln\pfrac{m_W^2-m_\ell^2}{\bar\mu^2}
  -i\pi\right)+O(m_\nu^2),\nonumber\\
B'_f(m_W^2;m_\nu,m_\ell)&=&-\frac1{m_W^2}+\frac{m_\ell^2}{m_W^4}
  \ln\pfrac{m_\ell^2}{\bar\mu^2}-\frac{m_\ell^2}{m_W^4}
  \left(\ln\pfrac{m_W^2-m_\ell^2}{\bar\mu^2}-i\pi\right)+O(m_\nu^2).\nonumber\\
\end{eqnarray}
This is sufficient for the parts coming from the $W$ boson self energy.
However, for the parts from the $Z$ boson self energy in $\delta s_W$ we have
to expand $B_f(m_Z^2;m_\nu,m_\nu)$ and $B'_f(m_Z^2;m_\nu,m_\nu)$ in $m_\nu$.
With
\begin{equation}
\sqrt{\lambda(m_Z^2,m_\nu^2,m_\nu^2)}=\sqrt{m_Z^2(m_Z^2-4m_\nu^2)}
  =m_Z^2-2m_\nu^2-\frac{2m_\nu^4}{m_Z^2}+O(m_\nu^6)
\end{equation}
one obtains
\begin{eqnarray}
  B_f(m_Z^2;m_\nu,m_\nu)&=&2-\ln\pfrac{m_\nu^2}{\bar\mu^2}
  +\sqrt{1-\frac{4m_\nu^2}{m_Z^2}}
  \left(\ln\pfrac{1-\sqrt{1-4m_\nu^2/m_Z^2}}{1+\sqrt{1-4m_\nu^2/m_Z^2}}+i\pi
  \right),\\
  B'_f(m_Z^2;m_\nu,m_\nu)&=&-\frac1{m_Z^2}
  +\frac{2m_\nu^2}{m_Z^4\sqrt{1-4m_\nu^2/m_Z^2}}
  \left(\ln\pfrac{1-\sqrt{1-4m_\nu^2/m_Z^2}}{1+\sqrt{1-4m_\nu^2/m_Z^2}}+i\pi
  \right).\qquad
\end{eqnarray}
For the $G_\mu$ scheme dealt with later in this Appendix, we have to calculate
$B_f(0;m_1,m_2)$, As $p^2=0$, we have to use the first case. Without loss of
generality, we assume that $m_1^2>m_2^2$ (the opposite inequality will lead to
the same result). We have
\begin{equation}
\sqrt{\lambda(p^2,m_1^2,m_2^2)}=\sqrt{(m_1^2-m_2^2)^2-2p^2(m_1^2+m_2^2)+p^4}
  =m_1^2-m_2^2-\frac{m_1^2+m_2^2}{m_1^2-m_2^2}p^2+O(p^4).
\end{equation}
In this case $p^2<(m_1-m_2)$. Therefore
\begin{eqnarray}
B_f(p^2;m_1,m_2)&=&2-\frac12\left(\ln\pfrac{m_1^2}{\bar\mu^2}
  +\ln\pfrac{m_2^2}{\bar\mu^2}\right)-\frac{m_1^2-m_2^2}{2p^2}
  \ln\pfrac{m_1^2}{m_2^2}+\strut\nonumber\\&&\strut
  -\frac1{2p^2}\left(m_1^2-m_2^2-p^2\frac{m_1^2+m_2^2}{m_1^2-m_2^2}\right)
  \left(\ln\pfrac{m_2^2}{m_1^2}+\frac{2p^2}{m_1^2-m_2^2}\right)\ =\nonumber\\
  &=&1-\frac1{m_1^2-m_2^2}\left(m_1^2\ln\pfrac{m_1^2}{\bar\mu^2}
  -m_2^2\ln\pfrac{m_2^2}{\bar\mu^2}\right)+O(p^2)
\end{eqnarray}
and
\begin{equation}
B_f(0;m_1,m_2)=\frac{m_1^2A_f(m_1)-m_2^2A_f(m_2)}{m_1^2-m_2^2},
\end{equation}
were the finite part $A_f(m)$ of the one-point function is defined via
\begin{equation}
A(m)=\frac{im^2\bar\mu^{-2\eps}}{(4\pi)^2}\left(\frac1\eps+A_f(m)\right),\qquad
  A_f(m)=1-\ln\pfrac{m^2}{\bar\mu^2}.
\end{equation}
The same approximation we need for equal masses. Starting with
\begin{equation}
\sqrt{\lambda(p^2,m^2,m^2)}=\sqrt{(4m^2-p^2)p^2}
  =\sqrt{4m^2p^2}\sqrt{1-\frac{p^2}{4m^2}}
  =p^2\sqrt{\frac{4m^2}{p^2}}\left(1-\frac{p^2}{8m^2}\right)+O(p^2),
\end{equation}
with a short Taylor series expansion one obtains
\begin{equation}
B_f(p^2;m,m)=2-\ln\pfrac{m^2}{\bar\mu^2}-2\sqrt{\frac{4m^2}{p^2}}
  \arctan\left(\sqrt{\frac{p^2}{4m^2}}\right)+O(p^2)
  =-\ln\pfrac{m^2}{\bar\mu^2}+O(p^2)
\end{equation}
and, therefore, $B_f(0;m,m)=A_f(m)-1$. With a longer expansion one has
\begin{eqnarray}
\lefteqn{B'_f(m_A^2;m,m)\ =\ -\frac1{p^2}+\frac{4m^2}{p^2\sqrt{(4m^2-p^2)p^2}}
    \arctan\left(\sqrt{\frac{p^2}{4m^2-p^2}}\right)\ =}\nonumber\\
  &=&-\frac1{p^2}+\frac1{p^2}\sqrt{\frac{4m^2}{p^2}}
  \left(1+\frac{p^2}{8m^2}\right)\sqrt{\frac{p^2}{4m^2}}
  \left(1+\frac{p^2}{24m^2}\right)+O(p^2)\ =\ \frac1{6m^2}+O(p^2)\qquad
\end{eqnarray}
and $B'_f(0;m,m)=1/(6m^2)$.

In the $\alpha(m_W^2)$ scheme, all light fermion loops occuring in the counter
terms $\delta Z_e$ and $\delta s_W/s_W$ are resummed into the coupling
constant, as explained in Refs.~\cite{Denner:1991kt,Bohm:2001,Denner:2019vbn}.

\subsection{Applying the $G_\mu$ scheme}
In the quite detailed review article in Ref.~\cite{Denner:2019vbn} is is
argued that for processes involving the electroweak vector bosons, the $G_F$
(or $G_\mu$) scheme based on the Fermi constant (in muon decay) and discussed
in Ref.~\cite{Consoli:1989fg} is preferable. Therefore, for our work we will
deal with this scheme. From Eqs.~(426) and~(423) of Ref.~\cite{Denner:2019vbn}
we deduce that
\begin{eqnarray}\label{delZeGmu}
\delta Z_e\Big|_{G_\mu}&=&\frac12\Pi_{AA}^T(m_A^2)-\frac{s_W}{c_W}
  \frac{\Sigma_{AZ}^T(m_A^2)}{m_Z^2}-\frac12\Delta r\Big|_{G_\mu}\ =\nonumber\\
  &=&\frac12\Bigg(\frac{c_W^2}{s_W^2}\left(\frac{\Sigma_{ZZ}^T(m_Z^2)}{m_Z^2}
  -\frac{\Sigma_{WW}^T(m_W^2)}{m_W^2}\right)-\frac{\Sigma_{WW}^T(0)
    -\Sigma_{WW}^T(m_W^2)}{m_W^2}+\strut\nonumber\\&&\strut
  -\frac{2\Sigma_{AZ}^T(m_A^2)}{s_Wc_Wm_Z^2}
  -\frac{\alpha(0)}{4\pi s_W^2}\left(6+\frac{7-4s_W^2}{2s_W^2}
  (A_f(m_Z)-A_f(m_W)\right)\Bigg)\qquad
\end{eqnarray}
(cf.\ also Eq.~(4.6.11) in Ref.~\cite{Bohm:2001}).
For the self energy of the $W$ boson at $p^2=0$ one obtains
\begin{eqnarray}
\lefteqn{\Sigma_{WW}^T(0)\ =\ \frac\alpha{4\pi}\Bigg[
  \frac{-1}{2m_W^2s_W^2\eps}\left(4m_W^2-2m_Z^2
  +\sum_i(m_{\nu_i}^2+m_{\ell_i}^2)+N_c\sum_{i,j}|V_{ij}|^2(m_i^2+m_j^2)\right)
  +\strut}\nonumber\\&&\strut
  -\frac1{6m_W^2s_W^2}\left(18m_W^2+m_Z^2+m_H^2
  -2\sum_i(m_{\nu_i}^2+m_{\ell_i}^2)-2N_c\sum_{i,j}|V_{ij}|^2(m_i^2+m_j^2)
  \right)+\strut\nonumber\\&&\strut
  -\frac{(14m_W^2+m_Z^2)m_W^2-(16m_W^2-m_Z^2)m_H^2}{3(m_H^2-m_W^2)m_Z^2s_W^4}
  A_f(m_W)+\strut\nonumber\\&&\strut
  -\frac{(4m_W^2-m_Z^2)(2m_W^2+3m_Z^2)}{3m_W^2m_Z^2s_W^4}A_f(m_Z)
  +\frac{2m_H^2A_f(m_H)}{3(m_H^2-m_W^2)s_W^2}+\strut\nonumber\\&&\strut
  -\frac1{3m_W^2s_W^2}\sum_i\left(m_{\nu_i}^2A_f(m_{\nu_i})+m_{\ell_i}^2
  A_f(m_{\ell_i})+\frac{m_{\nu_i}^2+m_{\ell_i}^2}{2(m_{\nu_i}^2-m_{\ell_i}^2)}
  \left(m_{\nu_i}^2A_f(m_{\nu_i})-m_{\ell_i}^2A_f(m_{\ell_i})\right)\right)
  +\strut\kern-28pt\nonumber\\&&\strut
  -\frac{N_c}{3m_W^2s_W^2}\sum_{ij}|V_{ij}|^2\left(m_i^2A_f(m_i)+m_j^2A_f(m_j)
  +\frac{m_i^2+m_j^2}{2(m_i^2-m_j^2)}\left(m_i^2A_f(m_i)-m_j^2A_f(m_j)\right)
  \right)\Bigg]+\strut\kern-32pt\nonumber\\&&\strut\kern-24pt
  +\frac\alpha{4\pi}\Bigg[\frac{-1}{4m_H^2m_W^2s_W^2\eps}\left(6(2m_W^4+m_Z^4)
  +(2m_W^2+m_Z^2)m_H^2+3m_H^4-8\sum_fm_f^4\right)+\strut\nonumber\\&&\strut
  +\frac1{4m_H^2m_W^2s_W^2}\Bigg(4(2m_W^4+m_Z^4)-2m_W^2(m_H^2+6m_W^2)A_f(m_W)
  +\strut\nonumber\\&&\strut\qquad\qquad\qquad
  -m_Z^2(m_H^2+6m_Z^2)A_f(m_Z)-3m_H^4A_f(m_H)+8\sum_fm_f^4A_f(m_f)\Bigg)\Bigg].
\end{eqnarray}
It can be shown that
\begin{eqnarray}
\Delta r\Big|_{G_\mu}&=&\Pi_{AA}^T(m_A^2)+\frac{2c_W}{s_W}
  \frac{\Sigma_{AZ}^T(m_A^2)}{m_Z^2}+\frac{\Sigma_{WW}^T(0)
    -\Sigma_{WW}^T(m_W^2)}{m_W^2}+\strut\nonumber\\&&\strut
  -\frac{c_W^2}{s_W^2}\left(\frac{\Sigma_{ZZ}^T(m_Z^2)}{m_Z^2}
  -\frac{\Sigma_{WW}^T(m_W^2)}{m_W^2}\right)
  +\frac\alpha{4\pi s_W^2}\left(6+\frac{7-4s_W^2}{2s_W^2}\ln c_W^2\right)
\end{eqnarray}
is UV-finite. This has been done by using $\sum_f(1-4I_f^3Q_f)=0$, which in
detail results in $Q_c-Q_b=Q_W$ for all quark and lepton generations. In
replacing the $\alpha(0)$-scheme result
$\delta Z_e=-\delta Z_{AA}-\delta Z_{ZA}s_W/c_W$ by Eq.~(\ref{delZeGmu}) means
that the contribution $-\delta Z_{AA}$ producing the large mass logarithms no
longer appears. We can use Eq.~(\ref{delZeGmu}) to see that in this new scheme
$\delta Z_e$ together with the counter term for the sine of the Weinberg angle
simplifies to
\begin{eqnarray}
\lefteqn{\left[-\frac{\delta s_W}{s_W}+\delta Z_e\right]_{G_\mu}
    \ =\ \frac{c_W^2}{2s_W^2}\left(\frac{\Sigma_{WW}^T(m_W^2)}{m_W^2}
    -\frac{\Sigma_{ZZ}^T(m_Z^2)}{m_Z^2}\right)+\strut}\nonumber\\&&\strut
  +\frac12\Bigg[\frac{c_W^2}{s_W^2}\left(\frac{\Sigma_{ZZ}^T(m_Z^2)}{m_Z^2}
  -\frac{\Sigma_{WW}^T(m_W^2)}{m_W^2}\right)-\frac{\Sigma_{WW}^T(0)
    -\Sigma_{WW}^T(m_W^2)}{m_W^2}+\strut\nonumber\\&&\strut
  -\frac{2\Sigma_{AZ}^T(m_A^2)}{s_Wc_Wm_Z^2}
  -\frac{\alpha(0)}{4\pi s_W^2}\left(6+\frac{7-4s_W^2}{2s_W^2}
  (A_f(m_Z)-A_f(m_W)\right)\Bigg]\ =\\
  &=&-\frac12\Bigg[\frac{2\Sigma_{AZ}^T(m_A^2)}{s_Wc_Wm_Z^2}
  +\frac{\Sigma_{WW}^T(0)-\Sigma_{WW}^T(m_W^2)}{m_W^2}
  +\frac{\alpha(0)}{4\pi s_W^2}\left(6+\frac{7-4s_W^2}{2s_W^2}
  (A_f(m_Z)-A_f(m_W)\right)\Bigg],\kern-21pt\nonumber
\end{eqnarray}
which means that in this scheme, for practical reasons, we can set
\begin{eqnarray}
  \frac{\delta s_W}{s_W}\Big|_{G_\mu}&=&\frac12\Bigg[
  \frac{\Sigma_{WW}^T(0)-\Sigma_{WW}^T(m_W^2)}{m_W^2}
  +\frac{\alpha(0)}{4\pi s_W^2}\left(6+\frac{7-4s_W^2}{2s_W^2}
  (A_f(m_Z)-A_f(m_W)\right)\Bigg],\nonumber\\
  \delta Z_e|_{G_\mu}&=&-\frac{\Sigma_{AZ}^T(m_A^2)}{s_Wc_Wm_Z^2}
\end{eqnarray}
with UV singular contributions
\begin{eqnarray}
  \frac{\delta s_W^s}{s_W}\Big|_{G_\mu}
  =\frac{e^2}{12s_W^2\eps}\left(19-\sum_f1\right),\qquad
  \delta Z_e^s\Big|_{G_\mu}=-\frac{2e^2}{s_W^2\eps}
\end{eqnarray}
and UV finite contributions
\begin{eqnarray}
\lefteqn{\frac{\delta s_W^f}{s_W}\Big|_{G_\mu}=e^2\Bigg[
  \frac{-72m_W^2+182m_Z^2+\sum_fm_Z^2}{36m_Z^2s_W^2}+\strut}\nonumber\\&&\strut
  -\frac1{24(m_H^2-m_W^2)m_W^2m_Z^4s_W^4}\Big(m_W^2(48m_W^6-46m_W^4m_Z^2
  +17m_W^2m_Z^4-m_Z^6)+\strut\nonumber\\&&\strut\qquad\qquad\qquad
  -m_H^2(48m_W^6-39m_W^4m_Z^2+10m_W^2m_Z^4-m_Z^6)+m_H^4m_Z^4s_W^2\Big)A_f(m_W)
  +\strut\nonumber\\&&\strut
  +\frac{16m_W^6-51m_W^4m_Z^2+16m_W^2m_Z^4+m_Z^6}{24m_W^4m_Z^2s_W^4}A_f(m_Z)
  +\strut\nonumber\\&&\strut
  +(4m_W^2-m_Z^2)\frac{12m_W^4+20m_W^2m_Z^2+m_Z^4}{24m_W^4m_Z^2s_W^2}
  B_f(m_W^2;m_W,m_Z)+\strut\nonumber\\&&\strut
  +m_H^2\frac{11m_W^4-4m_W^2m_H^2+m_H^4}{24(m_H^2-m_W^2)m_W^4s_W^2}A_f(m_H)
  -\frac{12m_W^4-4m_W^2m_H^2+m_H^4}{24m_W^4s_W^2}B_f(m_W^2;m_W,m_H)
  +\strut\kern-8pt\nonumber\\&&\strut
  +\frac1{12m_W^4s_W^2}\sum_i
  \Bigg(\left((m_{\nu_i}^2-m_{\ell_i}^2)^2+(m_\mu^2+m_{\ell_i}^2)m_W^2-2m_W^4
  \right)B_f(m_W^2;m_{\nu_i},m_{\ell_i})
  +\strut\nonumber\\&&\strut\qquad\qquad\qquad\qquad
  -\left((m_{\nu_i}^2-m_{\ell_i}^2)^2+(m_{\nu_i}^2+m_{\ell_i}^2)m_W^2\right)
  \frac{m_{\nu_i}^2A_f(m_{\nu_i})-m_{\ell_i}^2A_f(m_{\ell_i})}{m_{\nu_i}^2
  -m_{\ell_i}^2}\Bigg)+\strut\nonumber\\&&\strut
  +\frac1{12m_W^4s_W^2}\sum_{i,j}|V_{ij}|^2
  \Bigg(\left((m_i^2-m_j^2)^2+(m_i^2+m_j^2)m_W^2-2m_W^4\right)
  B_f(m_W^2;m_i,m_j)+\strut\nonumber\\&&\strut\qquad\qquad\qquad\qquad
  -\left((m_i^2-m_j^2)^2+(m_i^2+m_j^2)m_W^2\right)
  \frac{m_i^2A_f(m_i)-m_j^2A_f(m_j)}{m_i^2-m_j^2}\Bigg)\Bigg],\\
\lefteqn{\delta Z_e^f\Big|_{G_\mu}
    \ =\ \frac{2e^2}{s_W^2}\left(1-A_f(m_W)\right).}
\end{eqnarray}
Together with
\begin{equation}
\delta Z_{WW}^{Ts}=\frac{e^2}{12s_W^2\eps}\left(19-\sum_f1\right)
\end{equation}
from above, one has
$\delta Z_{WW}^{Ts}-\delta s_W^s/s_W+\delta Z_e^s|_{G_\mu}=-2e^2/(s_W^2\eps)$
as for the other schemes, only that the first two UV singularities cancel
each other and the only singularity that is left is the one from
$\delta Z_e|_{G_\mu}$.

\section{Cross check for the unpolarised decay rate}
\setcounter{equation}{0}\def\theequation{D\arabic{equation}}
By integrating Eq.~(\ref{Gamma}) over the polar angle $\theta$, we cross-check
our result with the unpolarised result given in
Refs.~\cite{Denner:1990tx,Denner:1991kt}. One has
\begin{equation}
\Gamma=\Gamma_0^0(\rho_{++}+\rho_{00}+\rho_{--})(H^{++}+H^{00}+H^{--})
  =:\Gamma_0^0H,
\end{equation}
where
\begin{eqnarray}
H({\it tree\/})&=&-\Frac12\Big[Q_1^2+Q_2^2+Q_W^2\Big]\sla(2-\mu_1-\mu_1^2
  -\mu_2+2\mu_1\mu_2-\mu_2^2)\ell_\zeta+\strut\nonumber\\&&\strut\kern-24pt
  -\Frac12\Big[(1-\mu_2)Q_1^2-\mu_1(Q_2^2-Q_W^2)\Big]
  (2-\mu_1-\mu_1^2-\mu_2+2\mu_1\mu_2-\mu_2^2)(t_\zeta^\ell-2t_z^{\ell+})
  +\strut\nonumber\\&&\strut\kern-24pt
  +\Frac12\Big[Q_1^2-Q_2^2+(\mu_1-\mu_2)Q_W^2\Big]
  (2-\mu_1-\mu_1^2-\mu_2+2\mu_1\mu_2-\mu_2^2)
  (t_{\zeta W}^\ell+2t_{zW}^{\ell+})+\strut\nonumber\\&&\strut\kern-24pt
  -\Frac14\Big[(2+9\mu_1+4\mu_1^2-3\mu_2-6\mu_1\mu_2+2\mu_1^2\mu_2
  +3\mu_1\mu_2^2+\mu_2^3)Q_1^2+\strut\nonumber\\&&\strut
  +(4-2\mu_1-8\mu_1^2-\mu_1^3-6\mu_2+6\mu_1\mu_2+9\mu_1^2\mu_2-4\mu_1\mu_2^2
  +2\mu_2^3)Q_2^2+\strut\nonumber\\&&\strut
  +2(2-\mu_1+3\mu_1^2+\mu_1^3-3\mu_2+3\mu_1\mu_2-2\mu_1^2\mu_2-2\mu_1\mu_2^2
  +\mu_2^3)Q_W^2\Big]\ell_1+\strut\nonumber\\&&\strut\kern-24pt
  -\Frac14\Big[(2-15\mu_1-4\mu_1^2+2\mu_1^3+\mu_2+12\mu_1\mu_2-6\mu_1^2\mu_2
  -8\mu_2^2+6\mu_1\mu_2^2-2\mu_2^3)Q_1^2+\strut\nonumber\\&&\strut
  -(2+\mu_1-8\mu_1^2-2\mu_1^3-15\mu_2+12\mu_1\mu_2+6\mu_1^2\mu_2-4\mu_2^2
  -6\mu_1\mu_2^2+2\mu_2^3)Q_2^2+\strut\nonumber\\&&\strut
  -2(\mu_1-\mu_2)(2+3\mu_1+3\mu_2)Q_W^2\Big]\ell_{1W}
  +\strut\nonumber\\&&\strut\kern-24pt
  +\Frac1{24}\Big[3(22-37\mu_1-15\mu_1^2+3\mu_2+38\mu_1\mu_2-11\mu_2^2)Q_1^2
  +\strut\nonumber\\&&\strut
  +3(22+3\mu_1-11\mu_1^2-37\mu_2+38\mu_1\mu_2-15\mu_2^2)Q_2^2
  +\strut\nonumber\\&&\strut
  +2(44-43\mu_1-31\mu_1^2-43\mu_2+50\mu_1\mu_2-31\mu_2^2)Q_W^2\Big]\sla,
  \nonumber\\
H({\it loop\/})&=&\Frac12(2-\mu_1-\mu_1^2-\mu_2+2\mu_1\mu_2-\mu_2^2)\sla V_-
  +3\mu_1\mu_2\sla V_+-\Frac14\sqrt{\lambda^3}(\mu_1V_1+\mu_2V_2),
  \kern-6pt\nonumber\\
H(\alpha)&=&\Frac12(2-\mu_1-\mu_1^2-\mu_2+2\mu_1\mu_2-\mu_2^2)\sla V_-^*
  +3\mu_1\mu_2\sla V_+-\Frac14\sqrt{\lambda^3}(\mu_1V_1+\mu_2V_2)
  +\strut\kern-15pt\nonumber\\&&\strut\kern-24pt
  -\Frac12\Big[(1-\mu_2)Q_1^2-\mu_1(Q_2^2-Q_W^2)\Big]
  (2-\mu_1-\mu_1^2-\mu_2+2\mu_1\mu_2-\mu_2^2)(t_\zeta^{\ell*}-2t_z^{\ell+})
  +\strut\nonumber\\&&\strut\kern-24pt
  +\Frac12\Big[Q_1^2-Q_2^2+(\mu_1-\mu_2)Q_W^2\Big]
  (2-\mu_1-\mu_1^2-\mu_2+2\mu_1\mu_2-\mu_2^2)
  (t_{\zeta W}^{\ell*}+2t_{zW}^{\ell+})+\strut\nonumber\\&&\strut\kern-24pt
  -\Frac14\Big[(2+9\mu_1+4\mu_1^2-3\mu_2-6\mu_1\mu_2+2\mu_1^2\mu_2
  +3\mu_1\mu_2^2+\mu_2^3)Q_1^2+\strut\nonumber\\&&\strut
  +(4-2\mu_1-8\mu_1^2-\mu_1^3-6\mu_2+6\mu_1\mu_2+9\mu_1^2\mu_2-4\mu_1\mu_2^2
  +2\mu_2^3)Q_2^2+\strut\nonumber\\&&\strut
  +2(2-\mu_1+3\mu_1^2+\mu_1^3-3\mu_2+3\mu_1\mu_2-2\mu_1^2\mu_2-2\mu_1\mu_2^2
  +\mu_2^3)Q_W^2\Big]\ell_1+\strut\nonumber\\&&\strut\kern-24pt
  -\Frac14\Big[(2-15\mu_1-4\mu_1^2+2\mu_1^3+\mu_2+12\mu_1\mu_2-6\mu_1^2\mu_2
  -8\mu_2^2+6\mu_1\mu_2^2-2\mu_2^3)Q_1^2+\strut\nonumber\\&&\strut
  -(2+\mu_1-8\mu_1^2-2\mu_1^3-15\mu_2+12\mu_1\mu_2+6\mu_1^2\mu_2-4\mu_2^2
  -6\mu_1\mu_2^2+2\mu_2^3)Q_2^2+\strut\nonumber\\&&\strut
  -2(\mu_1-\mu_2)(2+3\mu_1+3\mu_2)Q_W^2\Big]\ell_{1W}
  +\strut\nonumber\\&&\strut\kern-24pt
  +\Frac1{24}\Big[3(22-37\mu_1-15\mu_1^2+3\mu_2+38\mu_1\mu_2-11\mu_2^2)Q_1^2
  +\strut\nonumber\\&&\strut
  +3(22+3\mu_1-11\mu_1^2-37\mu_2+38\mu_1\mu_2-15\mu_2^2)Q_2^2
  +\strut\nonumber\\&&\strut
  +2(44-43\mu_1-31\mu_1^2-43\mu_2+50\mu_1\mu_2-31\mu_2^2)Q_W^2\Big]\sla.
\end{eqnarray}
We find full agreement, where $\ln w_1+\ln w_\mu=\ell_{1W}$,
$\ln w_1-\ln w_\mu=\ell_1-\ell_{1W}$, and
\begin{equation}
8m_W^2(I_{01}+I_{02})=t_\zeta^\ell-2t_z^{\ell+},\qquad
8m_W^2I_{02}=t_{\zeta W}^\ell+2t_{zW}^{\ell+}.
\end{equation}
The result for $H({\it tree\/})$ turns out to coincide with Eq.~(35) in
Ref.~\cite{Denner:1990tx} and with Eq.~(9.16) in Ref.~\cite{Denner:1991kt}.
As the vertex corrections are already checked against Eqs.~(9.7)--(9.9) in
Ref.~\cite{Denner:1991kt}, we are able to claim that also $H({\it loop\/})$
and our formulas in general are consistent with the results for the decay of
the unpolarised $W$ decay as given in the literature. In addition, also
numerically we are consistent with previous results, as we could reproduce
the corresponding line of Table~9.2 in Ref.~\cite{Denner:1991kt} for the
parameters given in this reference.

\section{Adaption of the light quark masses}
\setcounter{equation}{0}\def\theequation{E\arabic{equation}}
According to Ref.~\cite{Denner:1990tx} and using techniques from
Ref.~\cite{Jegerlehner:1985gq}, the light quark masses are adapted to the
hadronic contribution to the vacuum polarisation. As the old value
$0.0286\pm 0.0007$~\cite{Jegerlehner:1985gq} is updated by a more recent one,
in Sec.~10 of Ref.~\cite{ParticleDataGroup:2024cfk} cited as
\begin{equation}
\Delta\alpha_{\rm had}^{(5)}(m_Z)=0.02783\pm 0.00006,
\end{equation}
a reiteration of the adaption procedure is in order at this point. According
to the Appendix~A of Ref.~\cite{Jegerlehner:1985gq}, written in a more compact
form, one has
\begin{equation}
\Delta\alpha_{\rm had}^{(5)}(m_Z)
  =-e^2N_c\sum_{i=1}^5Q_i^2s\Delta\pi_V^{q_iq_i}(m_Z^2)
\end{equation}
with
\begin{equation}
\Delta\pi_V^{q_1q_2}(q^2):=\pi_V^{q_1q_2}(q^2)-\pi_V^{q_1q_2}(0)
  =q^2\int_0^\infty\frac{\rho_V^{q_1q_2}(s)ds}{s(s-q^2-i\epsilon)},
\end{equation}
where the spectral density for $s\ge(m_1+m_2)^2$ is given by
\begin{equation}
\rho_V^{q_1q_2}(s)=\frac{\sqrt{\lambda(s,m_1^2,m_2^2)}}{12\pi^2s}\left[
  1-\frac{m_1^2+m_2^2}{2s}+3\frac{m_1m_2}s-\frac{(m_1^2-m_2^2)^2}{2s^2}
  \right].
\end{equation}
For $q_1=q_2$ the integration results in
\begin{equation}
\Delta\pi_V^{q_iq_i}(q^2)=\frac1{36\pi^2q^2}\left(5q^2+12m_i^2
  -3\frac{(q^2+2m_i^2)(q^2-4m_i^2)}{\sqrt{q^2(q^2-4m_i^2)}}
  \ln\pfrac{\sqrt{q^2}+\sqrt{q^2-4m_i^2}}{\sqrt{q^2}-\sqrt{q^2-4m_i^2}}\right).
\end{equation}
Using this formula, for the mass settings of Ref.~\cite{Denner:1990tx} given
by $m_u=m_d=0.041\GeV$, $m_s=0.15\GeV$, $m_c=1.5\GeV$ and $m_b=4.5\GeV$, the
result of $\Delta\alpha_{\rm had}^{(5)}(m_Z)=0.0285$ is reproduced within the
error bars. Using instead $m_u=m_d=0.06\GeV$ and $m_s=0.17\GeV$ and the values
for the heavy quarks from PDG~\cite{ParticleDataGroup:2024cfk}, one obtains
$\Delta\alpha_{\rm had}^{(5)}(m_Z)=0.027827$ that fits to the actual value.
Therefore, in the main text we use these adapted light quark mass values.

\end{appendix}

\end{document}